\documentclass[graybox, secnum]{svmult}

\usepackage{mathptmx}       % selects Times Roman as basic font
\usepackage{helvet}         % selects Helvetica as sans-serif font
\usepackage{courier}        % selects Courier as typewriter font
\usepackage{type1cm}        % activate if the above 3 fonts are
\usepackage{makeidx}         % allows index generation
\usepackage{graphicx}        % standard LaTeX graphics tool
\usepackage{multicol}        % used for the two-column index
\usepackage[bottom]{footmisc}% places footnotes at page bottom
\usepackage{hyperref}        %for hyperlinks
\usepackage{soul}            % for high-lighting of text
\hypersetup{colorlinks=true,urlcolor=blue}
\usepackage[square,numbers]{natbib}
\makeindex             % used for the subject index
\usepackage{latexsym}
\usepackage{amssymb}
\usepackage{amsfonts}
\usepackage{amsmath}
\usepackage{graphicx} 
\usepackage{indentfirst}
\usepackage{bbm}
\usepackage{verbatim}
\usepackage{mathrsfs}
\usepackage{dsfont}
\usepackage{xcolor}
\usepackage{ytableau}
\newcommand {\cD}{{\cal D}}
\newcommand {\cE}{{\cal E}}
\newcommand {\cF}{{\cal F}}

\newcommand {\cH}{{\cal H}}

\newcommand {\cK}{{\cal K}}
\newcommand {\cL}{{\cal L}}
\newcommand {\cM}{{\cal M}}
\newcommand {\cN}{{\cal N}}
\newcommand {\cO}{{\cal O}}

\newcommand {\cT}{{\cal T}}
\newcommand {\cU}{{\cal U}}

\newcommand {\cW}{{\cal W}}

\newcommand {\cZ}{{\cal Z}}

\def\a{\alpha}

\def\b{\beta}
\def\c{\chi}
\def\d{\delta}
\def\e{\epsilon}
\def\f{\phi}
\def\g{\gamma}
\def\G{\Gamma}

\def\j{\psi}
\def\k{\kappa}
\def\l{\lambda}
\def\m{\mu}

\def\o{\omega}
\def\p{\pi}
\def\q{\theta}
\def\r{\rho}
\def\s{\sigma}
\def\t{\tau}

\def\x{\xi}
\def\z{\zeta}
\def\D{\Delta}
\def\F{\Phi}
\def\J{\Psi}
\def\L{\Lambda}
\def\O{\Omega}

\def\S{\Sigma}
\def\U{\Upsilon}
\def\X{\Xi}

\def\rd{{\rm d}}
\def\ri{{\rm i}}
\def\re{{\rm e}}

\newcommand {\mfD}{\mathfrak{D}}
\newcommand {\mfDB}{\bar{\mathfrak{D}}}

\newcommand{\ad}{{\dot{\alpha}}}                           %new
\newcommand{\bd}{{\dot{\beta}}}                            %new
\newcommand{\ve}{\varepsilon}                            %new
\newcommand{\cDB}{{\bar\cD}}                            %new

\renewcommand{\aa}{{\a\ad}}

\newcommand{\pa}{\partial}                           %new
\newcommand{\hf}{\frac12}
\newcommand{\vf}{\varphi}
\newcommand{\be}{\begin{equation}}
\newcommand{\ee}{\end{equation}}
\newcommand{\bea}{\begin{eqnarray}}
\newcommand{\eea}{\end{eqnarray}}
\newcommand{\non}{\nonumber}
\newcommand{\1}{{\underline{1}}}
\newcommand{\2}{{\underline{2}}}

\def\double #1{#1{\hbox{\kern-2pt $#1$}}}

\newcommand{\gd}{{\dot\g}}
\newcommand{\dd}{{\dot\d}}

\renewcommand{\ts}{{\tilde{\s}}}

\newcommand{\sba}{{\bar{\s}}}

\newif\ifdtup

\def\de{{\nabla}}                                         % del
\def\deb{{\bar{\de}}}

\newcommand{\bsubeq}{\begin{subequations}}
\newcommand{\esubeq}{\end{subequations}}
\newcommand{\eol}{\notag \\}

\newcommand{\sSU}{\mathsf{SU}}
\newcommand{\sSL}{\mathsf{SL}}
\newcommand{\sGL}{\mathsf{GL}}
\newcommand{\sSO}{\mathsf{SO}}
\newcommand{\sU}{\mathsf{U}}

\newcommand{\scT}{{\mathscr{T}}}
\newcommand{\bbD}{{\mathbb {D}}}

\newcommand{\loco}{\vert}
\newcommand{\doubar}{{{\loco}\!{\loco}}}
\allowdisplaybreaks
\begin{document}
%\tableofcontents{}
\title*{Covariant superspace approaches to $\cN=2$ supergravity}
\author{S.
%Sergei 
M. Kuzenko,\thanks{corresponding author}
%Emmanouil 
E. S. N. Raptakis
 and 
 %Gabriele 
 G.
 Tartaglino-Mazzucchelli}
\institute{Sergei M. Kuzenko and Emmanouil S. N. Raptakis\at Department of Physics M013, The University of Western Australia 35 Stirling Highway, Perth W.A. 6009, Australia, \email{sergei.kuzenko@uwa.edu.au},
\email{emmanouil.raptakis@research.uwa.edu.au}
%\and Emmanouil S. N. Raptakis  \at Department of Physics M013, The University of Western Australia 35 Stirling Highway, Perth W.A. 6009, Australia, \email{emmanouil.raptakis@research.uwa.edu.au}
\and Gabriele Tartaglino-Mazzucchelli \at School of Mathematics and Physics, University of Queensland, St Lucia, Brisbane, Queensland 4072, Australia,
\email{g.tartaglino-mazzucchelli@uq.edu.au}}
\maketitle
\abstract{We provide a unified description of the three covariant superspace approaches to  ${\cal N}=2$ conformal supergravity in four dimensions: (i) conformal superspace; (ii) $\mathsf{U}(2)$ superspace; and (iii) $\mathsf{SU}(2)$ superspace. Each of them can be used 
to formulate general supergravity-matter systems, although conformal superspace has the largest structure group and is intimately related to the superconformal tensor calculus. We review the structure of covariant projective multiplets and demonstrate how they are used to describe pure and matter-coupled supergravity, including  locally superconformal off-shell sigma models. Higher-derivative invariants, topological invariants and super-Weyl anomalies are also briefly discussed. 
}
\section*{Keywords} 
Superconformal symmetry, Supergravity, Superspace, Projective multipets
\vspace{0.5cm}
\begin{flushright}
%{\it Dedicated to Daniel Butter who created Conformal Superspace}
{\it To Daniel Butter with gratitude and admiration}
\end{flushright}

\numberwithin{equation}{section}

%%%%%%%%%%%%%%%%%%%%%%%%%%%%%%%

\section{Introduction} \label{Section1}

Pure $\cN=2$ supergravity in four dimensions was constructed by Ferrara and van Nieuwenhuizen in 1976 \cite{FvN}, some six months after the creation of $\cN=1$ supergravity
\cite{FvNF,DZ}. It fulfilled Einstein's dream of unifying gravity and electromagnetism, albeit using a symmetry principle that was not known to Einstein --  local supersymmetry.

In 1979, Fradkin and Vasiliev \cite{Fradkin:1979as} and, independently, de Wit and van Holten \cite{deWit:1979xpv} proposed an off-shell formulation for linearised $\cN=2$ supergravity. Shortly thereafter, these linearised results were extended to 
 the first off-shell formulation for $\cN=2$ supergravity \cite{Fradkin:1979cw, deWvHVP}.
In \cite{deWvHVP} de Wit, van Holten and Van Proeyen  
made use of the so-called $\cN=2$ superconformal tensor calculus, a natural extension of the $\cN=1$ superconformal method \cite{KTvN1,KakuTownsend, KTvN2, Townsend:1979ki, Ferrara:1978rk}.  
Since then, the $\cN=2$ superconformal tensor calculus of 
\cite{deWvHVP} has been further developed \cite{BdeRdeW,deWvHVP2, deWPV} and applied 
 \cite{deWit:1983xhu, deWit:1984rvr}
to derive many important results for $\cN=2$ supergravity-matter systems. 
 For comprehensive reviews of this method, see \cite{FVP,Lauria:2020rhc}.

In parallel with the progress achieved in \cite{Fradkin:1979as, deWit:1979xpv, Fradkin:1979cw, deWvHVP}, 
there appeared several works \cite{BS1,CvNG, Gates:1980eu, Gates:1980ky, BS2, GS82, Howe} devoted  to $\cN=2$ superfield supergravity. 
In these works, the component results were recast  in a superspace setting. 
More importantly, 
these publications pursued an ambitious goal of
developing superspace formulations to describe general supergravity-matter systems, including the construction of an off-shell  {\it charged} hypermultiplet that can be coupled to a $\sU(1)$ vector multiplet. 

 Within the superconformal tensor calculus, hypermultiplets are either on-shell or involve a gauged central charge.  As is well known,
such hypermultiplet realisations cannot be used to provide an off-shell formulation 
for the most general locally supersymmetric sigma model. 
It is also known that such a sigma model formulation, if it exists, must
use off-shell hypermultiplets possessing an infinite number of auxiliary fields \cite{Siegel:1981dx, Howe:1985ar,Stelle:1985jh}.
The latter feature makes the off-shell hypermultiplets extremely difficult
to work with at the component level, and a superfield setting is required.

The problem of constructing an off-shell  charged hypermultiplet 
(in short, the {\it charged hypermultiplet problem})
remained unsolved until 1984. Nevertheless, the early works on $\cN=2$ superfield supergravity 
\cite{BS1,CvNG, Gates:1980eu, Gates:1980ky, BS2, GS82, Howe} have yielded several important results. It suffices to mention the linear multiplet action 
originally discovered by Sohnius in the rigid supersymmetric case \cite{Sohnius78}.  
Since the linear multiplet was lifted to $\cN=2$ supergravity \cite{BS1}, and then 
reformulated  \cite{deWvHVP3} within the $\cN=2$ superconformal tensor calculus
\cite{deWvHVP,BdeRdeW,deWvHVP2}, it has become a universal tool to construct 
the component actions for supergravity-matter systems. 

The construction of the relaxed hypermultiplet in 1983 \cite{HST-relaxed} was perhaps the pinnacle of conventional $\cN=2$ superspace techniques, but it did not solve the charged hypermultiplet problem.\footnote{There exist infinitely many off-shell formulations for the neutral hypermultiplet  \cite{GIO,G-RRWLvU}, in addition to the relaxed hypermultiplet.} 
In spite of being off-shell, 
this hypermultiplet is neutral and cannot couple to a $\sU(1)$ vector multiplet. 
It became apparent that 
the conventional $\cN=2$ superspace 
${\mathbb M}^{4|8}$ is not suitable, say,  for off-shell $\s$-model 
constructions. 
The correct superspace setting was found in 1983--1984 independently by three groups who pursued somewhat different goals \cite{Rosly,GIKOS,KLR}, which is:
\bea
{\mathbb M}^{4|8}\times {\mathbb C}P^1={\mathbb M}^{4|8}\times S^2~.
\label{1.1}
\eea
This superspace was introduced for the first time by Rosly \cite{Rosly} who used it to derive an interpretation of the $\cN=2$ super Yang-Mills constraints \cite{GSW} as integrability conditions. Rosly and Schwarz \cite{RS} called \eqref{1.1} {\it isotwistor superspace}.

The starting point of the analysis in \cite{Rosly}
was the observation that, given an isotwistor $v^{i} \in {\mathbb C}^2 \setminus \{0\} $,
 the set of eight spinor covariant derivatives $D_{\a i} $ and $ \bar D_{\ad i}$ for ${\mathbb M}^{4|8}$ contains  a subset of four operators, 
$ D^{(1)}_\a := -v^i D_{\a i} $ and ${\bar D}^{(1)}_{ \ad} := -v^i {\bar D}_{\ad i}$,
 which strictly anticommute with each other.
Therefore, one can introduce a new family of supersymmetric multiplets constrained by 
\bea
D^{(1)}_\a \f=0~, \qquad  {\bar D}^{(1)}_{\ad} \f=0~, \qquad
\f= { \f(z,v ,\bar v)}~, \qquad {\bar v}_i :=  \overline{v^i} ~.
\label{AnaCo}
\eea
In order for these constraints to be invariant under arbitrary re-scalings  of $v$, 
$\f$ should be homogeneous,
\bea
\f(z,c\, v, {\bar c}\, \overline{v})= c^{n_+} \, {\bar c}^{n_-} \,\f(z,v, \overline{v})~,
\qquad c\in  {\mathbb C} \setminus \{0\} \equiv {\mathbb C}^* ~,
\eea
for some parameters $n_\pm$ such that $n= n_+ - n_-$ is an integer. 
Redefining $ \f(z,v, \bar v) \to \f(z, v, \bar v)/  (v^\dagger v  ) ^{n_-}$
allows one  to choose ${ n_-=0}$. 
Any superfield with the homogeneity property 
\bea
\f^{(n)} (z,c\, v, {\bar c}\, \overline{v})= c^{n}  \,\f^{(n)}(z,v, \overline{v})~,
\qquad c\in {\mathbb C}^*
\label{HomCon}
\eea
is said to have the weight $n \in {\mathbb Z}$. This  superfield 
lives in the superspace \eqref{1.1}, since the isotwistor
$ v^i \in {\mathbb C}^2 \setminus\{0\}$ is   defined modulo the equivalence relation
$ v^i \sim c\,v^i$,  with $c\in {\mathbb C}^*$, {hence it parametrises ${\mathbb C}P^1$}.
A weight-$n$ superfield 
$\f^{(n)}(z,v, \bar v)$ is called {\it isotwistor} if it obeys the constraints \eqref{AnaCo}.

A new approach to $\cN=2$ supersymmetric field theory was put forward by Galperin {\it et al.} \cite{GIKOS}. Using  {\it harmonic superspace} 
${\mathbb M}^{4|8}\times S^2$, they proposed the first off-shell formulation of charged hypermultiplet (the so-called $q^+$ hypermultiplet) and its self-couplings. Moreover,  unconstrained prepotential descriptions of $\cN=2$ super Yang-Mills and supergravity theories were also provided. Since then the harmonic superspace approach has developed into a powerful branch of supersymmetric field theory, see \cite{GIOS} for a review. In the harmonic superspace approach, one deals with those isotwistor superfields $\f^{(n)}(z,v, \overline{v})$ which are globally defined smooth functions on ${\mathbb C}P^1$. In the literature, they are known as {\it harmonic analytic superfields}.

{\it Projective superspace} ${\mathbb M}^{4|8}\times {\mathbb C}P^1$ 
was originally employed in \cite{KLR}
to provide a manifestly $\cN=2$ supersymmetric description for the general self-couplings
of $\cN=2$ tensor multiplets constructed earlier \cite{LR83}
 in terms of $\cN=1$ superfields. Since then, this approach has  been extended
to include some other interesting multiplets \cite{LR1,LR2}. In particular, a new off-shell formulation for the charged hypermultiplet was derived \cite{LR1} and used to construct off-shell nonlinear  $\s$-models, see \cite{LR2008,K2010} for reviews. 
The name `projective superspace' was coined in 1990 \cite{LR2}.
In the projective superspace approach, one deals with those isotwistor superfields $\f^{(n)}(z,v)$ which are holomorphic functions on an open domain of ${\mathbb C}P^1$.
In the literature, they are known as {\it projective superfields}.

Both harmonic and projective superspace make use of the same superspace
\eqref{1.1}.
Without going into technical details, which are  spelled out in \cite{K98} (see also \cite{K2010,JainSiegel,Butter:2012ta}),
they differ in (i) the structure of off-shell supermultiplets used; 
and (ii) the supersymmetric action principle chosen. 
Due to these conceptual differences, the two approaches prove to be  complementary 
to each other in many respects.
In particular, harmonic superspace offers powerful prepotential formulations for 
$\cN=2$ supergravity \cite{SUGRA-har1,SUGRA-har2} (reviewed in \cite{GIOS})
which are similar in spirit  to the Ogievetsky-Sokatchev approach to $\cN=1$ supergravity \cite{Ogievetsky:1978mt}.
Projective superspace proves to be ideal for developing covariant geometric formulations for supergravity-matter systems  with eight supercharges.
The harmonic superspace approach to $\cN=2$ supergravity is reviewed 
in this volume by Ivanov \cite{Ivanov}.

The formalism of curved projective superspace was originally developed in 2008 for $\cN=1$ supergravity-matter systems in five dimensions \cite{KT-M08,KT-M08-Weyl}
using the structure of superconformal projective multiplets \cite{K06}.
Shortly thereafter, these constructions were generalised to develop the projective superspace approach for $\cN=2$ matter-coupled supergravity in four dimensions
\cite{KLRT-M1,K2008,KT-M2009, KLRT-M2}.\footnote{It was subsequently extended to 
supergravity-matter  theories in two \cite{T-M}, three \cite{KLT-M11} and six \cite{LT-M12} dimensions.}  
With the advent of $\cN=2$ conformal superspace \cite{ButterN=2}, and its applications to component reduction \cite{Butter:2012xg},
a novel formulation of curved projective superspace has been given in \cite{Butter:2014gha,Butter:2014xua}. 
This approach has also been extended to a novel covariant harmonic superspace framework  in \cite{Butter:2015nza}.
All of these publications followed the philosophy of the $\cN=2$ superconformal tensor calculus to realise supergravity-matter systems as conformal supergravity coupled to superconformal matter multiplets. 

There are three superspace formulations for $\cN=2$ conformal supergravity that have found numerous applications in the recent years, specifically: (i) $\sU(2)$ superspace \cite{Howe}; (ii) $\sSU(2)$ superspace 
\cite{KLRT-M1}; and (iii) conformal superspace \cite{ButterN=2}.
The  $\cN=2$ conformal superspace of \cite{ButterN=2} is an ultimate formulation for $\cN=2$ 
conformal supergravity in the sense that any different off-shell formulation is either equivalent to it or is obtained from it by partially fixing the gauge freedom. 
In particular,  the $\sU(2)$ and $\sSU(2)$ superspaces can be derived from conformal superspace by 
imposing partial gauge fixing conditions.\footnote{The relationship between  the  $\sU(2)$ and $\sSU(2)$ superspaces is described 
	in \cite{KLRT-M2}.}
At the component level, $\cN=2$ conformal superspace reduces to 
the $\cN=2$ superconformal tensor calculus.

The $\cN=2$ conformal superspace of \cite{ButterN=2, Butter:2012xg} is a natural extension of the $\cN=1$ formulation \cite{ButterN=1}. Conformal superspace approaches have also been developed for extended supergravity-matter systems in three \cite{BKNT-M1,BKNT-M2,KNT-M14}, five \cite{BKNT-M15} and six \cite{BKNT16, BNT-M17} dimensions. These references include various applications.

Recently, new supertwistor formulations were discovered for 
conformal supergravity theories  in diverse dimensions \cite{HL20}.
In the four-dimensional $\cN=2$ case, the supertwistor formulation is expected to be related to conformal superspace, however relevant technical details have not yet been worked out in the literature.

Our two-component spinor notation and conventions follow 
\cite{BK}, and are similar to those adopted in  \cite{WB}.  The only difference is that the spinor Lorentz generators $(\s_{ab})_\a{}^\b$ and 
$({\tilde \s}_{ab})^\ad{}_\bd$  used in \cite{BK} have an extra minus sign as compared with \cite{WB}, 
specifically $\s_{ab} = -\frac{1}{4} (\s_a \tilde{\s}_b - \s_b \tilde{\s}_a)$ and 
 $\tilde{\s}_{ab} = -\frac{1}{4} (\tilde{\s}_a {\s}_b - \tilde{\s}_b {\s}_a)$.

%%%%%%%%%%%%%%%%%%%%%%%%%%%%%%%%%%
%%%%%%%%%%%%%%%%%%%%%%%%%%%%%%%%%%

\section{Rigid superconformal transformations} 
\label{section2}

We denote by $z^A =(x^a, \q^\a_i , \bar \q_\ad^i)$ the Cartesian coordinates
for Minkowski superspace ${\mathbb M}^{4|8}$
and use the notation $D_A =(\pa_a, D_\a^i  , \bar D^\ad_i)$ 
for the  superspace covariant derivatives. 
The only non-zero graded commutation relation is 
\bea
\{ D_\a^i , \bar D_{\bd j}\}= 
-2\ri \d^i_j (\s^c)_{\a\bd}\pa_c
=-2\ri \d^i_j\pa_{\a\bd}~, \qquad i,j=\1, \2 ~.
\label{flatcd}
\eea
The $\cN=2$ super-Poincar\'e algebra has an outer automorphism group 
$\sSU(2)_R \times \sU(1)_R$, which is also called the $R$-symmetry group. 
The $\sSU(2)_R$ indices are raised and lowered using the antisymmetric tensor 
$\ve^{ij} = -\ve^{ji} $ and its inverse $\ve_{ij}$ normalised by $\ve^{{\1} \2} =1$. 

%%%%%%%%%%%%%%%%%%%%%%%%%%%%%%%%%%%%%%%
%%%%%%%%%%%%%%%%%%%%%%%%%%%%%%%%%%%%%%%

\subsection{Conformal Killing supervector fields}

 Superconformal transformations in ${\mathbb M}^{4|8}$ were first studied by Sohnius \cite{Sohnius}. Our presentation follows \cite{KT}.

An infinitesimal superconformal  transformation 
$ z^A \to    z^A  + \d z^A$, with 
$\d z^A =\x  \,z^A = \Big(\x^a +\ri (\x_i\s^a \bar \q^i - \q_i \s^a \bar \x^i), \x^\a_i, \bar \x_\ad^i \Big), $ is generated by 
a {\it conformal Killing supervector field}
\bea
\x = \x^b  \pa_b + \x^\b_j D_\b^j
+ {\bar \x}_{\bd}^j  {\bar D}^{\bd}_j = {\overline \x}
~.
\eea 
The defining property of $\x$ is 
\begin{align}
	[\x , D_\a ^i] = - (D_\a^{i} \x^\b_j) D_\b^j~.
\end{align}
This condition implies the relations 
\bea
{\bar D}^{ \ad }_i \x^\b_j =0~, \qquad
{\bar D}^{ \ad }_i \x^{ \bd \b} = 4{\rm i} \, \ve^{\ad \bd} \x^\b_i \quad \implies \quad
\x^\a_i = -\frac{\ri }{8} \bar D_{\ad i} \x^{\ad \a }  
\eea
and their complex conjugates, 
and therefore
\bea
{\bar D}_{(\a i} \x_{\b) \bd } =0~, \qquad {\bar D}_{(\ad }^i \x_{\b \bd )} =0 
\quad \implies \quad \pa_{(\a (\ad} \x_{\b) \bd)}=0~.
\eea
It then follows that
\bea
[\x , D_\a^i ]
= - K_\a{}^\b [\x] D_\b^i - \hf \bar{\s}[\xi] D_\a^i - \L^{i}{}_{j}[\xi] D_{\a}^j~.
\label{4Dmaster2N=2} 
\eea
Here we have introduced the chiral Lorentz
$K_{\b\g}[\x]$ and 
super-Weyl $\s[\x]$ parameters, 
as well as the $\sSU(2)_R$ parameter $K^{ij}[\xi]$ defined by 
\begin{subequations} 
\label{2.7}
\bea
K_{\a\b}[\x]&=& \hf D_{(\a}^i \x_{\b)i}=K_{\b \a}[\xi]~, \qquad \bar D^\ad_i K_{\a\b}[\x] =0~, \\
\s [\x] &= & \frac{1}{2} \bar{D}^\ad_i \bar{\xi}_\ad^i~, \qquad \qquad \qquad \quad \;\;\;\;\bar{D}^\ad_i \s[\xi] = 0~, \label{2.7b} \\
\L^{ij}[\xi] &=& - \frac{\ri}{16} [D^{(i}_\a , \bar{D}_\ad^{j)}] \xi^{\aa} = \L^{ji}[\xi]~,
\quad 
\overline{\L^{ij}[\xi] } = \L_{ij}[\xi] ~. \label{2.7c}
\eea
\end{subequations}
We recall that the Lorentz parameters with vector and spinor indices are related to each other as follows:
$K^{bc}[\x] = (\s^{bc})_{\b\g}K^{\b\g}[\x] - (\tilde{\s}^{bc})_{\bd\gd}\bar K^{\bd\gd}[\x] $.
The parameters in \eqref{2.7} obey several first-order differential properties:
\begin{subequations}
\bea
D^i_\a {\L}^{jk} [\x]&=&
\ve^{i(j} D_\a^{k)} 
\s [\x]~,
\label{an1N=2} \\
D^i_\a K_{\b\g} [\x] &=& -\ve_{\a(\b} D^i_{\g)} \s[\x]~,
\eea
\end{subequations}
and therefore 
\begin{subequations}
\bea
D^{(i}_\a \L^{jk)} [\x]&=& {\bar D}^{(i}_{\ad} \L^{jk)} [\x]=0~, 
\label{L-an} \\
D^i_\a D^j_\b \s[\x] &=& 0~.
\eea
\end{subequations}

The most general conformal Killing supervector field has the form
\begin{subequations} \label{chiraltra}
	\bea
	\x_+^{\ad\a} &=& a^{\ad\a}  +\hf (\s +{\bar \s})\, y^{\ad\a}
	+{\bar K}^\ad{}_\bd  \,y^{\bd \a} +y^{\ad\b}K_\b{}^\a 
	-y^{\ad \b} b_{\b \bd} y^{\bd \a} \non \\
	&& \qquad +4{\rm i}\, {\bar \e}^{\ad i}  \q^{\a}_i - 4 y^{\ad \b} \eta_\b^i \q^\a_i ~,
	\\
	\x^\a_i &=& \e^\a_i + \hf \bar{\s} \q^\a_i + \q^\b_i K_\b{}^\a + \L_{i}{}^{j} \q^{\a}_j
	-  \q^\b_i b_{\b\bd}   y^{\bd \a} \non \\	
	&& \qquad -{\rm i}\,{\bar \eta}_{\bd i} y^{\bd \a} - 4 \q^\a_i \eta_\b^j \q^\b_j~,~~~~
	\eea
\end{subequations}
where we have introduced the complex four-vector
\be
\x_+^a = \x^a + 2\ri \x_i \s^a  {\bar \q}^i~, 
\qquad \bar \x^a =\x^a~,
\ee
along with the complex bosonic coordinates $y^a = x^a +\ri \q_i \s^a \bar \q^i$ 
of the chiral subspace of ${\mathbb M}^{4|8}$. 
The constant bosonic parameters in \eqref{chiraltra}
correspond to the spacetime translation ($a^{\ad \a}$), 
Lorentz transformation ($K_\b{}^\a,~{\bar K}^{\ad}{}_{\bd})$,
$\sSU(2)_R$ transformation ($\L^{ij} = \L^{ji}$),
special conformal transformation
($ b_{\a \bd}$), and  combined scale and $\sU(1)_R$ transformations 
($\s =\t - 2 \ri \vf$). The constant fermionic parameters in \eqref{chiraltra}
correspond to the $Q$-supersymmetry ($\e^\a_i$) and $S$-supersymmetry 
($\eta_i^\a$) transformations. The constant parameters $K_{\a\b}$, $\L^{ij}$ and $\s$ are obtained 
from $K_{\a\b}[\x]$, $\L^{ij}[\x]$ and $\s[\x]$, respectively,  by setting $z^A=0$.

It is useful to introduce a condensed notation for the superconformal parameters 
\bea
\l^{\tilde a} = (a^A, K^{ab} , \L^{ij}, \t, \vf , b_A)~, \;\;
a^A:= (a^a, \e^\a_i , \bar \e_\ad^i)~, \;\; b_A :=(b_a, \eta_\a^i , \bar \eta^\ad_i)~,
\eea 
as well as for the generators of the superconformal group
\bea
X_{\tilde a} = (P_A, M_{ab} , J_{ij}, {\mathbb D}, \mathbb{Y} , K^A)~, \;\; 
P_A:= (P_a, Q_\a^i , \bar Q^\ad_i)~, \;\; K^A:=(K^a, S^\a_i , \bar S_\ad^i)~. 
\label{generators}
\eea 
The general conformal Killing supervector field on ${\mathbb C}^{4|2}$,
\bea
\x= \x^a_+ (y,\q) \frac{\pa}{\pa y^a} +\x^\a_i (y, \q)  \frac{\pa}{\pa \q^\a_i} 
\equiv \x^a_+ \partial/\partial y^a + \x^\a_i \pa_\a^i~,
\eea
may be written in the form:
\bea
\x &=& \l^{\tilde a} \x_{\tilde a} (X) = a^A \x_A (P) +\hf K^{ab} \x_{ab}(M) + \L^{ij} \xi_{ij}(J)
+\t \x({\mathbb D}) \non\\
&& \qquad \qquad \qquad + \ri \vf \x(\mathbb{Y}) +b_A \x^A (K)~.~~
\eea
We read off the relevant supervector fields:
\begin{subequations}
	\bea
	\x_a(P) &=& \partial/\partial y^a ~, \quad \x_\a^i (P) = \pa_\a^i~, \quad 
	\bar \x^\ad_i (P) = -2\ri (\tilde{\s}{}^c \q_i )^\ad \partial/\partial y^c~, \\
	\x_{ab}(M) &=& y_a \partial/\partial y^b - y_b \partial/\partial y^a + (\q_i \s_{ab} )^\g \pa_\g^i~, \quad \x_{ij}(J) = \q^\a_{(i} \partial_{\a j)} ~, \\
	\x({\mathbb D}) &=& y^c \partial/\partial y^c + \hf \q^\g_i \pa_\g^i~, \quad \x(\mathbb{Y}) = \q^\g_i \pa_\g^i~, \\
	\x^a (K) &=& 2 y^a y^c \partial/\partial y^c -y^2 \partial/\partial y^a - (\q_i \s^a \tilde{\s}{}^c)^\g y_c \pa_\g^i ~,\\
	\x^\a_i (K) &=& 2 (\q\s^c \tilde{\s}{}^d)^\a y_d \partial/\partial y^c + 4 \q^\a_i \q ^\b_j \pa_\b^j ~, \\
	{\bar \x}_\ad^i (K) &=& \ri (\s^c)^\g{}_\ad y_c \pa_\g^i~.
	\eea
\end{subequations}

Making use of the above operators, 
 the graded commutation relations for the superconformal algebra, 
 $\big[X_{\tilde a} , X_{\tilde b} \big\}= - f_{\tilde a \tilde b}{}^{\tilde c} X_{\tilde c}$,
 can be derived
 keeping in mind the relation
 \bea
 \x = \l^{\tilde a} \x_{\tilde a} (X)~ \to~ \d_\x =  \l^{\tilde a} X_{\tilde a} ~, \qquad 
 \big[\x_1 , \x_2 \big] ~\to ~- \big[\d_{\x_1} , \d_{\x_2}\big] ~.
 \eea

%%%%%%%%%%%%%%%%%%%%%%%%%%%%%%%%%%%

\subsection{Superconformal algebra} 

Here we describe the graded commutation relations for the $\cN=2$ superconformal algebra ${\mathfrak su} (2,2|2)$.
We start with the commutation relations for the conformal algebra:
\begin{subequations} \label{confal}
\begin{align}
&[M_{ab},M_{cd}]=2\eta_{c[a}M_{b]d}-2\eta_{d[a}M_{b]c}~, \phantom{inserting blank space inserting}\\
&[M_{ab},P_c]=2\eta_{c[a}P_{b]}~, \qquad \qquad \qquad \qquad ~ [\mathbb{D},P_a]=P_a~,\\
&[M_{ab},K_c]=2\eta_{c[a}K_{b]}~, \qquad \qquad \qquad \qquad [\mathbb{D},K_a]=-K_a~,\\
&[K_a,P_b]=2\eta_{ab}\mathbb{D}+2M_{ab}~.
\end{align}
\end{subequations}
The $R$-symmetry generators $\mathbb{Y}$ and $J_{ij}$ commute with all the generators of the conformal group. Amongst themselves, they obey the algebra:
\bea
[J^{ij},J^{kl}] = \ve^{k(i}J^{j)l} + \ve^{l(i}J^{j)k}~.
\eea
The superconformal algebra is then obtained by extending the translation generator to $P_A$ and the special conformal generator to $K^A$. The commutation relations involving the $Q$-supersymmetry generators with the bosonic ones are:
 \begin{subequations} \label{superconfal1}
 \bea
 \big[M_{ab}, Q_\g^i \big] &=& (\s_{ab})_\g{}^\d Q_\d^i ~,\quad 
\big[M_{ab}, \bar Q^\gd_i \big] = (\tilde{\s}_{ab})^\gd{}_\dd \bar Q^\dd_i~,\\
\big[\mathbb{D}, Q_\a^i \big] &=& \hf Q_\a^i ~, \quad
\big[\mathbb{D}, \bar Q^\ad_i \big] = \hf \bar Q^\ad_i ~, \\
\big[\mathbb{Y}, Q_\a^i \big] &=&  Q_\a^i ~, \quad
\big[\mathbb{Y}, \bar Q^\ad_i \big] = - \bar Q^\ad_i ~,  \\
\big[K^a, Q_\b^i \big] &=& -\ri (\s^a)_\b{}^\bd \bar{S}_\bd^i ~, \quad 
\big[K^a, \bar{Q}^\bd_i \big] = 
-\ri ({\s}^a)^\bd{}_\b S^\b_i ~.
 \eea
 \end{subequations}
The commutation relations involving the $S$-supersymmetry generators 
with the bosonic operators are: 
\begin{subequations} \label{superconfal2}
 \bea
\big [M_{ab} , S^\g_i \big] &=& - (\s_{ab})_\b{}^\g S^\b_i ~, \quad
\big[M_{ab} , \bar S_\gd^i \big] = - (\ts_{ab})^\bd{}_\gd \bar S_\bd^i~, \\
\big[\mathbb{D}, S^\a_i \big] &=& -\hf S^\a_i ~, \quad
\big[\mathbb{D}, \bar S_\ad^i \big] = -\hf \bar S_\ad^i ~, \\
\big[\mathbb{Y}, S^\a_i \big] &=&  -S^\a_i ~, \quad
\big[\mathbb{Y}, \bar S_\ad^i \big] =  \bar S_\ad^i ~,  \\
\big[ S^\a_i , P_b \big] &=& \ri (\s_b)^\a{}_\bd \bar{Q}^\bd_i ~, \quad 
\big[\bar{S}_\ad^i , P_b \big] = 
\ri ({\s}_b)_\ad{}^\b Q_\b^i ~.
 \eea
 \end{subequations}
Finally, the anti-commutation relations of the fermionic generators are: 
\begin{subequations}\label{superconfal3}
\bea
\{Q_\a^i , \bar{Q}^\ad_j \} &=& - 2 \ri \d^i_j (\s^b)_\a{}^\ad P_b=- 2 \ri \d^i_j  P_\a{}^\ad~, \\
\{ S^\a_i , \bar{S}_\ad^j \} &=& 2 \ri  \d_i^j (\s^b)^\a{}_\ad K_b=2 \ri \d_i^j  K^\a{}_\ad
~, 
\label{superconfal3-2}\\
\{ S^\a_i , Q_\b^j \} &=& \d_i^j \d^\a_\b (2 \mathbb{D} - \mathbb{Y}) - 4 \d_i^j  M^\a{}_\b 
+ 4 \d^\a_\b  J_i{}^j ~, \\
\{ \bar{S}_\ad^i , \bar{Q}^\bd_j \} &=& \d_j^i \d^\bd_\ad (2 \mathbb{D} + \mathbb{Y}) + 4 \d_j^i  \bar{M}_\ad{}^\bd 
- 4 \d_\ad^\bd  J^i{}_j  ~,
\eea
\end{subequations}
where $M_{\a\b}=\hf(\s^{ab})_{\a\b}M_{ab}$ and $\bar{M}_{\ad\bd}=-\hf(\ts^{ab})_{\ad\bd}M_{ab}$. Note that all remaining (anti-)commutators not explicitly listed above vanish identically.

The graded commutation relations
\eqref{confal} -- \eqref{superconfal3}
constitute
the $\cN=2$ superconformal algebra, ${\mathfrak{su}}(2,2|2)$. Its generators obey
the graded Jacobi identity
\be
(-1)^{\varepsilon_{\tilde{a}}  \varepsilon_{\tilde{c}}}[X_{\tilde{a}}, [X_{\tilde{b}}, X_{\tilde{c}} \} \} 
~+~ \text{(two cycles)}
= 0 \ ,
\label{Jacobi-0}
\ee
where $\varepsilon_{\tilde{a}} = \varepsilon(X_{\tilde{a}})$ is the Grassmann parity of the generator $X_{\tilde{a}}$.
Making use of 
$\big[X_{\tilde a} , X_{\tilde b} \big\}= - f_{\tilde a \tilde b}{}^{\tilde c} X_{\tilde c}$,
the Jacobi identities are equivalently written as
\be 
f_{[\tilde{a}\tilde{b}}{}^{\tilde{d}} f_{|\tilde{d}| \tilde{c} \} }{}^{\tilde{e}} = 0 \ .
\ee

%%%%%%%%%%%%%%%%%%%%%%%%%%%%%%

\subsection{Superconformal primary multiplets} \label{section2.3}

The superconformal transformation law of a primary tensor superfield (with suppressed indices) is
\bea
\d_\x U &=& \cK[\x] U, \non \\
 \cK[\x] &=& \x +\hf K^{ab}[\x]M_{ab} + \L^{ij}[\x] J_{ij} 
+p\s[\x] +q\bar \s[\x] ~.
\label{FlatPrimaryMultiplet}
\eea
Here the generators $M_{ab}$ and $J_{ij}$ act on the Lorentz and $\sSU(2) $ indices of $U$, respectively. The parameters $p$ and $q$ are related to the dimension (or Weyl weight) $w$ and $\sU(1)_R $ charge $c$ of $U$ as $ p+q =w$ and $p-q = - \hf c$. 

As an example, let us consider a chiral tensor superfield $\f$, $\bar D^\ad_i \f =0$. 
Requiring it to be primary leads to the conditions
\bea
\bar M_{\ad\bd} \f = 0~, \quad J_{ij} \f =0~, \quad q=0~.
\eea
These conditions imply that (i) $\f$ can carry only undotted spinor indices; (ii) $\f$ must be neutral under the group $\sSU(2)_R$; and (iii) the dimension $w$ and the $\sU(1)_R $ charge $c$ of $\f$ are related as $c= - 2w$. A chiral scalar $W$ is called reduced if it obeys the reality condition 
\bea
D^{ij} W = \bar D^{ij} \bar W~, \qquad D^{ij} := D^{\a(i} D_\a^{ j)} ~, \quad
\bar D_{ij} := \bar{D}_{\ad (i} \bar{D}^\ad_{j)}~,
\eea
which uniquely fixes the dimension of $W$ to be $+1$.
Chiral multiplets exist both in the $\cN=1$ and $\cN=2$ supersymmetric cases. New types of multiplets are offered by $\cN=2$ supersymmetry, as will be discussed below. 

An $\cO(n)$ multiplet $H^{i_1\dots i_n} = H^{(i_1\dots i_n)}$ 
obeys the analyticity constraints\footnote{The $\cO(n)$ multiplets are well-known in the
literature on $\cN=2$ supersymmetric field theories. The $n = 1$ case corresponds to the on-shell Fayet-Sohnius hypermultiplet \cite{Sohnius78,Fayet}. The field strength of the $\cN=2$ tensor multiplet  \cite{Wess} is described by a real $\cO(2)$ multiplet \cite{BS1,BS2,SSW}.
General $\cO(n)$ multiplets, with $n>2$, were introduced in \cite{KLT,GIO,LR1}.
The case $n=4$ was first studied in \cite{SSW}.} 
\bea
D^{(i_1}_\a H^{i_2\cdots i_{n+1})} = 0~, \qquad {\bar D}^{(i_1}_{\ad} H^{i_2\cdots i_{n+1})} = 0~.
\label{ana2}
\eea
In the super-Poincar\'e case, these constraints are  consistent with $H^{i_1\dots i_n} $
 carrying Lorentz indices in addition to the $\sSU(2)$ ones. However, this is no longer allowed if $H^{i_1\dots i_n}$ is a  superconformal primary multiplet. Then, the superconformal transformation law of $H$ is uniquely determined by the constraints \eqref{ana2} to be 
\bea
\d_\x H^{i_1\dots i_n} = \x H^{i_1\dots i_n} +n \L_j{}^{(i_1} [\x] H^{i_2 \dots i_n) j}
+\frac{n}{2} \big(\s[\x] + \bar \s [\x] \big)H^{i_1\dots i_n} ~.
\label{O(n)transformation}
\eea
In the case that  $n$ is even, $n=2m$, this transformation law is compatible with the reality condition 
$\overline{H^{i_1\dots i_{2m}} } = H_{i_1 \dots i_{2m}} = \ve_{i_1 j_1} \dots 
\ve_{i_{2m} j_{2m}} H^{j_1 \dots j_{2m}}$.

%%%%%%%%%%%%%%%%%%%%%%%%%%%%%%%

\subsection{Superconformal projective multiplets}

The constraints \eqref{ana2} can be given a more transparent interpretation if one makes use of an isotwistor $v^{i} \in {\mathbb C}^2 \setminus \{0\} $
that allows one to introduce  
new spinor covariant derivatives,
\bea
D^{(1)}_\a = v_i D^i_\a  ~, \qquad {\bar D}^{(1)}_{ \ad} = v_i {\bar D}^i_{\ad}~, \qquad
v_i:= \ve_{ij}\,v^j~,
\label{2.28}
\eea
and to convert $H^{i_1 \dots i_n}(z) $ into an index-free homogeneous polynomial  of degree $n$,
 \bea
 H^{(n)}(z,v)=v_{i_1}\cdots v_{i_n}\,H^{i_1\cdots i_n}(z)~.
 \label{O(n)multiplet}
 \eea
In accordance with \eqref{flatcd}, the operators \eqref{2.28}
 strictly anticommute with each other,  
 \bea
 \{ D^{(1)}_\a , D^{(1)}_\b\}=\{ {\bar D}^{(1)}_{\ad} , {\bar D}^{(1)}_{ \bd} \}
=\{ D^{(1)}_\a , {\bar D}^{(1)}_{\bd} \} = 0~,
\label{StrictlyZero} 
\eea 
and annihilate $H^{(n)}$,
\bea
D^{(1)}_\a H^{(n)}=0~, \qquad  {\bar D}^{(1)}_{\ad} H^{(n)}=0~.
\label{ana3}
\eea
These constraints do not change if we replace ${ v^i \to {\mathfrak c} \,v^i}$, 
with ${\mathfrak c} \in {\mathbb C}\setminus \{0\} \equiv {\mathbb C}^*$, 
in the definition of the operators \eqref{2.28} and the superfield \eqref{O(n)multiplet}.
We see that the isotwistor
$ v^i \in {\mathbb C}^2 \setminus\{0\}$ is defined modulo the equivalence relation
$ v^i \sim {\mathfrak c}\,v^i$,  with ${\mathfrak c}\in {\mathbb C}^*$, hence it provides homogeneous coordinates for 
${\mathbb C}P^1$.
The  superfield \eqref{O(n)multiplet} can be interpreted to be a holomorphic tensor field on the superspace \eqref{1.1}.

The superconformal transformation law \eqref{O(n)transformation} can be recast in term of
$H^{(n)}$. For this it is useful to introduce a new isotwistor $u^i$ such that $v^i$ and $u^i$ form a basis for ${\mathbb C}^2$, that is $(v,u):=v^iu_i \neq 0$.
\bea
\d_\x H^{(n)} =  \Big(  \x -  \L^{(2)}[\x] \pa^{(-2)} \Big)  H^{(n)} 
+n \S [\x] H^{(n)} ~.
\label{harmult1}
\eea
Here we have
introduced the differential operator
\bea
{\pa}^{(-2)} :=\frac{1}{(v,u)}u^{i}\frac{\pa}{\pa v^{i}}
~,
\label{5.28}
\eea
as well as  the parameters 
\begin{subequations}
\bea
\L^{(2)} [\x]&:=& v_i v_j \L^{ij} [\x] ~, \qquad 
\L^{(0)} [\x]:=\frac{v_i u_j }{(v,u)}\L^{ij}[\x]~,
\\
 \S[\x] &:=& \L^{(0)}[\x] + \hf (\s[\x]+\bar \s[\x])  ~,
\label{W2t3}
\eea
\end{subequations}
which have the following properties
\begin{subequations}
\bea
D^{(1)}_\a \L^{(2)}[\x]&=&0~, \quad  {\bar D}^{(1)}_{\ad} \L^{(2)}[\x]=0~, \\
D^{(1)}_\a \S[\x]&=&0~, \quad  {\bar D}^{(1)}_{\ad} \S[\x]=0~.
\eea
\end{subequations}
The variation \eqref{harmult1} obeys the analyticity constraints \eqref{ana3}
due to the identity 
\bea
\Big[  \x -  \L^{(2)}[\x] \pa^{(-2)} ~, ~D^{(1)}_\a \Big] = -K_\a{}^\b[\x] D^{(1)}_\b 
- \Big( \hf \s [\x]+ \L^{(0)} [\x] \Big) D^{(1)}_\a~,
\eea
and a similar relation for $\bar D^{(1)}_\ad$.

The above discussion can be extended to more general superconformal projective multiplets \cite{K06,K07}. 
A superconformal projective multiplet of weight $n$,
$Q^{(n)}(z,v)$, is a superfield that 
lives on  ${\mathbb R}^{4|8}$ with respect to the superspace variables $z^A$, 
is holomorphic with respect to 
the isotwistor variables $v^i $ on an open domain of 
${\mathbb C}^2 \setminus  \{0\}$, 
and is characterised by the following conditions:

(a) it obeys the analyticity constraints 
\begin{subequations} \label{RigidProjectiveM}
\bea
D^{(1)}_{\a} Q^{(n)} =0~, \qquad {\bar D}^{(1)}_{\ad} Q^{(n)} =0~;
\label{ana}
\eea  

(b) it is  a homogeneous function of $v$ of degree $n$,  
\bea
Q^{(n)}(z, {\mathfrak c} \,v)={\mathfrak c}^n\,Q^{(n)}(z,v)~, \qquad {\mathfrak c} \in \mathbb{C}^*~;
\label{weight}
\eea

(c) it possesses the  superconformal transformation law
\bea
\d_\x Q^{(n)} = 
\Big(  \x -  \L^{(2)}[\x] \pa^{(-2)} \Big)  Q^{(n)} 
+n \S [\x] Q^{(n)} ~.
\label{harmultQ}
\eea
\end{subequations}
By construction, the superfield $Q^{(n)}$ is independent of the isotwistor $u^i$, 
\bea
\pa^{(2)}   Q^{(n)}  =0~,\qquad
\pa^{(2)} := (v,u) v^{i} \frac{\pa}{\pa u^{i}} ~.
\eea
One may check that the variation $\d_\x Q^{(n)} $, eq.
 \eqref{harmultQ}, is characterised by 
the same property, $\pa^{(2)}\d_\x Q^{(n)} =0$,
due to the homogeneity condition (\ref{weight}). 

There exists 
a real structure on the space of projective multiplets \cite{Rosly,GIKOS,LR1}.
Given a  weight-$n$ projective multiplet $ Q^{(n)} (v^{i})$, 
its {\it smile conjugate} $ \breve{Q}^{(n)} (v^{i})$ is defined by 
\bea
 Q^{(n)}(v^{i}) \longrightarrow  {\bar Q}^{(n)} ({\bar v}_i) 
  \longrightarrow  {\bar Q}^{(n)} \big({\bar v}_i \to -v_i  \big) =:\breve{Q}^{(n)}(v^{i})~,
\label{smile-iso}
\eea
with ${\bar Q}^{(n)} ({\bar v}_i)  :=\overline{ Q^{(n)}(v^{i} )}$
the complex conjugate of  $ Q^{(n)} (v^{i})$, and ${\bar v}_i$ the complex conjugate of 
$v^{i}$. One can show that $ \breve{Q}^{(n)} (v)$ is a weight-$n$ projective multiplet.
In particular,   $ \breve{Q}^{(n)} (v)$
obeys the analyticity constraints $D_\a^{(1)}\breve{Q}^{(n)} =0$ and $\bar D_\ad^{(1)}\breve{Q}^{(n)} =0$,
unlike the complex conjugate of $Q^{(n)}(v) $.
One can also check that 
\bea
\breve{ \breve{Q}}^{(n)}(v) =(-1)^n {Q}^{(n)}(v)~.
\label{smile-iso2}
\eea
Therefore, if  $n$ is even, one can define real projective multiplets, 
which are constrained by $\breve{Q}^{(2n)} = {Q}^{(2n)}$.
Note that geometrically, the smile-conjugation is complex conjugation composed
with the antipodal map on the projective space ${\mathbb C}P^1$.

The $\cO(n)$ multiplets,  $H^{(n)} (v)$,  are well defined on the entire projective space ${\mathbb C}P^1 $.
There also exist important projective multiplets that are defined only on an open domain 
of ${\mathbb C}P^1 $. Before introducing them, let us give a few definitions. 
We define the {\it north chart} of ${\mathbb C}P^1$ to consist of those points for which 
the first component  of $v^i = (v^{\1}, v^{\2})$ is non-zero,  $v^{\1} \neq 0$.
The north chart of ${\mathbb C}P^1$ may be parametrised by the inhomogeneous complex coordinate 
$\z= v^{\2}/v^{\1} \in \mathbb C$. The only point of ${\mathbb C}P^1 $ outside the north 
chart is characterised by $v_\infty^i = (0, v^{\2})$ and describes an infinitely separated point.
Thus we may think of the projective space ${\mathbb C}P^1 $
as  ${\mathbb C}P^1 ={\mathbb C} \cup \{\infty \}$. 
The {\it south chart} of ${\mathbb C}P^1$ is defined to consist of those points for which 
the second component  of $v^i = (v^{\1}, v^{\2})$ is non-zero,  $v^{\2} \neq 0$.
The south chart is naturally parametrized by $1/\z$.
The intersection of the north and south charts is ${\mathbb C} \setminus \{ 0\}$.

An off-shell (charged) hypermultiplet can be described in terms of the so-called {\it arctic} 
weight-$n$ multiplet $\U^{(n)} (v)$ which is defined to be 
holomorphic  in the north chart of ${\mathbb C}P^1$: 
\bea
\U^{(n)} ( v) &=&  (v^{\1})^n\, \U^{[n]} ( \z) ~, \qquad 
\U^{ [n] } ( \z) = \sum_{k=0}^{\infty} \U_k  \z^k 
~. 
\label{arctic1}
\eea
Its smile-conjugate {\it antarctic} multiplet $\breve{\U}^{(n)} (v) $, has the explicit form 
 \bea
\breve{\U}^{(n)} (v) &=& 
(v^{\2}  \big)^{n}\, \breve{\U}^{[n]}(\z) =
(v^{\1} \,\z \big)^{n}\, \breve{\U}^{[n]}(\z) ~, \quad
\breve{\U}^{[n]}( \z) = \sum_{k=0}^{\infty}  {\bar \U}_k \,
\frac{(-1)^k}{\z^k}
\label{antarctic1}
\eea
and is holomorphic in the south chart of ${\mathbb C}P^1$.
The arctic multiplet can be  coupled to a Yang-Mills multiplet
in a complex representation of the gauge group \cite{LR2}.
The pair consisting of $\U^{[n]} ( \z)$ and $\breve{\U}^{[n]}(\z) $ 
constitutes the so-called polar weight-$n$ multiplet.

Our last example is a real {\it tropical} multiplet $\cU^{(2n)} (v) $ of weight $2n$ defined by 
\bea
\cU^{(2n)} (v) &=&\big({\rm i}\, v^{1} v^{2}\big)^n \cU^{[2n]}(\z) =
\big(v^{1}\big)^{2n} \big({\rm i}\, \z\big)^n \cU^{[2n]}(\z)~,  \non \\
\cU^{[2n]}(\z) &=& 
\sum_{k=-\infty}^{\infty} \cU_k  \z^k~,
\qquad  {\bar \cU}_k = (-1)^k \cU_{-k} ~.
\label{2n-tropica1}
\eea
This multiplet is holomorphic in the intersection of the north and south charts 
of the projective space ${\mathbb C}P^1$.

It should be pointed out that the modern projective superspace terminology was introduced in \cite{G-RRWLvU}.

%%%%%%%%%%%%%%%%%%%%%%%%%%%%%%%%%%%%

\subsection{Non-superconformal projective multiplets}

In the original papers \cite{LR1,LR2}, general projective multiplets were introduced for the case of $\cN=2$ Poincar\'e supersymmetry, while definition \eqref{RigidProjectiveM} corresponds to superconformal projective multiplets. 
To define the former, 
the conformal Killing supervector field $\x$ in \eqref{RigidProjectiveM} should be replaced by a Killing supervector field
\bea
\X = \X^b  \pa_b + \X^\b_j D_\b^j
+ {\bar \X}_{\bd}^j  {\bar D}^{\bd}_j= {\overline \X} 
~.
\label{2.46}
\eea 
By definition, $\X$  is a conformal Killing supervector field such that the parameters 
\eqref{2.7b} and \eqref{2.7c} vanish.
Its components are obtained from \eqref{chiraltra} by switching several parameters off: 
\begin{subequations} 
	\bea
	\X_+^{\ad\a} &=& a^{\ad\a}  
	+{\bar K}^\ad{}_\bd  \,y^{\bd \a} +y^{\ad\b}K_\b{}^\a 
	 +4{\rm i}\, {\bar \e}^{\ad i}  \q^{\a}_i ~,
	\\
	\X^\a_i &=& \e^\a_i  + \q^\b_i K_\b{}^\a 
	~,
	\eea
\end{subequations}
where the complex four-vector $\X^a_+$ is related to the vector component $\X^a$ 
in \eqref{2.46}  by the rule
$\X_+^a = \X^a + 2\ri \X_i \s^a  {\bar \q}^i.$
The super-Poincar\'e transformation law of a weight-$n$ projective multiplet
$Q^{(n)} (z,v)$  is obtained 
from \eqref{harmultQ} by replacing $\x \to \X$:
\bea
\d_\X Q^{(n)} = 
  \X   Q^{(n)}  ~.
  \label{248}
\eea
It is seen that the weight $n$ of $Q^{(n)}$ becomes irrelevant from the point of view of the Poincar\'e supersymmetry. In particular, for the arctic  \eqref{arctic1} 
and antarctic \eqref{antarctic1} multiplets we can use  the simplified notation
\bea
\U^{(n)} ( v) =  (v^{\1})^n\, \U ( \z)~, \qquad
\breve{\U}^{(n)} (v) =
(v^{\1} \,\z \big)^{n}\, \breve{\U}(\z)~.
\eea

%%%%%%%%%%%%%%%%%%%%%%%%%%%%%%%%%%%%
%%%%%%%%%%%%%%%%%%%%%%%%%%%%%%%%%%%%

\section{Rigid supersymmetric sigma models}

In order to get a better understanding of the opportunities provided by the projective multiplets, in this section we briefly discuss off-shell $\cN=2$ supersymmetric sigma models in Minkowski superspace. 
 We recall that the target spaces of $\cN=2$ supersymmetric sigma models 
are hyperk\"ahler manifolds in the super-Poincar\'e case \cite{A-GF} and hyperk\"ahler cones in the superconformal case \cite{deWKV1,deWKV2} (see also \cite{GR}
for the mathematical framework and \cite{Sezgin:1994th} for a discussion in dimensions $3\leq d \leq 6$).

The $\cN=2$ {\it supersymmetric action principle} in projective superspace is formulated in terms of 
a Lagrangian $\cL^{(2)}(z,v)$ which is  a  real weight-2 projective superfield.
The action  is
\bea
S:=  \frac{1}{2\p} \oint_{\g}  (v, \rd v)
\int {\rm d}^4x \, D^{(-4)} \cL^{(2)} (z,v) \Big|_{\q={\bar \q} =0}~, 
 \qquad  (v, \rd v) := v^i \rd v_i ~,
\label{PAP}
\eea
where $\g$ denotes a closed integration contour, 
and $D^{(-4)}$ is the  fourth-order differential operator:
\bea
D^{(-4)} := \frac{1}{16} (D^{(-1) })^2  (\bar D^{(-1)})^2~, \quad
D^{(-1)}_\a := \frac{{ u_i} D^i_\a  }{ (v,u)} 
~, \quad 
\bar D^{(-1)}_{\ad} := \frac{u_i{\bar D}^i_{ \ad} }{(v,u)} 
 ~.~~
\eea
We recall that  $u_i$ is a {fixed} isotwistor chosen to be arbitrary modulo 
the condition $(v,u) \neq 0$ along the integration contour.
The action  is invariant under arbitrary
{\it projective transformations} of the form:
 \bea
(u^i\,,\,v^i)~\to~(u^i\,,\, v^i )\,{\mathfrak R}~,\qquad {\mathfrak R}=
\left(\begin{array}{cc}{\mathfrak a}~&0\\ {\mathfrak b}~&{\mathfrak c}~\end{array}\right)\,\in\,\sGL(2,\mathbb{C})~.
\label{projectiveGaugeVar}
\eea
This gauge-like  symmetry implies that the action is independent of $u_i$, 
$\d_u S=0$.
It is also invariant under $\cN=2$ supersymmetry transformations 
\bea
 \d_{\rm SUSY} \cL^{(2)} =
 \big(  \e^\a_i \, { Q^i_\a} 
+ {\bar \e}^i_{ \ad} \, {{\bar Q}^{ \ad}_i } \big)\cL^{(2)} ~.
\eea
The projective superspace action was originally given in \cite{KLR} 
in a form that differs slightly from \eqref{PAP}.  
The latter representation appeared first in  \cite{Siegel85}.

The action \eqref{PAP} is superconformal if the Lagrangian $ \cL^{(2)} $ is a superconformal 
weight-2 projective multiplet, see \cite{K06,K07} for the proof. As an example, we consider an off-shell nonlinear $\s$-model described by $n$ superconformal weight-1 arctic multiplets 
$\U^{(1) I} $ and their smile-conjugates $\breve{\U}^{(1) \bar I} $ with Lagrangian \cite{K07}
\begin{subequations} \label{superconformal_sigma}
\bea
 \cL^{(2)} &=&  \ri  \cL( \U^{(1)} , \breve{\U}^{(1)} ) ~, \\
 2 \cL( \U^{(1)} , \breve{\U}^{(1)} ) &=&
 \Big( \U^{ (1)I} \frac{\pa}{\pa \U^{(1) I} } 
 +  \breve{\U}^{(1) \bar I}  \frac{\pa}{\pa \breve{\U}^{(1) \bar I} } \Big) 
 \cL( \U^{(1)} , \breve{\U}^{(1)} ) ~.
 \eea
 \end{subequations}
 In order for $\cL^{(2)}$ to be real, it suffices to choose  
 \bea
 \cL( \U^{(1)} , \breve{\U}^{(1)} ) =  \cK( \U^{(1)} , \breve{\U}^{(1)} ) ~, \quad
 \F^I \frac{\pa}{\pa \F^I} \cK(\F, \bar \F) =  \cK( \F,   \bar \F)~,
 \label{Kkahler2}
 \eea
where 
 $\cK (\F , \bar \F) $ is a real analytic function of $n$ complex variables $\F^I$ and their complex conjugates $\bar \F^{\bar I}$. The homogeneity properties of $\cK$ imply that it can be interpreted as the K\"ahler potential of a K\"ahler cone \cite{GR}.
 
 The Lagrangian $\cL^{(2)}$ in the general case of \eqref{superconformal_sigma}
is real if $\cL(\F , \bar \O)$  obeys the reality condition 
$\bar{\cL}(\bar \F , - \O) = - \cL(\F, \bar \O)$, 
where $\bar {\cL} (\bar \F ,  \O)$ denotes the complex conjugate of  ${\cL} ( \F , \bar \O )$.  
 A detailed study of the superconformal $\s$-model 
\eqref{Kkahler2} was carried out  in  \cite{K07,K-duality}. That analysis was extended and generalised in \cite{Butter:2014xua} to the case of the most general superconformal $\s$-model \eqref{superconformal_sigma}.

Without loss of generality,  we can assume  that the integration contour $\g$
does not pass through the ``{north pole}'' $v^{i} \sim (0,1)$.
%Then, making use of 
This chart is naturally parametrised by 
the inhomogeneous complex coordinate $ \z$ 
%on  ${\mathbb C}P^1 \setminus\{\infty\}$ 
defined by $ v^i = v^{\1} \,(1, \z)$.
Since the action is independent of $u_i$, the latter can be chosen to be   $ u_i =(1,0)$, 
such that $(v,u) = v^{\1}\neq 0$. 
We also represent the Lagrangian in the form: 
\bea
\cL^{(2)}(z,v)={\rm i} \,v^{\1}v^{\2}\cL(z,\z)
= {\rm i} (v^{\1})^2 \,\z\,{ \cL(z,\z)~, 
\qquad \breve{\cL} =\cL}~. 
\eea
Now, the action takes the form:
\bea
S= \frac{1}{ 16}\oint_\g  \frac{\rd\z }{ 2\pi\ri}
\int\rd^4 x\,\z\,
({D}^{\1})^2({\bar D}_{\2})^2\cL(z,\z)\Big|_{\q_i={\bar \q}^i =0} ~.
\label{ac2}
\eea
${}$Finally, the analyticity constraints \eqref{ana} on $\cL\propto \cL^{(2)}$
 are equivalent to 
\bea
D^{\2}_{\a}\cL(\z)=\z\,D^{\1}_{\a}\cL(\z)~, \qquad 
{\bar D}_{\2}^{\ad} \cL(\z)=-\frac{1}{\z}\,{\bar D}_{\1}^{\ad}\cL(\z)~,
\label{39}
\eea
hence the action turns into 
\bea
S &=&
 \frac{1 }{ 2\pi\ri }
 \oint_\g 
 \frac{\rd\z }{  \z}
\int\rd^{4|4} z \,
\cL(z,\z)\Big|_{\q_{\2}={\bar \q}^{\2} =0}~, \qquad
\rd^{4|4}z:=\rd^4x\,\rd^2\q \rd^2 \bar \q~.
\eea
Here the integration is carried out over the $\cN=1$ Minkowski superspace with Grassmann coordinates $\q^\a \equiv \q^\a_\1$ and $\bar \q_\ad \equiv \bar \q_\ad^\1$.
The action is now formulated entirely in terms of $\cN=1$ superfields. 
By construction, it is off-shell $\cN=2$ supersymmetric! 
This is one of the most powerful features of the projective superspace approach.
In what follows, we assume that $\q^\a_{\2}=0$ and ${\bar \q}_\ad^{\2} =0$.

The most general off-shell $\cN=2$ supersymmetric nonlinear $\s$-model 
in projective superspace \cite{LR1} can be realised in terms of polar supermultiplets
\bea
S[\U, \breve{\U}]  =  
\frac{1}{2\pi {\rm i}} \, 
\oint_\g \frac{{\rm d}\z}{\z} \,  
 \int 
 \rd^{4|4} z
 \, 
\cL \big( \U^I  , \breve{\U}^{\bar{J}} , \z  \big) ~,
\label{nact} 
\eea
where 
the arctic $\U (\z)$ and  antarctic $\breve{\U} (\z)$ dynamical variables  
are generated by an infinite set of ordinary $\cN=1$ superfields:
\begin{subequations}
\bea
 \U (\z) &=& \sum_{n=0}^{\infty}  \, \U_n \z^n = 
\F + \S \,\z+ O(\z^2) ~,\\
\breve{\U} (\z) &=& \sum_{n=0}^{\infty}  \, {\bar
\U}_n
 (-\z)^{-n}=\bar \F -\frac{1}{\z} \bar \S +O(\z^{-2}) ~.
\label{exp}
\eea
\end{subequations}
As follows from the analyticity conditions \eqref{ana}
(see also \eqref{39}),  $\F:=\U_0 $ is chiral, $\bar D_{\ad} \F =0$, $\S:=\U_1 $  is complex linear, $\bar D^2 \S=0$, 
while the remaining components, $\U_2, \U_3, \dots, $ are unconstrained complex 
$\cN=1$ superfields.  
The latter superfields are auxiliary, since they  appear in the action without derivatives. 

Although the $\s$-model  \eqref{nact} was first
introduced in 1988  \cite{LR1},
for some ten years it remained  a purely formal construction, because  
there existed no technique to eliminate the auxiliary superfields contained in $\U^I$, 
except in the case of  Lagrangians quadratic in $\U^I $ and $\breve{\U}^{\bar I}$.
This situation changed  in the late 1990s when  Refs. \cite{K98,GK1,GK2} 
identified a subclass of  models (\ref{nact}) with interesting geometric properties.
They correspond to the special case
\bea
\cL \big( \U^I  , \breve{\U}^{\bar{J}} , \z  \big) = K \big( \U^I  , \breve{\U}^{\bar{J}}  \big) ~,
\label{nozeta}
\eea
where $K \big( \F^I  , \bar \F^{\bar{J}}  \big)$ is the K\"ahler potential of a K\"ahler manifold $\cM$.
The K\"ahler invariance 
$K(\F, \bar \F) \to K(\F, \bar \F) +
\L(\F)+  {\bar \L} (\bar \F) $
of the $\cN=1$ supersymmetric $\s$-model \cite{Zumino79},
\bea
S[\F, \bar \F]  =  
 \int 
 \rd^{4|4} z
 \, 
K \big( \F^I  , \bar \F^{\bar{J}}   \big) ~,
\label{N=1sigma} 
\eea
turns into 
\be
K(\U, \breve{\U})  \quad \longrightarrow \quad K(\U, \breve{\U}) ~+~
\L(\U) \,+\, {\bar \L} (\breve{\U} ) 
\label{kahl2}
\ee
for the model 
\bea
S[\U, \breve{\U}]  =  
\frac{1}{2\pi {\rm i}} \, 
\oint_\g \frac{{\rm d}\z}{\z} \,  
 \int 
 \rd^{4|4} z
 \, 
K \big( \U^I  , \breve{\U}^{\bar{J}}   \big) ~.
\label{nact-Kahl} 
\eea
A holomorphic reparametrisation of the K\"ahler manifold,
$ \F^I  \to \F'{}^I=f^I \big( \F \big) $,
has the following
counterpart
$
\U^I (\z) \to \U'{}^I(\z)=f^I \big (\U(\z) \big)
$
in the $\cN=2$ case. Therefore, the physical
superfields of the 
${\cal N}=2$ theory, $\F^I$ and $\S^I$, 
should be regarded as  coordinates of a point in the K\" ahler
manifold and a tangent vector at  the same point, respectively.
Thus the variables $(\F^I, \S^J)$ parametrise the holomorphic tangent 
bundle $T\cM$ of the K\"ahler manifold $\cM$ \cite{K98}. 

We assume that the auxiliary superfields in the model  (\ref{nact}) have been eliminated. 
Then,
the action  (\ref{nact})  turns into
\bea
S= \int  \rd^{4|4} z \, {\mathbb L} (\F ,  \bar \F ,\S , \bar \S )~,
\label{on-shell-action}
 \eea
 for some Lagrangian ${\mathbb L}$. In the case of the model \eqref{nact-Kahl}, 
${\mathbb L} $ has the form \cite{GK2}
\bea
{\mathbb L} (\F,  \bar \F,  \S ,\bar \S )
=
K \big( \F, \bar{\F} \big)+  
\sum_{n=1}^{\infty}  \cL_{I_1 \cdots I_n {\bar J}_1 \cdots {\bar 
J}_n }  \big( \F, \bar{\F} \big) \S^{I_1} \dots \S^{I_n} 
{\bar \S}^{ {\bar J}_1 } \dots {\bar \S}^{ {\bar J}_n }
~,~~~~~
\label{act-tab}
\eea
where $\cL_{I {\bar J} }=  - g_{I \bar{J}} \big( \F, \bar{\F}  \big) $ 
and the coefficients $\cL_{I_1 \cdots I_n {\bar J}_1 \cdots {\bar 
J}_n }$, for  $n>1$, 
are tensor functions of the K\"ahler metric
$g_{I \bar{J}} \big( \F, \bar{\F}  \big) 
= \pa_I 
\pa_ {\bar J}K ( \F , \bar{\F} )$,  the Riemann curvature $R_{I {\bar 
J} K {\bar L}} \big( \F, \bar{\F} \big) $ and its covariant 
derivatives. Each term in the action contains equal powers
of $\S$ and $\bar \S$, since the original model (\ref{nact-Kahl}) 
is invariant under rigid $\sU(1)$  transformations
 \cite{GK1}
\be
\U(\zeta) ~~ \mapsto ~~ \U({\rm e}^{{\rm i} \a} \zeta) 
\quad \Longleftrightarrow \quad 
\U_n(z) ~~ \mapsto ~~ {\rm e}^{{\rm i} n \a} \U_n(z) ~.
\label{rfiber}
\ee

To uncover the explicit structure of the hyperk\"ahler target space associated with the $\s$-model
(\ref{on-shell-action}), we should construct a dual formulation 
of the theory (\ref{on-shell-action}), 
obtained by dualising each complex linear superfield
$\S^I$ and its conjugate $\bar \S^{\bar I}$ into a chiral--antichiral pair $\J_I$ and $\bar \J_{\bar I}$. In accordance with \cite{LR1},
this is accomplished through the use of the first-order action
\bea
S_{\text{first-order}}=   \int \rd^{4|4} z\, 
\Big\{ {\mathbb L}\big(\F, \bar \F, \S , \bar \S \big)
+\J_I \,\S^I + {\bar \J}_{\bar I} {\bar \S}^{\bar I} 
\Big\}~.
\label{f-o}
\eea
Here  $\S^I$ is  an unconstrained complex superfield, 
while  $\Psi_I$ is chiral, 
${\bar D}_{ \ad} \J_I =0$.
This model is equivalent to  (\ref{on-shell-action}). Indeed, varying $S_{\text{first-order}}$
with respect to $\J^I$  gives $\bar D^2 \S^I =0$ 
and then (\ref{f-o}) reduces to the original theory, eq (\ref{on-shell-action}). 
On the other hand, we can integrate out $\S$'s and their conjugates using their equations of motion
\bea
\frac{\pa  }{\pa \S^I}  {\mathbb L}\big(\F, \bar \F, \S , \bar \S \big)+ \J_I =0~,
\eea 
which can be used to express the variables $\S^I$ and their conjugates in terms of the other superfields, 
$\S^I = \S^I (\F, \J, \bar \F, \bar \J)$.
This leads to the dual action
\bea
S_{\text{dual}}  
&=&   \int \rd^{4|4}z \, {\mathbb K} \big(\F,  \J , \bar \F ,\bar \J \big)~.
\label{act-ctb}
\eea
Since this $\cN=2$ supersymmetric $\s$-model is formulated in terms of chiral $\cN=1$ superfields, its Lagrangian ${\mathbb K} \big(\F,  \J , \bar \F ,\bar \J \big)$ is the K\"ahler potential of a hyperk\"ahler manifold \cite{HKLR} (or simply the {\it hyperk\"ahler potential}).

It may be shown \cite{Kuzenko:2011ib}
that the dual theory \eqref{act-ctb} is invariant under the second supersymmetry transformation 
\bea
\d \F^I &=&\hf {\bar D}^2 \Big\{ \bar{\e} \bar{\q} \, \frac{\pa \mathbb K}{\pa \J_I} \Big\} ~, 
\qquad
\d \J_I =- \hf {\bar D}^2 \Big\{ \bar{\e} \bar{\q} \, 
\frac{\pa \mathbb K}{\pa \F^I}   \Big\}~.
\label{SUSY-ctb4}
\eea
These transformation laws follow from the linear supersymmetry  \eqref{248} 
of the off-shell theory \eqref{nact}. 
If we introduce the condensed notation 
$\f^a := (\F^I\,, \J_I)$ and  ${\bar \f}^{\,\bar a} = ({\bar \F}^{\bar I}\,, {\bar \J}_{\bar I})$, as well as
the symplectic matrices
\bea
{\mathbb J}^{a b} = {\mathbb J}^{\bar a \bar b} = 
\left(
\begin{array}{rc}
0 ~ &  {\mathbbm 1} \\
-{\mathbbm 1} ~ & 0  
\end{array}
\right)~,  
\eea
then the supersymmetry transformation (\ref{SUSY-ctb4}) can be rewritten as 
\bea 
\d \f^a &=&\hf 
{\bar D}^2 \Big\{ \bar{\e} \bar{\q} \,  {\mathbb J}^{ab}  \frac{\pa \mathbb K}{\pa \f^b} \Big\} ~,
\label{SUSY-ctb5}
\eea
which agrees with the general results of \cite{HKLR}. A remarkable result of 
Lindstr\"om and Ro\v{c}ek \cite{LR2008} is the observation that 
the $\cN=2$ superfield Lagrangian in  (\ref{nact}) can be identified 
with the generating function of a twisted symplectomorphism. 

In the case of the model \eqref{nact-Kahl}, the hyperk\"ahler potential 
has the form 
\be
{\mathbb K} (\F, \J, \bar \F,  \bar \J )
=K \big( \F, \bar{\F} \big)+  
\sum_{n=1}^{\infty} \cH^{I_1 \cdots I_n {\bar J}_1 \cdots {\bar 
J}_n }  \big( \F, \bar{\F} \big) \J_{I_1} \dots \J_{I_n} 
{\bar \J}_{ {\bar J}_1 } \dots {\bar \J}_{ {\bar J}_n } 
\label{323}
\ee
where 
$\cH^{I {\bar J}} \big( \F, \bar{\F} \big) 
= g^{I {\bar J}} \big( \F, \bar{\F} \big) $. By construction, $(\S^I , {\bar \S}^{\bar I} ) $ is a  tangent vector at the point $(\F^I, \bar \F^{\bar I})$ of
$\cM$, therefore $(\J_I , {\bar \J}_{\bar I} )$ is a one-form at the same point.
The variables $\f^{\rm a} = (\F^I, \J_I)$ parametrise the holomorphic cotangent 
bundle $T^* \cM$ of the K\"ahler manifold $\cM$ \cite{GK1,GK2}.
The hyperk\"ahler potential \eqref{323} was computed for all Hermitian symmetric spaces, 
see \cite{AKL1,AKL2,KN} and references therein.

To conclude this section, we consider one more example of an off-shell $\s$-model, introduced in \cite{GHK}.  It is described by several real $\cO(2)$ multiplets $H^{(2)I} (v)$, where $I=1,\dots, n$, which we represent as 
 \bea
 H^{(2)I} (v) = \ri (v^\1)^2 H^I (\z) ~, \qquad H^I(\z) = \F^I +\z G^I -\z^2 \bar \F^I~.
 \eea
 The action functional is defined as follows
 \bea
S &=&-
 \frac{1 }{ 2\pi\ri }
 \oint_\g 
 \frac{\rd\z }{  \z}
\int\rd^{4|4} z \, \frac{ F\big(H^I (\z)\big) }{\z^2 } +{\rm c.c}~,
\label{328}
\eea
 where $\g$ is a closed contour around the origin, and $F(z^I)$ is a holomorphic function of $n$ variables. In accordance with the analyticity conditions \eqref{RigidProjectiveM}, 
the $\cN=1$ superfield   $\F^I$ is chiral, $\bar D_{\ad} \F^I =0$, 
while the real superfield $G^I =\bar G^I $  is  linear, $\bar D^2 G^I=0$. The contour integral in 
\eqref{328} is easy to evaluate if we take into account that 
\bea
F \big(H (\z)\big) = F(\F) + \z F_I(\F) G^I  -\z^2 \Big( F_I (\F) \bar \F^I - \hf F_{IJ} G^I G^J \Big) 
+O(\z^3)~.
\eea
Only the $\z^2$ term in this expression contributes to the contour integral. Thus we get 
\begin{subequations}
\bea
S[\F, \bar \F , G] =  \int\rd^{4|4} z \, \Big\{ K(\F, \bar \F) - \hf G_{IJ} (\F, \bar \F) G^IG^J\Big\}~,
\label{330}
\eea
where we have defined 
\bea
K(\F, \bar \F) = \bar \F^I F_I (\F) + \F^I \bar F_I (\bar \F)~,\qquad
g_{IJ} (\F, \bar \F) = F_{IJ} (\F) + \bar F_{IJ} (\bar \F) ~.~~~
\eea
\end{subequations}
We can interpret $K(\F, \bar \F) $ and $g_{IJ}(\F, \bar \F) $ as the K\"ahler potential of a K\"ahler manifold and the corresponding K\"ahler metric. Each linear  superfield $G^I$ in 
\eqref{330} may be dualised into into a chiral superfield $\J_I$ and its conjugate $\bar \J_I$. As a result, the action turns into
\begin{subequations}
\bea
S[\F, \J, \bar \F, \bar \J] &=&  \int\rd^{4|4} z \, {\mathbb K} [\F, \J, \bar \F, \bar \J] ~,
\label{331a} \\
 {\mathbb K} [\F, \J, \bar \F, \bar \J] &=&  K (\F,  \bar \F) +\hf g^{IJ} (\F , \bar \F) 
 ( \J_I + \bar \J_I ) ( \J_J + \bar \J_J) ~.
 \label{331b}
 \eea
 \end{subequations}
 Since the original action \eqref{328} is $\cN=2$ supersymmetric, its dual \eqref{331a} is also 
 $\cN=2$ supersymmetric. Since the latter is formulated in terms of $\cN=1$ superfields, 
 \eqref{331b} is the K\"ahler potential of a hyperk\"ahler manifold. The correspondence
$K(\F, \bar \F)  \to  {\mathbb K} [\F, \J, \bar \F, \bar \J] $ constitutes the so-called 
{\it rigid c-map} \cite{cmap1,cmap2}.
 
%%%%%%%%%%%%%%%%%%%%%%%%%%%%%%%
%%%%%%%%%%%%%%%%%%%%%%%%%%%%%%%

\section{Conformal superspace}\label{N1_conformal-superspace}

In section \ref{section2} we have reviewed a simple approach to obtain 
the $\mathcal{N}=2$ superconformal algebra by employing the conformal Killing supervector fields of flat superspace. 
The goal of this section is to construct the gauge theory of the latter, known in the literature
as conformal superspace. It was introduced in \cite{ButterN=2} as a generalisation of the $\mathcal{N}=1$ case analysed in \cite{ButterN=1}.
This approach,  which will be reviewed in the present section, is of particular importance as it provides powerful tools to construct manifestly gauge-invariant supergravity actions and to engineer general couplings of supergravity to matter.

%%%%%%%%%%%%%%%%%%%%%%%%%%%%%%

\subsection{Gauging the superconformal algebra in superspace} \label{Gauging}

Conformal superspace is a gauge theory of the superconformal algebra. It can be identified with a pair $(\cM^{4|8}, \nabla)$. Here $\mathcal{M}^{4|8}$ denotes a supermanifold parametrised by local  coordinates 
$z^M = (x^m, \q^\m_\imath, \bar \q_{\dot \m}^\imath)$, and $\nabla$ is a covariant derivative associated with the superconformal algebra. We recall that the generators $ X_{\tilde a}$ of the superconformal algebra are given by eq. \eqref{generators}. They can be grouped in two disjoint subsets,
\bea 
X_{\tilde a} = (P_A, X_{\underline{a}} )~, \qquad
X_{\underline{a}} = (M_{ab} , \mathbb{Y}, J_{ij}, {\mathbb D} , K^A)~,
\eea  
each of which constitutes a superalgebra:
\bsubeq\label{RigidAlgebra}
\begin{align}
	[P_{ {A}} , P_{ {B}} \} &= -f_{{ {A}} { {B}}}{}^{{ {C}}} P_{ {C}}
	\ , \\
	[X_{\underline{a}} , X_{\underline{b}} \} &= -f_{\underline{a} \underline{b}}{}^{\underline{c}} X_{\underline{c}} \ , \\
	[X_{\underline{a}} , P_{{B}} \} &= -f_{\underline{a} { {B}}}{}^{\underline{c}} X_{\underline{c}}
	- f_{\underline{a} { {B}}}{}^{ {C}} P_{ {C}}
	\ . \label{mixing}
\end{align}
\esubeq
Here the structure constants $f_{{ {A}}{ {B}}}{}^{ {C}}$ contain only one non-zero component, which is $f_\a^i{\,}^\bd_j{\,}^c = 2 \ri \d^i_j \,(\s^c)_\a{}^\bd$.

In order to define the covariant derivatives, $\nabla_A= (\nabla_a, \nabla_\a^i, \bar \nabla^\ad_i)$, we associate with each generator $X_{\underline{a}} =
(M_{ab} , \mathbb{Y}, J_{ij}, {\mathbb D} , K^A) = ( M_{ab} , \mathbb{Y}, J_{ij} ,{\mathbb D},K^a, S^\a_i , \bar S_\ad^i)$
a connection one-form 
$\omega^{\underline{a}} = (\O^{ab},\Phi,\Theta^{ij},B,\mathfrak{F}_{A})= (\O^{ab},\Phi,\Theta^{ij},B,\mathfrak{F}_{a},\mathfrak{F}_{\a},
\bar{\mathfrak{F}}^{\ad})= \rd z^M \omega_M{}^{\underline{a}}$,  
and with $P_{ {A}}$ a supervielbein one-form
$E^{ {A}} = (E^a, E^\a_i,\bar{E}_\ad^i) = \rd z^{ {M}} E_M{}^A$ (the latter will be often referred to as the vielbein). 
It is assumed that the supermatrix $E_M{}^A$ is nonsingular, 
$E:= {\rm Ber} (E_M{}^A) \equiv {\rm sdet} (E_M{}^A)\neq 0$, 
and hence there exists a unique inverse supervielbein. The latter is given by 
the supervector fields $E_A = E_A{}^M (z)\pa_M $, with 
$\pa_M = \pa /\pa z^M$, which constitute a new basis for the tangent space at each
point $z^M \in \cM^{4|8}$. The supermatrices $E_A{}^M $ and $E_M{}^A$ satisfy the 
properties $E_A{}^ME_M{}^B=\d_A{}^B$ and $E_M{}^AE_A{}^N=\d_M{}^N$.
With respect to  the basis $E^A$,  the connection is expressed as 
$\omega^{\underline{a}} =E^B\omega_B{}^{\underline{a}}$, 
where $\omega_B{}^{\underline{a}}=E_B{}^M\omega_M{}^{\underline{a}}$. 
The {\it covariant derivative} is given by 
\bea\label{nablaA}
\nabla_A 
&=& E_A  - \o_A{}^{\underline b} X_{\underline b}=
E_A -  \hf \Omega_A{}^{bc} M_{bc} - \ri \Phi_A {\mathbb Y} - \Theta_A{}^{jk} J_{jk} - B_A \mathbb{D}
- \mathfrak{F}_{AB} K^B~.~~~
\eea
It can be recast as a super one-form
\be\label{eq:covD}
\nabla = \rd - \omega^{\underline{a}} X_{\underline{a}} \ , \quad \nabla = E^A \nabla_A \ .
\ee
The translation generators $P_B$ do not show up in \eqref{nablaA} and 
\eqref{eq:covD}. It is assumed that the operators $\nabla_A$ replace $P_A$ and obey the graded commutation relations
\be
[ X_{\underline{b}} , \nabla_A \} = -f_{\underline{b} A}{}^C \nabla_C
- f_{\underline{b} A}{}^{\underline{c}} X_{\underline{c}} \ ,
\ee
compare with \eqref{mixing}.
In particular, the algebra of $K^A$ with $\nabla_B$ is given by
\begin{subequations}\label{Knabla}
	\begin{align}
		[K^a, \nabla_b] &= 2 \delta^a_b  \mathbb{D} + 2 M^{a}{}_b
		~, \label{Knabla.a}\\
		\{ S^\a_i , \nabla_\b^j \} &=  \d_i^j \d^\a_\b (2 \mathbb{D} - \mathbb{Y}) - 4 \d_i^j  M^\a{}_\b 
		+ 4 \d^\a_\b  J_i{}^j
		~,\label{Knabla.b}\\
		\{ \bar{S}_\ad^i , \bar{\nabla}^\bd_j \} &=   \d_j^i \d^\bd_\ad (2 \mathbb{D} + \mathbb{Y}) + 4 \d_j^i  \bar{M}_\ad{}^\bd 
		- 4 \d_\ad^\bd  J^i{}_j
		~, \label{Knabla.c} \\
		[K^a, \nabla_\b^i] &= -\ri (\s^a)_\b{}^\bd \bar{S}_\bd^i \ , \qquad \qquad \qquad[K^a, \bar{\nabla}^\bd_i] = 
		-\ri ({\s}^a)^\bd{}_\b S^\b_i ~, \label{Knabla.d} \\
		[S^\a_i , \nabla_b] &= \ri (\s_b)^\a{}_\bd \bar{\nabla}^\bd_i \ , \qquad \qquad \quad \qquad[\bar{S}_\ad^i , \nabla_b] = 
		\ri ({\s}_b)_\ad{}^\b \nabla_\b^i \ , \label{Knabla.e}
	\end{align}
\end{subequations}
where all other graded commutators vanish.

By definition, the gauge group of conformal supergravity  is generated by local transformations of the form
\begin{subequations}\label{SUGRAtransmations}
	\bea
	\delta_\cK \nabla_A &=& [\cK,\nabla_A] \ , \\
	\cK &=& \xi^B \nabla_B +  \L^{\underline{b}} X_{\underline{b}} ~, \non \\
	&=&  \xi^B \nabla_B+ \hf K^{bc} M_{bc} + \S \mathbb{D} + \ri \rho Y 
	+ \theta^{jk} J_{jk}
	+ \L_B K^B \ ,
	\eea
\end{subequations}
where  the gauge parameters satisfy natural reality conditions. 
In applying
eq. \eqref{SUGRAtransmations}, 
we interpret that
\begin{subequations}
	\bea
	\nabla_A \xi^B &:=& E_A \xi^B + \omega_A{}^{\underline{c}} \xi^D f_{D \underline{c}}{}^{B} \ , \\
	\nabla_A \L^{\underline{b}} &:=& E_A \L^{\underline{b}}
	+ \omega_A{}^{\underline{c}} \xi^D f_{D \underline{c}}{}^{\underline{b}}
	+ \omega_A{}^{\underline{c}} \L^{\underline{d}} f_{\underline{d}\underline{c}}{}^{\underline{b}} \ .
	\eea
\end{subequations}
Then it follows from \eqref{SUGRAtransmations} that 
\begin{subequations}\label{3.8}
	\begin{align}
		\delta_\cK E^A &= \rd \xi^A + E^B \L^{\underline c} f_{\underline c B}{}^A
		+ \omega^{\underline b} \xi^{C} f_{C \underline b}{}^{A}
		+ E^B \xi^{C} \scT_{C B}{}^A~, \\
		\delta_\cK \omega^{\underline a} &= \rd \L^{\underline a}
		+ \omega^{\underline b} \L^{\underline c} f_{\underline c \underline b}{}^{\underline a}
		+ \omega^{\underline b} \xi^{C} f_{C \underline b}{}^{\underline a}
		+ E^B \L^{\underline c} f_{\underline c B}{}^{\underline a}
		+ E^B \xi^{C}  \mathscr{R}_{C B}{}^{\underline a}~.
	\end{align}
\end{subequations}
Here we have made use of the graded commutation relations
\be
\label{3.9}
[ \nabla_A , \nabla_B \} = - \scT_{AB}{}^C \nabla_C - \mathscr{R}_{AB}{}^{\underline{c}} X_{\underline{c}} 
\ ,
\ee
where $ \scT_{AB}{}^C $ and $ \mathscr{R}_{AB}{}^{\underline{c}} $ denote the torsion and the curvature, respectively. 
They can be recast in terms of two-forms
\begin{subequations} \label{TRexpComp}
	\begin{align}
		\scT^A &:= \hf E^C \wedge E^B \scT_{BC}{}^A = \rd E^A - E^C \wedge \omega^{\underline{b}} \,f_{\underline{b} C}{}^A \ , \\
		\mathscr{R}^{\underline{a}} &:= \hf E^C \wedge E^B \mathscr{R}_{BC}{}^{\underline{a}} = \rd \omega^{\underline{a}}
		- E^C \wedge \omega^{\underline{b}} \, f_{\underline{b} C}{}^{\underline{a}}
		- \hf \omega^{\underline{c}} \wedge \omega^{\underline{b}} \,
		f_{\underline{b} \underline{c}}{}^{\underline{a}} \ .
	\end{align}
\end{subequations}
Making use of the graded Jacobi identity
\be 
0 = (-1)^{\ve_{\underline{a}} \ve_C } 
[ X_{\underline{a}} , [ \nabla_B , \nabla_C \} \} ~+~\text{(two cycles)}
\ee
we derive the action of $X_{\underline{a}} $ on the geometric objects
\begin{subequations}
	\begin{align}
		X_{\underline{a}} \scT_{BC}{}^D =&
		- (-1)^{\varepsilon_{\underline{a}} (\varepsilon_{B}+\varepsilon_{C})} 
		\scT_{BC}{}^{ {E}} f_{ {E} \underline{a}}{}^D
		- 2 f_{\underline{a} [B}{}^{ {E}} \scT_{| {E}| C\}}{}^D
		- 2 f_{\underline{a} [B}{}^{\underline{e}} f_{|\underline{e}| C\}}{}^D \ , \\
		X_{\underline{a}} \mathscr{R}_{BC}{}^{\underline{d}} =&
		- (-1)^{\varepsilon_{\underline{a}} (\varepsilon_{B}+\varepsilon_{C})} 
		\Big(\scT_{BC}{}^{ {E}} f_{ {E} \underline{a}}{}^{\underline{d}}
		+ \mathscr{R}_{BC}{}^{\underline{e}} f_{\underline{e}\underline{a} }{}^{\underline{d}}\Big)
		- 2 f_{\underline{a} [B}{}^{ {E}} \mathscr{R}_{| {E}| C \}}{}^{\underline{d}} \non\\
		&- 2 f_{\underline{a} [B}{}^{\underline{e}} f_{|\underline{e}| C \}}{}^{\underline{d}} \ .
	\end{align}
\end{subequations}

The supergravity gauge group acts on a conformal tensor superfield $U$ (with indices suppressed) as 
\bea 
\label{3.14}
\d_{\cK} U = \cK U \ .
\eea
The torsion
$ \scT_{AB}{}^C $ and the curvature $ \mathscr{R}_{AB}{}^{\underline{c}} $ are conformal tensor superfields.
Of special significance are primary superfields. 
A tensor superfield $U$ (with suppressed indices) is said to be \emph{primary} if it is 
 characterised by the properties 
 \bea
K^A U = 0~, \quad \mathbb D U = w U~,  \quad {\mathbb Y} U = c U~,
\eea
for some real constants $w$ and $c$, 
which are called the 
dimension (or Weyl weight)  and $\sU(1)_R$ charge of $U$ respectively.
From the algebra \eqref{superconfal3-2}, it is seen that if a superfield is annihilated by the $S$-supersymmetry generators, then it is necessarily primary. 

Let us summarise some important features of the gauging procedure. In curved superspace, the 
superconformal algebra \eqref{RigidAlgebra} is replaced with 
\begin{subequations}\label{gaugedSCA}
	\begin{align}
		[X_{\underline{a}} , X_{\underline{b}} \} &
		= -f_{\underline{a} \underline{b}}{}^{\underline{c}} X_{\underline{c}} \ ,
		\label{eq:XwithX} \\
		[X_{\underline{a}} , \nabla_B \} &= - f_{\underline{a} B}{}^C \nabla_C -f_{\underline{a} B}{}^{\underline{c}} X_{\underline{c}} \ , \label{eq:XwithNabla} \\
		[\nabla_A , \nabla_B \} &= -\scT_{AB}{}^C \nabla_C - \mathscr{R}_{AB}{}^{\underline{c}} X_{\underline{c}} \ .
	\end{align}
\end{subequations}
Here the torsion and curvature tensors obey Bianchi identities which follow from
\be 
0 = (-1)^{\ve_A \ve_C} [ \nabla_A , [ \nabla_B , \nabla_C \} \} +
\text{(two cycles)}\ .
\label{3.15}
\ee
Unlike \eqref{RigidAlgebra}, which is determined by the structure constants,  the graded commutation relations \eqref{gaugedSCA} involve
structure functions $\scT_{AB}{}^C $ and $ \mathscr{R}_{AB}{}^{\underline{c}} $.

\subsection{Conventional constraints for Weyl multiplet} 
\label{Appendix A.1}

The framework described in the previous subsection 
defines a geometric set-up to obtain a multiplet of conformal supergravity containing the metric. However, in general, the resulting multiplet is reducible. To obtain an irreducible multiplet 
it is necessary to impose constraints on the torsion and curvatures appearing in eq.~\eqref{3.9}. This is a standard task in geometric superspace approaches to supergravity, and it is pedagogically reviewed in \cite{BK,GGRS}. One beautiful feature of the construction of \cite{ButterN=2} is the simplicity of the superspace constraints needed to obtain the Weyl multiplet of conformal supergravity. In fact, to obtain a sufficient set of constraints, one requires the algebra \eqref{3.9} to have a Yang-Mills structure. Specifically, one imposes 
	\bsubeq \label{CSSAlgebra-00}
	\bea
	\label{CSSAlgebra-0}
	\{ \nabla_{\a}^i , \nabla_{\b}^j \}  &=&  -2 \ve^{ij} \ve_{\a \b} \bar{\cW} ~, \quad \{ \bar{\nabla}^{\ad}_i , \bar{\nabla}^{\bd}_j \} = 2 \ve_{ij} \ve^{\ad \bd} \cW ~, \\ &&\quad \;\; \{\nabla_{\a}^i , \bar{\nabla}^{\bd}_{j} \} = - 2 \ri \d^i_j \nabla_{\a}{}^{\bd} ~,
	\eea
	\esubeq
	where the operator $\bar{\cal{W}}$ is the complex conjugate of $\cal{W}$. The latter takes the form
	\bea
	{\cal{W}}
	&=&
	\frac{1}{2}{\cal{W}}(M)^{ab} M_{ab}
	+\ri {\cal{W}}(\mathbb Y)\mathbb Y
	+\cW(J)^{ij} J_{ij}
	+{\cal{W}}(\mathbb D)\mathbb D
	\non\\
	&&
	+{\cal{W}}(S)_\a^i S^\a_i
	+{\cal{W}}(\bar S)_{i}^\ad \bar{S}_\ad^i
	+{\cal{W}}(K)_a K^a
	~.
	\eea
	Having imposed the constraints \eqref{CSSAlgebra-0}, the Bianchi identities \eqref{3.15} become non-trivial and now play the role of consistency conditions which may be used to determine the torsion and curvature. Their solution, up to mass dimension-3/2
	is as follows
	\begin{subequations}
	\label{CSSAlgebra}
	\begin{align}
	\{ \nabla_\a^i , \nabla_\b^j \} &= 2 \ve^{ij} \ve_{\a\b} \big( \bar{W}_{\gd\dd} \bar{M}^{\gd\dd} + \frac 1 4 \bar{\nabla}_{\gd k} \bar{W}^{\gd\dd} \bar{S}^k_\dd - \frac 1 4 \nabla_{\g\dd} \bar{W}^\dd{}_\gd K^{\g \gd} \big)~, \\
	\{ \bar{\nabla}^\ad_i , \bar{\nabla}^\bd_j \} &= -2 \ve_{ij} \ve^{\ad\bd} \big( W^{\g\d} M_{\g\d} - \frac 1 4 \nabla^{\gd k} W_{\g\d} S_k^\d + \frac 1 4 \nabla^{\g\gd} W_\g^\d K_{\d \gd} \big)~, \\
	\{ \nabla_\a^i , \bar{\nabla}^\bd_j \} &= - 2 \ri \d_j^i \nabla_\a{}^\bd~, \\
	[\nabla_{\a\ad} , \nabla_\b^i ] &= - \ri \ve_{\a\b} \bar{W}_{\ad\bd} \bar{\nabla}^{\bd i} - \frac{\ri}{2} \ve_{\a\b} \bar{\nabla}^{\bd i} \bar{W}_{\ad\bd} \mathbb{D} - \frac{\ri}{4} \ve_{\a\b} \bar{\nabla}^{\bd i} \bar{W}_{\ad\bd} {\mathbb Y} + \ri \ve_{\a\b} \bar{\nabla}^\bd_j \bar{W}_{\ad\bd} J^{ij}
	\eol & \quad
	- \ri \ve_{\a\b} \bar{\nabla}_\bd^i \bar{W}_{\gd\ad} \bar{M}^{\bd \gd} - \frac{\ri}{4} \ve_{\a\b} \bar{\nabla}_\ad^i \bar{\nabla}^\bd_k \bar{W}_{\bd\gd} \bar{S}^{\gd k} + \frac{1}{2} \ve_{\a\b} \nabla^{\g \bd} \bar{W}_{\ad\bd} S^i_\g
	\eol & \quad
	+ \frac{\ri}{4} \ve_{\a\b} \bar{\nabla}_\ad^i \nabla^\g{}_\gd \bar{W}^{\gd \bd} K_{\g \bd}~, \\
	[ \nabla_{\a\ad} , \bar{\nabla}^\bd_i ] &=  \ri \d^\bd_\ad W_{\a\b} \nabla^{\b}_i + \frac{\ri}{2} \d^\bd_\ad \nabla^{\b}_i W_{\a\b} \mathbb{D} - \frac{\ri}{4} \d^\bd_\ad \nabla^{\b}_i W_{\a\b} {\mathbb Y} + \ri \d^\bd_\ad \nabla^{\b j} W_{\a\b} J_{ij}
	\eol & \quad
	+ \ri \d^\bd_\ad \nabla^{\b}_i W^\g{}_\a M_{\b\g} + \frac{\ri}{4} \d^\bd_\ad \nabla_{\a i} \nabla^{\b j} W_\b{}^\g S_{\g j} - \hf \d^\bd_\ad \nabla^\b{}_\gd W_{\a\b} \bar{S}^{\gd}_i
	\eol & \quad
	+ \frac{\ri}{4} \d^\bd_\ad \nabla_{\a i} \nabla^\g{}_\gd W_{\b\g} K^{\b\gd} ~.
	\end{align}
	\end{subequations}
	
	We note that the conformal superspace algebra is expressed in terms of a single superfield $W_{\alpha \beta }= W_{(\a\b)}$, its conjugate $\bar W_{\ad \bd}$, and their covariant derivatives. This superfield is an $\mathcal{N}=2$ extension of the Weyl tensor, and is called the super-Weyl tensor. It proves to be a primary chiral superfield of dimension 1,
\bea
K^C W_{\a \b } =0~, \quad \bar \nabla^\gd_k W_{\a\b}=0 ~, \quad
{\mathbb D} W_{\a\b} = W_{\a\b}~,\quad
{\mathbb Y} W_{\a\b} = -2 W_{\a\b}~,
\label{constr-W-1}
\eea
	and it obeys the Bianchi identity
	\bsubeq
	\bea
	{\mathfrak B} &:=&  \nabla_{\a \b} W^{\a\b}
	= \bar \nabla^{\ad \bd} \bar W_{\ad \bd}
	= \bar {\mathfrak B}~,
	\label{super-Bach} \\
	\nabla_{\a \b} &:=& \nabla_{(\a}^{i} \nabla_{\b) i} ~, \qquad \bar{\nabla}^{\ad \bd} := \bar{\nabla}^{(\ad}_i \bar{\nabla}^{\bd) i} ~.
	\eea
	\esubeq
	The real scalar superfield $\mathfrak B$ is
	the $\cN=2$ supersymmetric generalisation of the Bach tensor.
	This {\it super-Bach multiplet} proves to be primary, $K^A {\mathfrak B} =0$, carries weight $2$, ${\mathbb D} {\mathfrak B}= 2 {\mathfrak B}$, and satisfies the conservation equation \cite{BdeWKL}
\begin{subequations}
	\bea 
	\nabla^{ij} {\mathfrak B}&=&0 \quad \Longleftrightarrow \quad
	\bar \nabla_{ij} {\mathfrak B} =0
	~, \label{ConsEq} \\
	\nabla^{ij} &:=& \nabla^{\a(i} \nabla_\a^{ j)} ~, \qquad
\bar \nabla_{ij} := \bar{\nabla}_{\ad (i} \bar{\nabla}^\ad_{j)}~.
	\eea
	\end{subequations}
	
	The structure of the conformal superspace algebra leads to highly non-trivial implications. In particular, eq. \eqref{Knabla.c} implies that 
	primary covariantly chiral superfields, $\bar \nabla^\bd_j U=0$,  can carry neither isospinor nor dotted spinor indices.
	Given such a superfield, $\f_{\a(n)} :=\f_{\a_1 \dots \a_n} =\f_{(\a_1 \dots \a_n)}$, eq. \eqref{Knabla.c} further implies that 
	the  $\sU(1)_R$ charge of $\f_{\a(n)}$ 
	is determined in terms of its dimension, 
	\begin{align}
	K^B \f_{\a(n)} =0~, \quad \bar \nabla_j^\bd \f_{\a(n)} =0 \ ,\quad {\mathbb D} \f_{\a(n)} = w \f_{\a(n)} ~, \quad
	{\mathbb Y} \f_{\a(n)} = -2w \f_{\a(n)} 
	\label{chirals} 
	\end{align}
	and thus $c = - 2  w$.
	
There is a regular procedure to construct primary chiral multiplets and their conjugate antichiral ones. It is based on the use of operators
\bea
\nabla^4 :=\frac{1}{48}\nabla^{ij}\nabla_{ij}=-\frac{1}{48}\nabla^{\a\b}\nabla_{\a\b}~,\quad
\bar \nabla^4:=\frac{1}{48} \bar \nabla^{ij} \bar \nabla_{ij}=-\frac{1}{48}
\bar \nabla^{\ad\bd} \bar \nabla_{\ad\bd}~.
\label{3.25}
\eea
Let us consider a rank-$n$ spinor superfield $\j_{\a(n)}$ that is $\sSU(2)_R$ neutral and has the following superconformal properties:
\bea
K^B  \j_{\a(n)} = 0~, \quad {\mathbb D} \j_{\a(n)} = (w-2)  \j_{\a(n)}~, 
\quad {\mathbb Y}  \j_{\a(n)} = 2(2-w)  \j_{\a(n)}~.
\eea
Then its descendant 
\bea
\label{4.27}
 \f_{\a(n)} = \bar \nabla^4  \j_{\a(n)}
 \eea
 is a primary covariantly chiral superfield of the type \eqref{chirals}.
	
%%%%%%%%%%%%%%%%%%%%%%%%%%%%%%%%%%
%%%%%%%%%%%%%%%%%%%%%%%%%%%%%%%%%%

\subsection{Covariant projective multiplets}

The concept of rigid superconformal projective multiplets, which was reviewed in subsection 
\ref{section2.3}, naturally extends to conformal superspace. The operators 
\eqref{2.28} are replaced with 
\bea
\nabla^{(1)}_\a = v_i \nabla^i_\a  ~, \qquad {\bar \nabla}^{(1)}_{ \ad} = v_i {\bar \nabla}^i_{\ad}~, \eea
which strictly anti-commute with each other due to \eqref{CSSAlgebra-00}.
We recall that the rigid superconformal projective multiplet $Q^{(n)}(z,v) $ is defined by the relations \eqref{RigidProjectiveM}, of which the conditions \eqref{ana} and \eqref{weight} trivially extend to conformal superspace, 
\begin{subequations} \label{3.29}
\bea
K^A Q^{(n)} &=&0~, \qquad \nabla^{(1)}_{\a} Q^{(n)} =0~, \quad {\bar \nabla}^{(1)}_{\ad} Q^{(n)} =0~,
\label{ana-local} \\
Q^{(n)}(z,{\mathfrak c} \,v)&=&{\mathfrak c}^n\,Q^{(n)}(z,v)~, \qquad {\mathfrak c}\in \mathbb{C}^*~,
\label{weight-local}
\eea
\end{subequations}
while the rigid superconformal transformation law \eqref{harmultQ} is replaced with 
\bsubeq \label{3.30}
\bea
\d_\cK Q^{(n)} 
&=& \Big( \x^A \de_A + \L^{ij} J_{ij}+\S\bbD \Big) Q^{(n)} ~,  
\\ 
\L^{ij} J_{ij}  Q^{(n)}&=& -\Big(\L^{(2)} {\pa}^{(-2)} 
-n \, \L^{(0)}\Big) Q^{(n)} ~.
\label{harmult2}   
\eea
Making use of the graded commutation relations \eqref{Knabla.b} and \eqref{Knabla.c}
uniquely fixes the dimension of $Q^{(n)} $
\bea
\bbD  Q^{(n)} = n   Q^{(n)} ~.
\eea
\esubeq

We now list some projective multiplets that can be  used to describe superfield 
dynamical variables.
A complex $\cO(m) $ multiplet, 
with $m=1,2,\dots$,   is described by a  weight-$m$ projective superfield $H^{(m)} (v)$ 
of the form:
\begin{subequations} \label{54}
\bea
H^{(m)} (v) &=&v_{i_1} \dots v_{i_{m}}  H^{i_1 \dots i_{m}} 
~.
\eea
The analyticity constraint (\ref{ana}) is equivalent to 
\bea
\de_\a^{(i_1} H^{i_2 \dots i_{m+1} )} =0~, \qquad \bar \de_\ad^{(i_1} H^{i_2 \dots i_{m+1} )} =0~.
\eea
\end{subequations}
If $m$ is even, $m=2n$, we can define a real $\cO(2n) $ multiplet
obeying 
the reality condition $\breve{H}^{(2n)}  = {H}^{(2n)} $, or equivalently
\bea
\overline{ H^{i_1 \dots i_{2n}} } &=& H_{i_1 \dots i_{2n}}
=\ve_{i_1 j_1} \cdots \ve_{i_{2n} j_{2n} } H^{j_1 \dots j_{2n}} ~.
\label{55}
\eea
For $n>1$, 
the real $\cO(2n) $ multiplet can be used to describe an off-shell (neutral) hypermultiplet. 

There is a simple construction to generate covariant projective multiplets. 
It makes use of isotwistor superfields. 
By definition, a weight-$n$ isotwistor superfield  $U^{(n)} (z,v)$ is a primary tensor superfield (with suppressed Lorentz indices)
that  
 has the following properties:
(i) it is neutral with respect to the group $\sU(1)_R$;  
(ii) it is holomorphic with respect to 
the isospinor variables $v^i $ on an open domain of 
${\mathbb C}^2 \setminus  \{0\}$; (iii) it 
is a homogeneous function of $v^i$ of degree $n$,
\bsubeq
\bea
&&U^{(n)}({\mathfrak c} \,v)\,=\,{\mathfrak c}^n\,U^{(n)}(v)~, \qquad {\mathfrak c}\in 
{\mathbb C} \setminus  \{0\}
~;
\eea
and (iv) it is  characterised by the gauge transformation law
\bea
\d_\cK U^{(n)} 
&=& \Big( \x^A \de_A
+  \hf \L^{ab} M_{ab}
+\L^{ij} J_{ij} 
+\S\bbD
\Big) 
U^{(n)} ~,  ~~~
\non \\
J_{ij}  U^{(n)}&=& -\Big(v_{(i}v_{j)}{\pa}^{(-2)} 
-\frac{n}{(v,u)}v_{(i}u_{j)}\Big) U^{(n)}
~.
\label{iso2}
\eea 
\esubeq
It is clear that any weight-$n$ projective multiplet  is an isotwistor superfield, but not vice versa.
The main property in the definition of isotwistor superfields is their transformation rules under $\sSU(2)_R$.
In principle, the definition could be extended to consider non-primary superfields.

Let  $U^{(n-4)}$ be a Lorentz-scalar isotwistor superfield
such that 
\bea
\bbD  U^{(n-4)}=
(n-2)  U^{(n-4)}~.
\label{iso3}
\eea
Then the weight-$n$ isotwistor superfield 
\bea
Q^{(n)}:=\nabla^{(4)}U^{(n-4)}
\eea
satisfies all the properties of a covariant projective multiplet given by eqs. 
\eqref{3.29} and \eqref{3.30}.  Here we have introduced the operator 
\bea 
\nabla^{(4)} = \frac{1}{16} \nabla^{(2)} \bar \nabla^{(2)}~, \qquad
\nabla^{(2)} = v_i v_j \nabla^{ij} ~, \quad \bar \nabla^{(2)} = v_i v_j \bar \nabla^{ij} ~.
\label{3.42}
\eea

%%%%%%%%%%%%%%%%%%%%%%%%%%%%%%%%%%%%%%%%%%%%%%

\section{Component reduction and the Weyl multiplet} 
\label{section4}

Within the superconformal tensor calculus, the standard Weyl multiplet of conformal supergravity
is associated with the local off-shell gauging in spacetime of the
superconformal group $\sSU(2,2|2)$ \cite{deWvHVP,deWvHVP2, deWPV, deWit:1983xhu, deWit:1984rvr},
see also 
\cite{FVP,Lauria:2020rhc} for a review.
This multiplet comprises $24+24$ physical components described by a set of independent gauge fields:
the vielbein $e_m{}^a$ and a dilatation connection $b_m$;
the gravitino $\big(\psi_m{}^\alpha_i,\bar\psi_m{}_\ad^i\big)$, associated with the gauging of $Q$-supersymmetry;
a $\sU(1)_R$ gauge field $A_m$;
and $\sSU(2)_R$ gauge fields $\phi_m{}^{ij}=\phi_m{}^{ji}$.
The fields associated with the remaining generators of $\sSU(2,2|2)$,
specifically the Lorentz connections $\o_m{}^{cd}$, 
$S$-supersymmetry connection $\big(\phi_m{}_\alpha^i,\bar\phi_m{}^\ad_i\big)$
and the special conformal connection $\mathfrak{f}_m{}_a$, are composite fields.
To ensure that the local superconformal transformations of the standard Weyl multiplet close off-shell
it is necessary to add a set of covariant matter fields.
These are an anti-symmetric real tensor $T_{ab}=T_{ba}=T^+_{ab}+T^-_{ab}$, 
which decomposes
into its imaginary (anti-)self-dual components $T^{\pm}_{ab}$,
a real scalar field $D$,
and the fermions $(\Sigma^{\a i},\bar{\Sigma}_{\ad i})$.
 
As described
in the previous section, conformal superspace provides an off-shell gauging of the
superconformal group $\sSU(2,2|2)$ in superspace rather than spacetime. Apart from the fact that $Q$-supersymmetry is geometrically realised on superfields in a superspace setting, the conformal superspace and component approaches are very similar. In fact, it is straightforward to reduce the results of \ref{N1_conformal-superspace} from superspace to spacetime and obtain all the details of the standard Weyl multiplet \cite{ButterN=2}. 

The identification of the component gauge fields of the standard Weyl multiplet is straightforward.
The vielbein ($e_m{}^a$) and gravitini ($\psi_m{}^\a_i,\bar{\psi}_m{}_\ad^i$) appear as the $\q=0$ projections of the coefficients of 
$\rd x^m$ in the supervielbein $E^A$ one-form,
\begin{align}
e{}^a = \rd x^m e_m{}^a = E^a\doubar~,~~~
\psi^\a_i = \rd x^m \psi_m{}^\a_i =  2 E^\a_i \doubar
~,~~~
\bar{\psi}_\ad^i = \rd x^m \bar{\psi}_m{}_\ad^i =  2 E_\ad^i \doubar
 ~.
\end{align}
Here we have defined the double bar projection of a superform as $\O\doubar \equiv \O|_{\theta = \rd \theta = 0}$. On the other hand, a single bar next to a superfield denotes the usual bar projection $X| \equiv X|_{\q=0}$.
The remaining component one-forms are defined as 
\bea
&
A:=\Phi\doubar~,\quad
\phi^{kl} := \Theta{}^{kl} \doubar~, \quad
b := B\doubar ~, \quad
\omega^{cd} := \Omega{}^{cd} \doubar ~, 
\\
&{\phi}_\g^k := 2 \,{{\mathfrak F}}{}_\g^k\doubar~, \quad
\bar{\phi}^\gd_k := 2 \,{{\mathfrak F}}{}^\gd_k\doubar~, \quad
{\mathfrak{f}}{}_c := {\mathfrak{F}}{}_c\doubar~. 
\eea
The covariant matter fields $T_{ab}$, $D$, and $(\Sigma^{\a i},\bar{\Sigma}_{\ad i})$ arise as some of the components of the multiplet described by the super-Weyl tensor $W_{ab}=(\s_{ab})^{\a\b}W_{\a\b}-(\tilde{\s}_{ab})^{\ad\bd}\bar{W}_{\ad\bd}$, which satisfies the constraints \eqref{constr-W-1} and \eqref{super-Bach}. In particular, it holds that
\bsubeq\label{TWSSb}
\bea
T_{ab}&:=&W_{ab}\loco
~,\qquad
D=
\frac{1}{12}\de^{\a \b}W_{\a\b}\loco
=
\frac{1}{12}\deb^{\ad \bd}\bar{W}_{\ad\bd}\loco
~,
\\
&&\Sigma^{\a i}=\frac{1}{3}\de_\b^i W^{\a\b}\loco
~,~~~
\bar{\Sigma}_{\ad i}=-\frac{1}{3}\deb^\bd_i\bar{W}_{\ad\bd}\loco
~.
\eea
\esubeq
The local superconformal transformations of the gauge fields listed above can be straightforwardly derived by taking the $\q=0$ projection of the superspace transformations \eqref{3.8}. At the same time, the transformations of $T_{ab}$, $D$, and $(\Sigma^{\a i},\bar{\Sigma}_{\ad i})$ can be obtained by applying the transformation rule for covariant superfields, eq.~\eqref{3.14} and \eqref{SUGRAtransmations}, and the definition of the descendant fields in eq.~\eqref{TWSSb}. The resulting transformation laws are given in \cite{Gold:2022bdk}.

By taking the double bar projection of the superspace covariant derivative one-form $\de$, eq.~\eqref{eq:covD},
one defines a component vector covariant derivative  as follows\bsubeq\label{comp-der}
\bea
{\rm D}&=&e^a{\rm D_a}:=\de\doubar
~,
\\
e_m{}^a {\rm D}_a
 &=& 
 \pa_m
- \frac{1}{2} \psi_m{}^\a_i \de_{\a}^i\loco
- \frac{1}{2} \bar{\psi}_m{}_\ad^i \bar{\de}^{\ad}_i \loco
- \frac{1}{2}  {\omega}_m{}^{cd} M_{cd}
-\ri   A_m {\mathbb Y}
-  \phi_m{}^{kl} J_{kl}
\non\\
&&
-  {b}_m \mathbb D
- \frac{1}{2}  {\phi}_m{}_{\a}^{ i} S^{\a}_{ i}
- \frac{1}{2}  \bar{{\phi}}_m{}^{\ad}_{i} \bar{S}_{\ad}^{ i}
-  {{\mathfrak f}}_m{}_{c} K^c
~.
\eea
\esubeq
Provided we appropriately interpret the projected spinor covariant derivatives $\nabla_{\a}^i\loco$ and $\bar{\nabla}^{\ad}_i\loco$
as the generators of $Q$-supersymmetry,\footnote{Given a covariant superfield $U$, and its lowest component $\cU = U\vert$, one defines $Q_\a^i\cU =\nabla_\a^i| \cU := (\nabla_\a^i U) \loco$ and $\bar{Q}^\ad_i\cU =\bar{\nabla}^\ad_i| \cU := (\bar{\nabla}^\ad_i U) \loco$. The action of the other generators $X_{\underline{a}}$ on $\cU$ is simply given by  $X_{\underline{a}}\cU :=(X_{\underline{a}} U)\loco$.} $\rm D$ describes a gauging in space-time of the
superconformal group SU$(2,2|2)$, precisely as in \cite{deWvHVP}.
This means that local diffeomorphisms, and all other structure group transformations of the derivatives \eqref{comp-der}, including $Q$-supersymmetry, consistently descend from their corresponding rule in superspace.
With this interpretation, the algebra of component covariant derivatives acting on a covariant field is also completely determined
by the geometry of conformal superspace. All the component torsions and curvatures are simply the $\theta=0$ projections of the superspace ones. 
The algebra of ${\rm D}_a$ is\footnote{All fields and curvatures introduced so far satisfy natural conjugation properties. We refer the reader to \cite{ButterN=2} and, in particular, \cite{Gold:2022bdk} for results in our notation, with the only difference being that the field $W_{ab}$ in \cite{Gold:2022bdk} is denoted as $T_{ab}$ here.} 
\bea
{[}{\rm D}_a,{\rm D}_b{]}
&=&
-{R}(P)_{ab}{}^c{\rm D}_c
-{R}(Q)_{ab}{}^\a_i \de_\a^i\loco
-{R}(\bar{Q})_{ab}{}_\ad^i \bar{\de}^\ad_i\loco
\non\\
&&
- \frac{1}{2}  {R}(M)_{ab}{}^{cd} M_{cd}
-  {R}(\mathbb D)_{ab}\mathbb D
-\ri  R(Y)_{ab}{\mathbb Y}
-  R(J)_{ab}{}^{kl} J_{kl}
\non\\
&&
- R(S)_{ab}{}_{\a}^{ i} S^{\a}_{ i}
- R(\bar{S})_{ab}{}^{\ad}_{i} \bar{S}_{\ad}^{ i}
- R(K)_{ab}{}_{c} K^c
~.~~~~~~~~
\eea
By using the commutator of two superspace vector derivatives $\de_a$, see \cite{ButterN=2},
one can readily obtain all the component curvatures above. These prove to be determined by the lowest component of the super-Weyl tensor  $W_{ab}$ and its descendants. We do not present the results here but stress that the conformal superspace geometry implies the following conditions on the component superconformal curvatures
\bsubeq\label{conventional-constraints}
\bea
R(P)_{ab}{}^c
&=&
0~,
\\
R(Q)_{ab}{}^\b_j({\s}^b)_{\b\ad}
&=&
-\frac{3}{4}\S^\b_{j}(\s_a)_{\b\ad}
~,~~~
R(\bar{Q})_{ab}{}_\bd^{j}({\tilde\s}^b)^{\bd\a}
=
\frac{3}{4}\bar\S_\bd^{j}(\tilde\s_a)^{\bd\a}
~,
\\
R(M)^c{}_{a}{}_{cb}
&=&
R(\mathbb D)_{ab}
+3\eta_{ab}D
-\eta^{cd}
T^-_{ac}T^+_{bd}
~.
\eea
\esubeq
These are the conventional constraints that render the connections $\o_m{}^{cd}$, $(\phi_m{}_\alpha^i,\bar\phi_m{}^\ad_i)$,
and $\mathfrak{f}_m{}_a$ composite. We refer the reader to \cite{ButterN=2,Gold:2022bdk}  for the expressions of the composite connections and the superconformal curvatures expressed in terms of the independent physical fields of the standard Weyl multiplet. 
Note that the conventional constraints \eqref{conventional-constraints} are not the same as the ones originally employed in \cite{deWvHVP}. This is not surprising since there is large freedom in the choice of conventional constraints whenever it is necessary to add matter fields to achieve an off-shell representation. Different papers often make different choices. For example, the geometry of \cite{deWvHVP} is obtained through a shift of the special conformal connection $\mathfrak{f}_{ab}K^b$ proportional to $DK_a$, see \cite{ButterN=2}. A particularly useful choice of constraints for calculations by using component fields  is the ``traceless'' one employed in \cite{Butter:2017pbp} 
\begin{subequations}
\begin{align}
R(P)_{ab}{}^c&= 0~, \quad R(M)^c{}_{a}{}_{cb}= R(\mathbb D)_{ab}~, \\
R(Q)_{ab}{}^\b_j({\s}^b)_{\b\ad} &= 0 ~, \quad R(\bar{Q})_{ab}{}_\bd^{j}({\tilde\s}^b)^{\bd\a}=0.
\end{align}
\end{subequations}

%%%%%%%%%%%%%%%%%%%%%%%%%%%%%%%%%%%%%%
%%%%%%%%%%%%%%%%%%%%%%%%%%%%%%%%%%%%%%

\section{Other superspace formulations for conformal supergravity}
\label{Section5}

As pointed out in section \ref{Section1}, conformal superspace is not the only superspace setting to describe conformal supergravity. Here we
consider two other covariant formulations that have found applications in the recent years, specifically:
(i) $\sU(2)$ superspace \cite{Howe,KLRT-M2}; and (ii) $\sSU(2)$ superspace \cite{Grimm,KLRT-M1}. They differ by their structure groups, which are $\sSL( 2, {\mathbb C}) \times \sU(2)_R$ and $\sSL( 2, {\mathbb C}) \times \sSU(2)_R$, respectively.
Below we describe the relevant ``degauging'' procedures that lead to these geometries.

\subsection{$\sU(2)$ superspace}
\label{DegaugingSection}

According to \eqref{SUGRAtransmations}, under an infinitesimal special superconformal gauge transformation $\mathcal{K} = \Lambda_{B} K^{B}$, the dilatation connection transforms as follows
\bea
\d_{\mathcal{K}} B_{A} = - 2 \Lambda_{A} ~.
\eea
Thus, it is possible to choose a gauge condition
$B_{A} = 0$, which completely fixes 
the special superconformal gauge freedom.\footnote{There is a class of residual gauge transformations preserving the gauge $B_{A}=0$. These generate the super-Weyl transformations of $\sU(2)$ superspace, see the next subsection.} As a result, the corresponding connection is no longer required for the covariance of $\nabla_A$ under the residual gauge freedom and
may be
extracted from $\nabla_{A}$,
\bea
\nabla_{A} &=& \mathfrak{D}_{A} - \mathfrak{F}_{AB} K^{B} ~. \label{ND}
\eea
Here the operator $\mathfrak{D}_{A} $ involves only the Lorentz and $\sU(2)_R$ connections
\bea
\mathfrak{D}_A = E_A - \frac{1}{2} \O_A{}^{bc} M_{bc} - \F_A{}^{kl} J_{kl} - \ri \F_A \mathbb{Y}~.
\eea
It obeys the graded commutation relations
\bea
{[}\mathfrak{D}_{{A}},\mathfrak{D}_{{B}}\}&=&
- \mathfrak{T}_{{A}{B}}{}^{{C}} \mathfrak{D}_{{C}}
- \hf \mathfrak{R}_{{A}{B}}{}^{{c}{d}}M_{{c}{d}}
- \mathfrak{R}_{{A}{B}}{}^{kl}J_{kl}
- \ri \mathfrak{R}_{AB} \mathbb{Y}
~.
\label{U(2)algebra1}
\eea

The next step is to relate the special superconformal connection
$\mathfrak{F}_{AB}$  to the torsion tensor of $\sU(2)$ superspace. To do this, one can  make use of the relation
\bea
\label{4.3}
[ \nabla_{A} , \nabla_{B} \} &=& [ \mathfrak{D}_{A} , \mathfrak{D}_{B} \} - \big(\mathfrak{D}_{A} \mathfrak{F}_{BC} - (-1)^{AB} \mathfrak{D}_{B} \mathfrak{F}_{AC} \big) K^C - \mathfrak{F}_{AC} [ K^{C} , \nabla_B \} \non \\
&& + (-1)^{AB} \mathfrak{F}_{BC} [ K^{C} , \nabla_A \} + (-1)^{BC} \mathfrak{F}_{AC} \mathfrak{F}_{BD} [K^D , K^C \} ~.
\eea
In conjunction with \eqref{CSSAlgebra}, this relation leads to a set on consistency conditions that are equivalent to the Bianchi identities of $\sU(2)$ superspace \cite{Howe}. 
Their solution expresses the components of $\mathfrak{F}_{AB}$ in terms of the torsion 
tensor of $\sU(2)$ superspace and completely determines the geometry of the $\mathfrak{D}_{A}$ derivatives \cite{ButterN=2}. Here we will present results only up to mass dimension-3/2.
The outcome of the analysis is as follows:
\begin{subequations} \label{connections}
\bea
\mathfrak{F}_\a^i{}_\b^j  
&=&
-\hf\ve_{\a\b}S^{ij}
+\hf\ve^{ij}Y_{\a\b}
~,
 \\
 \mathfrak{F}^\ad_i{}^\bd_j  
&=&
-\hf\ve^{\ad\bd}\bar{S}_{ij}
+\hf\ve_{ij}\bar{Y}^{\ad\bd}
~,\\
 \mathfrak{F}_\a^i{}^\bd_j
&=&
- \mathfrak{F}^\bd_j{}_\a^i
=
-\d^i_jG_\a{}^\bd
 -\ri G_\a{}^\bd{}^i{}_j
 ~,
 \\
 \mathfrak{F}_{\a}^{i}{}_{b}
&=&-\hf(\tilde{\s}_b)^{\bd\b}\Big\{~
\frac{\ri}{4}\ve_{\a\b}\bar{\mathfrak{D}}^{\gd i} \bar{W}_{\bd\gd}  
-\frac{1}{6}\ve_{\a\b}\mathfrak{D}^{\g}_{j}G_{\g\bd}{}^{ij}
+\frac{\ri}{12}\ve_{\a\b}\bar{\mathfrak{D}}_{\bd j}S^{ij}
\non\\
&&~~~~~~~~~~~~~~~~~~~\,
-\frac{\ri}{4}\bar{\mathfrak{D}}_\bd^iY_{\a\b}
+\frac{1}{3}\mathfrak{D}_{(\a j}G_{\b)\bd}{}^{ij}\Big\}
~,
\\
\mathfrak{F}^{\ad}_{i}{}_b
&=&
-\hf({\s}_b)_{\b\bd}\Big\{~
\frac{\ri}{4}\ve^{\ad\bd}{\mathfrak{D}}_{\g i}W^{\b\g}  
+\frac{1}{6}\ve^{\ad\bd}\bar{\mathfrak{D}}_{\gd}^{ j}G^{\b\gd}{}_{ij}
+\frac{\ri}{12}\ve^{\ad\bd}{\mathfrak{D}}^{\b j}\bar{S}_{ij}
\non\\
&&~~~~~~~~~~~~~~~~~~~\,
-\frac{\ri}{4}\mathfrak{D}^\b_{i}\bar{Y}^{\ad\bd}
-\frac{1}{3}\bar{\mathfrak{D}}^{(\ad  j}G^{\b\bd)}{}_{ij}
\Big\}
~,
\\
\mathfrak{F}_a{}_\b^j 
&=&
-\hf(\tilde{\s}_a)^{\ad\a}\Big\{
-\frac{\ri}{12}\ve_{\a\b}\bar{\mathfrak{D}}_{\bd j}S^{kj}
-\frac{\ri}{4}\bar{\mathfrak{D}}_\ad^jY_{\a\b}
+\frac{1}{3}\mathfrak{D}_{\a k}G_{\b\ad}{}^{jk}
\Big\}
~,
\\
\mathfrak{F}_a{}^\bd_j 
&=&
-\hf({\s}_b)_{\bd\b}\Big\{
-\frac{\ri}{12}\ve^{\ad\bd}{\mathfrak{D}}^{\b j}\bar{S}_{kj}
-\frac{\ri}{4}{\mathfrak{D}}^{\a}_{j}\bar{Y}^{\ad\bd}
-\frac{1}{3}\bar{\mathfrak{D}}^{\ad k}G^{\a\bd}{}_{jk}
\Big\}
~.
\eea
\end{subequations}
The dimension-1 superfields have the following symmetry properties:  
\bea
S^{ij}=S^{ji}~, \qquad Y_{\a\b}=Y_{\b\a}~, 
\qquad W_{\a\b}=W_{\b\a}~, \qquad G_{\a\ad}{}^{ij}=G_{\a\ad}{}^{ji}~,
\eea
and the reality conditions
\bea
\overline{S^{ij}} =  \bar{S}_{ij}~,\quad
\overline{W_{\a\b}} = \bar{W}_{\ad\bd}~,\quad
\overline{Y_{\a\b}} = \bar{Y}_{\ad\bd}~,\quad
\overline{G_{\b\ad}} = G_{\a\bd}~,\quad
\overline{G_{\b\ad}{}^{ij}} = ~G_{\a\bd}{}_{ij}
~.~~~~~~
\eea
The ${\rm U}(1)_R$ charges of the complex fields are:
\bea
{\mathbb Y} \,S^{ij}=2S^{ij}~,\qquad
{\mathbb Y}  \,Y_{\a\b}=2Y_{\a\b}~, \qquad
{\mathbb Y} \, W_{\a\b}=-2W_{\a\b}~.
\eea
The algebra obeyed by ${\mathfrak{D}}_A$ takes the form:
\begin{subequations} \label{U(2)algebra}
\bea
\{ \mathfrak{D}_\a^i , \mathfrak{D}_\b^j \}
 &=&
 4 S^{ij}  M_{\a\b} 
 +2\ve_{\a\b}\ve^{ij}Y^{\g\d}  M_{\g\d}  
 +2\ve^{ij} \ve_{\a\b}  \bar{W}_{\gd\dd} \bar{M}^{\gd\dd} 
\non\\
&&
+2\ve_{\a\b}\ve^{ij}S^{kl}  J_{kl}
+ 4Y_{\a\b}  J^{ij}
~,
 \label{U(2)algebra.a}
 \\
\{ \mathfrak{D}_\a^i , \bar{\mathfrak{D}}^\bd_j \}
&=&
- 2 \ri \d_j^i\mathfrak{D}_\a{}^\bd
 +4\Big(
 \d^i_jG^{\g\bd}
 +\ri G^{\g\bd}{}^i{}_j
 \Big) 
 M_{\a\g} 
 +4\Big(
 \d^i_jG_{\a\gd}
 +\ri G_{\a\gd}{}^i{}_j
 \Big)  
 \bar{M}^{\bd\gd}
 \non\\
 &&
+8 G_\a{}^\bd J^i{}_j
-4\ri\d^i_j G_\a{}^\bd{}^{kl}  J_{kl}
-2\Big(
 \d^i_jG_\a{}^\bd
 +\ri G_\a{}^\bd{}^i{}_j
 \Big)
 \mathbb{Y} 
 ~,
 \label{U(2)algebra.b}
 \\
 {[} \mathfrak{D}_a, \mathfrak{D}_\b^j{]}&=& 
 -\ri 
(\ts_a)^{\ad\g}\Big(
\d^j_kG_{\b\ad}
+ \ri G_{\b\ad}{}^{j}{}_k\Big)
\mfD_\g^k
\non\\
&&
+{\frac\ri 2}\Big(({\s}_a)_{\b\gd}S^{jk}
-\ve^{jk}({\s}_a)_\b{}^{\dd}\bar{W}_{\dd\gd}
-\ve^{jk}({\s}_a)^{\a}{}_\gd Y_{\a\b}\Big)\mfDB^\gd_k
\non\\
&&
-\hf \mathfrak{R}_a{}_\b^j{}^{cd}M_{{c}{d}}
-\mathfrak{R}_a{}_\b^j{}^{kl}J_{kl}
-\ri \mathfrak{R}_a{}_\b^j\,{\mathbb Y}
~.
\label{U(2)algebra.c}
\eea
\esubeq
The dimension-3/2 components of the curvature appearing in (\ref{U(2)algebra.c}) 
are
\begin{subequations}
\bea
\mathfrak{R}_a{}_\b^j{}_{cd}&=&
-\ri(\s_d)_{\b}{}^{\dd} \mathfrak{T}_{ac}{}_\dd^j
+\ri(\s_a)_{\b}{}^{\dd} \mathfrak{T}_{cd}{}_\dd^j
-\ri(\s_c)_{\b}{}^{\dd} \mathfrak{T}_{da}{}_\dd^j
~,
\label{3/2curvature-1}
\\
\mathfrak{R}_a{}_{\b}^j{}^{kl}
&=&
-\hf(\tilde{\s}_a)^{\ad\a}\Big\{
{\ri}\ve^{j(k}\mfDB_\ad^{l)}Y_{\a\b}
+{\ri}\ve_{\a\b}\ve^{j(k}\mfDB^{\dd l)}\bar{W}_{\ad\dd}
+{\frac\ri 3}\ve_{\a\b}\ve^{j(k}\mfDB_{\ad q}S^{l)q}
\non\\
&&~~~~~~~~~~~~~~~~~~~~
-{\frac43}\ve^{j(k}\mfD_{(\a q}G_{\b)\ad}{}^{l)q}
-{\frac23}\ve_{\a\b}\ve^{j(k}\mfD^\d_{q}G_{\d\ad}{}^{l)q}
\Big\}
~,
\\
\mathfrak{R}_a{}_{\b}^j&=&
-\hf(\tilde{\s}_a)^{\ad\a}\Big\{
\mfD_{\b}^jG_{\a\ad}
-{\frac\ri 3}\mfD_{(\a k}G_{\b)\ad}{}^{jk}
-{\frac\ri 2}\ve_{\a\b}\mfD^{\g}_kG_{\g\ad}{}^{jk}
\Big\}
~,
\eea
\end{subequations}
together with their complex conjugates.
The right-hand side of  (\ref{3/2curvature-1}) involves the dimension-3/2 components 
of the torsion, which take the form
\begin{subequations}
\bea
&&\mathfrak{T}_{ab}{}_\gd^k\equiv(\s_{ab})^{\a\b}\mathfrak{T}_{\a\b}{}_{\gd}^{k}
-(\ts_{ab})^{\ad\bd}\mathfrak{T}_{\ad\bd}{}_{\gd}^{k}~,~~~
\\
&&\mathfrak{T}_{\a\b}{}_{\gd}^{k}
=
{\frac14}\mfDB_{\gd}^{k}Y_{\a\b}
-{\frac\ri 3}\mfD_{(\a}^lG_{\b)\gd}{}^{k}{}_l
~,
\\
&&\mathfrak{T}_{\ad\bd}{}_{\gd}^{ k}
=
{\frac14}\mfDB_{\gd}^k\bar{W}_{\ad\bd}
+{\frac16}\ve_{\gd(\ad}\mfDB_{\bd)l}S^{kl}
+{\frac\ri 3}\ve_{\gd(\ad}\mfD^{\d}_qG_{\d\bd)}{}^{kq}
~.
\eea
\end{subequations}
The consistency conditions arising from solving \eqref{4.3} and the constraints satisfied by $W_{\a\b}$ in conformal superspace lead to the following set of dimension-3/2 Bianchi identities:
\begin{subequations}\label{BI-U2}
\bea
\mfD_{\a}^{(i}S^{jk)}&=&0~,
 \\
\mfD_\a^i\bar{W}_{\bd\gd}&=&0~,\\
\mfD_{(\a}^{i}Y_{\b\g)}&=&0~,
 \\
\mfD_{(\a}^{(i}G_{\b)\bd}{}^{jk)}&=&0~, 
\\
\mfDB_{\ad}^{(i}S^{jk)} &=& \ri\mfD^{\b (i}G_{\b\ad}{}^{jk)}~,
\\
\mfD_{\a}^{i}S_{ij}
&=&
-\mfD^{\b}_{j}Y_{\b\a}
~,
\\
\mfD_\a^iG_{\b\bd}&=&
- \frac{1}{ 4}\mfDB_\bd^iY_{\a\b}
+ \frac{1}{ 12}\ve_{\a\b}\mfDB_{\bd j}S^{ij}
- \frac{1}{ 4}\ve_{\a\b}\mfDB^{\gd i}\bar{W}_{\gd\bd}
\non\\
&&
- \frac{\ri }{ 3}\ve_{\a\b}\mfD^{\g}_j G_{\g \bd}{}^{ij}
~,
\eea
\end{subequations}
and the dimension-2 constraint
\bea
\big( \mfD_{\a \b}
-4Y_{\a\b} \big) W^{\a\b}
&=& \big( \mfDB^{\ad \bd}
-4\bar{Y}^{\ad\bd} \big) \bar{W}_{\ad\bd}
~.
\label{BI-U2-2}
\eea
Here we have made the definitions
\begin{align}
	\mfD_{\a \b} = \mfD^i_{(\a} \mfD_{\b) i} ~, \qquad \bar{\mfD}_{\ad \bd} = \mfDB_i^{( \ad }\mfDB^{\bd ) i}~,
\end{align}
and it is useful to also define
\begin{align}
	\mfD^{ij} = \mfD^{\a (i} \mfD_{\a}^{j)} ~, \qquad \bar{\mfD}_{i j} = \mfDB_{\ad (i}\mfDB^{\ad}_{j)}~.
\end{align}

In closing, we note that, upon degauging, relation \eqref{4.27} takes the form \cite{KT-M2009,Muller}
\begin{align}
	\label{6.17}
	\f_{\a(n)} &= \Big( \frac{1}{96} \bar{\mathfrak{D}}^{ij}\bar{\mathfrak{D}}_{ij} - \frac{1}{96} \bar{\mathfrak{D}}_{\ad \bd}\bar{\mathfrak{D}}^{\ad \bd} + \frac 1 6 \bar{S}^{ij} \bar{\mathfrak{D}}_{ij} + \frac 1 6 \bar{Y}_{\ad \bd} \bar{\mathfrak{D}}^{\ad \bd} \Big) \psi_{\a(n)} \non \\
	&\equiv \bar{\D} \psi_{\a(n)} ~.
\end{align}

%%%%%%%%%%%%%%%%%%%%%%%%%%%%%%%%%%%%
%%%%%%%%%%%%%%%%%%%%%%%%%%%%%%%%%%%%

\subsection{The super-Weyl transformations of $\sU(2)$ superspace}

In the previous subsection we made use of the special conformal gauge freedom to degauge from conformal to $\sU(2)$ superspace. The goal of this subsection is to show that residual dilatation symmetry manifests in the latter as super-Weyl transformations.

To preserve the gauge $B_{A}=0$, every local dilatation transformation with parameter $ \S $ should be accompanied by a compensating special conformal one
\begin{align}
\mathcal{K}(\Sigma) = \L_{B} (\Sigma)K^{B} + \S \mathbb{D} \quad \implies \quad \d_{\mathcal{K}( \S )} B_A = 0~.
\end{align}
We then arrive at the following constraints
\bea
\L_{A}(\S) = \hf \nabla_A \S~.
\eea
As a result, we define the following transformation
\bea
\d_{\S} \nabla_{A} &=& \d_{\S} \mathfrak{D}_{A} - \d_{\S} \mathfrak{F}_{AB} K^{B} = [\mathcal{K}(\S) ~, \nabla_{A} ]~.
\eea

By making use of \eqref{connections}, one can obtain the following transformation laws for the $\sU(2)$ superspace covariant derivatives
\begin{subequations}
\bea
\d_{\S} \mathfrak{D}_\a^i&=&\hf\S\mathfrak{D}_\a^i+2(\mathfrak{D}^{\g i}\S)M_{\g\a}-2(\mathfrak{D}_{\a k}\S)J^{ki}
-\hf(\mathfrak{D}_\a^i\S) \,{\mathbb Y} 
~,
\label{Finite_D}\\
\d_{\S}\bar{\mathfrak{D}}_{\ad i}&=&\hf\S\bar{\mathfrak{D}}_{\ad i}
+2(\bar{\mathfrak{D}}^{\gd}_{i}\S)\bar{M}_{\gd\ad}
+2(\bar{\mathfrak{D}}_{\ad}^{k}\S)J_{ki}
+\hf(\bar{\mathfrak{D}}_{\ad i}\S)\,{\mathbb Y}~,
\label{Finite_Db} 
\\
\d_{\S}\mathfrak{D}_{\a\ad}
&=&
\S\mathfrak{D}_{\a\ad}
+\ri(\bar{\mathfrak{D}}_{\ad k}\S)\mathfrak{D}_\a^k
+\ri(\mathfrak{D}_\a^k\S)\bar{\mathfrak{D}}_{\ad k}
\non\\
&&
+(\mathfrak{D}^\g{}_\ad \S)M_{\g\a}
+(\mathfrak{D}_\a{}^\gd \S)\bar{M}_{\gd\ad}
~.
\label{Finite_D_c}
\eea
\end{subequations}
The dimension-1 components of the torsion transform as
\begin{subequations}
\bea
\d_{\S} W_{\a\b}&=&\S{W}_{\a \b}~,
\label{Finite_W}
\\
\d_{\S} Y_{\a\b}&=&\S Y_{\a\b}
-\hf\mathfrak{D}_{\a \b}\S
\label{Finite_Y}~,
\\
\d_{\S} S_{ij}&=&\S S_{ij}
-\hf\mathfrak{D}_{ij} \S
\label{Finite_S}~,
\\
\d_{\S} G_{\a\ad}&=&
\S G_{\a\ad}
-{\frac18}[\mathfrak{D}_\a^k,\bar{\mathfrak{D}}_{\ad k}]\S
~,
\label{Finite_G}
\\
\d_{\S} G_{\a\ad}{}^{ij}&=&\S G_{\a\ad}{}^{ij}
+{\frac\ri 4}[\mathfrak{D}_\a^{(i},\bar{\mathfrak{D}}_\ad^{j)}]\S
~.
\label{Finite_Gij}
\eea
\end{subequations}

%%%%%%%%%%%%%%%%%%%%%%%%%%%%%%%
%%%%%%%%%%%%%%%%%%%%%%%%%%%%%%%%

\subsection{$\sSU(2)$ superspace}

It can be proven that the torsion $G_{\a\ad}{}^{ij}$ of $\sU(2)$ superspace is a pure gauge degree of freedom \cite{Howe,KLRT-M2}.
One can use  super-Weyl gauge freedom \eqref{Finite_Gij} to choose
\bea
G_{\a\bd}{}^{ij}=0~.
\label{G2}
\eea
In this gauge, it is natural to  introduce new covariant derivatives $\cD_A$ defined by
\bea
\cD_\a^i = \mathfrak{D}_\a^i~, \qquad
\cD_a=\mathfrak{D}_a-\ri \,G_a \,{\mathbb Y}~.
\eea
Making use of \eqref{U(2)algebra}, we find that they obey the graded commutation relations
\begin{subequations} 
	\label{4.21}
	\bea
	\{\cD_\a^i,\cD_\b^j\}&=&
	4S^{ij}M_{\a\b}
	+2\ve^{ij}\ve_{\a\b}Y^{\g\d}M_{\g\d}
	+2\ve^{ij}\ve_{\a\b}\bar{W}^{\gd\dd}\bar{M}_{\gd\dd}
	\non\\
	&&
	+2 \ve_{\a\b}\ve^{ij}S^{kl}J_{kl}
	+4 Y_{\a\b}J^{ij}~, 
	\label{acr1} \\
	\{\cD_\a^i,\cDB^\bd_j\}&=&
	-2\ri\d^i_j(\s^c)_\a{}^\bd\cD_c
	+4\d^{i}_{j}G^{\d\bd}M_{\a\d}
	+4\d^{i}_{j}G_{\a\gd}\bar{M}^{\gd\bd}
	+8 G_\a{}^\bd J^{i}{}_{j}~,~~~~~~~~~
	\\
	{[}\cD_a,\cD_\b^j{]}&=&
	\ri(\s_a)_{(\b}{}^{\bd}G_{\g)\bd}\cD^{\g j}
	\non\\
	&&
	+{\frac{\ri}2}\Big(({\s}_a)_{\b\gd}S^{jk}
	-\ve^{jk}({\s}_a)_\b{}^{\dd}\bar{W}_{\dd\gd}
	-\ve^{jk}({\s}_a)^{\a}{}_\gd Y_{\a\b}\Big)\cDB^\gd_k
	\non\\
	&&
	+{\frac{\ri}2}\Big((\ts_a)^{\gd\g}\ve^{j(k}\cDB_\gd^{l)}Y_{\b\g}
	-(\s_a)_{\b\gd}\ve^{j(k}\cDB_{\dd}^{l)}\bar{W}^{\gd\dd}
	-{\frac12}(\s_a)_\b{}^{\gd}\cDB_{\gd}^{j}S^{kl}\Big)
	J_{kl}
	\non\\
	&&
	+{\frac{\ri}2}\Big((\s_a)_{\b}{}^{\dd} \hat{\cT}_{cd}{}_\dd^j
	+(\s_c)_{\b}{}^{\dd} \hat{\cT}_{ad}{}_\dd^j
	-(\s_d)_{\b}{}^{\dd} \hat{\cT}_{ac}{}_\dd^j\Big)M^{{c}{d}}
	~,
	\eea
where 
\bea
\hat{\cT}_{ab}{}_\gd^k&=&-{\frac14}(\s_{ab})^{\a\b}\cDB_\gd^{ k}Y_{\a\b}
+{\frac14}(\ts_{ab})^{\ad\bd}\cDB_{\gd}^{ k}\bar{W}_{\ad\bd}
-{\frac16}(\ts_{ab})_{\gd\dd}\cDB^{\dd}_{l}S^{kl}~.
\eea
\end{subequations}
The various torsion tensors in \eqref{4.21} obey the Bianchi identities \eqref{BI-U2} and \eqref{BI-U2-2} upon the replacement $\mathfrak{D}_A \rightarrow \cD_A$ and imposing \eqref{G2}. By examining equations \eqref{4.21} we see that the $\sU(1)_R$ curvature has been eliminated and therefore the corresponding connection is flat. 
Hence, by performing an appropriate local $\sU(1)_R$ transformation it may be gauged away
\be
\label{4.23}
\F_A=0~.
\ee
As a result, the gauge group reduces to $\sSL( 2, {\mathbb C}) \times \sSU(2)_R$ and the superspace geometry is the so-called $\sSU(2)$ superspace of \cite{Grimm,KLRT-M1}.

It turns out that the gauge conditions \eqref{G2} and \eqref{4.23} allow for residual super-Weyl transformations, which are described
by a parameter $\S$ constrained by
\be
 [\mfD_\a^{(i},\mfDB_\ad^{j)}]\S=0~.
 \label{Ucon}
 \ee
 The general solution of this condition is \cite{KLRT-M1}
\bea
\S = \frac{1}{2} (\s+\sba)~, \qquad {\bar \mfD}^\ad_i \s =0~,
\qquad {\mathbb Y}\, \s =0~,
\eea
where the parameter $\s$ is covariantly chiral, with zero $\sU(1)_R$ charge, but otherwise arbitrary.
To preserve the gauge condition $\F_A=0$, every super-Weyl transformation, see (\ref{Finite_D}) and (\ref{Finite_Db}), must be accompanied by the following compensating $\sU(1)_R$ transformation 
\be
\d\mfD_A
 =[\ri \rho {\mathbb Y}, \mfD_A]~, \qquad \r = \frac{\ri}{4} (\s - \bar{\s}) ~.
\ee
As a result, the $\sSU(2)$ geometry is left invariant by the following set of super-Weyl transformations \cite{KLRT-M1}:
\bsubeq
\bea
\d_{\s} \cD_\a^i&=&\hf\sba\cD_\a^i+(\cD^{\g i}\s)M_{\g\a}-(\cD_{\a k}\s)J^{ki}~, 
\\
 \d_{\s} \cDB_{\ad i}&=&\hf\s\cDB_{\ad i}+(\cDB^{\gd}_{i}\sba)\bar{M}_{\gd\ad}
 +(\cDB_{\ad}^{k}\sba)J_{ki}~, 
\label{super-Weyl1} \\
\d_{\s} \cD_a&=&
\hf(\s+\sba)\cD_a
+{\frac{\ri}4}(\s_a)^\a{}_{\bd}(\cD_{\a}^{ k}\s)\cDB^{\bd}_{ k}
+{\frac{\ri}4}(\s_a)^{\a}{}_\bd(\cDB^{\bd}_{ k}\sba)\cD_{\a}^{ k}
\non\\
&&
-{\frac12}\big(\cD^b(\s+\sba)\big)M_{ab}
~,
\\ 
\d_{\s} S^{ij}&=&\sba S^{ij}-{\frac14}\cD^{ij} \s~, 
\label{super-Weyl-S} \\
\d_{\s} Y_{\a\b}&=&\sba Y_{\a\b}-{\frac14}\cD_{\a \b}\s~,
\label{super-Weyl-Y} \\
\d_{\s} {W}_{\a \b}&=&\s {W}_{\a \b }~,\\
\d_{\s} G_{\a\bd} &=&
\hf(\s+\sba)G_{\a\bd} -{\frac{\ri}4}
\cD_{\a \bd} (\s-\sba)~.
 \label{super-Weyl-G} 
\eea
\end{subequations}
Here we have made use of the definitions
\begin{align}
	\cD_{\a \b} = \cD^i_{(\a} \cD_{\b) i} ~, \qquad \cD^{ij} = \cD^{\a (i} \cD_{\a}^{j)}~,
\end{align}
and it is useful to also define
\begin{align}
	\bar{\cD}_{\ad \bd} = \bar{\cD}_i^{( \ad }\bar{\cD}^{\bd ) i} ~, \qquad \bar{\cD}_{i j} = \bar{\cD}_{\ad (i}\bar{\cD}^{\ad}_{j)}~.
\end{align}
Due to these transformations, $\sSU(2)$ superspace provides a geometric description of the Weyl multiplet of $\cN=2$ conformal supergravity \cite{KLRT-M1}. It should be emphasised that the algebra of covariant derivatives \eqref{4.21} was derived 
originally
by Grimm \cite{Grimm}. However, no discussion of super-Weyl transformations was given in \cite{Grimm}

Let us fix a background curved superspace   $(\cM^{4|8}, \cD)$. 
A supervector field $\x= \x^B E_B$ on this superspace
is called {\it conformal Killing} if there exist a 
Lorentz parameter $K^{bc}[\x] $, $\sSU(2)_R $
parameter $\L^{ij}[\x]$  and a chiral super-Weyl parameter $\s [\x]$ such that 
\bea
\big[ \x^B\cD_B + \hf K^{bc} [\x] M_{bc} 
+\L^{ij}[\x] J_{ij} , \cD_A\big]  + \d_{\s[\x]} \cD_A =0~.
\label{conf_KillingN=2}
\eea
In other words, the  coordinate transformation generated by $\x$ is accompanied by certain Lorentz, $\sSU(2)_R$
and super-Weyl transformations such that the superspace geometry does not change. It can be shown that the equation \eqref{conf_KillingN=2} uniquely determines the spinor components of 
$\x^B= (\x^b, \x^\b_j , \bar \x_\bd^j) $ and the parameters 
$K^{bc}[\x] $, $\L^{ij}[\x]$ and  $\s [\x]$ in terms of $\x^b$, 
and the latter obeys the equation 
\bea
\cD_{(\a}^i \x_{\b) \bd} =0 \quad \Longleftrightarrow \quad 
\bar \cD_{(\ad}^i \x_{\b \bd)} =0 ~.
\eea
The set of all conformal Killing supervector fields on $(\cM^{4|8}, \cD)$ constitutes the superconformal algebra of $(\cM^{4|8}, \cD)$. Given a super-Weyl invariant theory on $(\cM^{4|8}, \cD)$ described by primary superfields $U$, its action is invariant under the superconformal transformations
\bea
\d_\x U &=&  \cK[\x] U~, \non \\
\quad \cK[\x] &=& \x^B\cD_B + \hf K^{bc}[\x] M_{bc}  
+\L^{ij}[\x] J_{ij} 
+  p \s[\x] + q \bar \s[\x] ~,
\label{primaryTL-curved}
\eea
for an arbitrary conformal Killing supervector field $\x$. In the case that  $(\cM^{4|8}, \cD)$
coincides with Minkowski superspace,  $({\mathbb M}^{4|8}, D)$, the superconformal Killing equation 
\eqref{conf_KillingN=2} is equivalent to \eqref{4Dmaster2N=2} and the transformation law  
\eqref{primaryTL-curved}  to \eqref{FlatPrimaryMultiplet}.

%%%%%%%%%%%%%%%%%%%%%%%%%%%%%%%%%%%%%%%%
%%%%%%%%%%%%%%%%%%%%%%%%%%%%%%%%%%%%%%%%

\section{Superconformal action principles} 

To construct supergravity-matter systems, a locally superconformal action principle is required. 
Here we review three types of superconformal actions in $\cN=2$ supergravity that have played important roles in the literature. 

%%%%%%%%%%%%%%%%%%%%%%%%%%%%

\subsection{Full superspace action} 

The simplest locally superconformal action involves a full superspace integral:
\begin{align}
S[\cL]= \int\rd^{4|8}z\,  E\, \cL~,\qquad\rd^{4|8}z:=\rd^4x\,\rd^4\q \rd^4 \bar \q~, \qquad
E := {\rm Ber}(E_M{}^A)~,
\end{align}
where $\cL$ is a primary real dimensionless scalar Lagrangian,
\bea
K^A \cL =0~, \qquad \bar \cL = \cL~, \qquad {\mathbb D}\cL=0~.
\eea

As an example, we consider a superconformal higher-derivative $\sigma$-model with action  \cite{BdeWKL,deWKvZ,GHKSST}
\bea
S =   \int  \rd^{4|8}z \,E \, \cK(X^I, \bar X^{\bar J} )~, \qquad 
K^A X^I =0~, \quad \bar \nabla_i^\ad X^I =0 \ ,\quad {\mathbb D} X^I=0
\label{73}
\eea
where $\cK$ is the K\"ahler potential of a K\"ahler manifold. The 
action is locally superconformal. It is also invariant under K\"ahler transformations
\bea
\cK(X, \bar X)~ \to ~\cK(X, \bar X) + \L(X) + \bar \L (\bar X) ~,
\eea
with $\L(X) $ an arbitrary holomorphic function. 

%%%%%%%%%%%%%%%%%%%%%%%%%%

\subsection{Chiral action}

More general  is the chiral action, which involves an integral
over the chiral subspace
\bea  
S_{\rm c} [\cL_{\rm c}]= \int  \rd^4 x \, \rd^4 \q \,\cE \, \cL_{\rm c} \ .
\label{chiralAc}
\eea
Here  $\cE$ is a
suitably chosen chiral measure,
and $\cL_{\rm c}$ is a primary covariantly chiral Lagrangian of dimension $+2$,
\bea
K^A \cL_{\rm c}=0~, \qquad
\bar\nabla^\ad_i \cL_{\rm c} = 0~, \qquad {\mathbb D}\cL_{\rm c} = 2 \cL_{\rm c}~.
\label{ChiralLagrangian}
\eea
The precise definition of $\cE$ in conformal superspace is somewhat technical \cite{ButterN=2}.
In $\sSU(2)$ superspace, $\cE$ was obtained by making use of normal coordinates 
\cite{KT-M2009}.

A different definition of $S_{\rm c}$ exists, which is based on the use of a primary complex superfield $\U$ with the following superconformal properties (for some constant $w$): 
\bea 
K^A \U =0~, \quad 
{\mathbb D} \U = (w-2) \U ~,\quad \mathbb{Y} \U = 2(2-  w)\U~, 
\eea
such that $\bar \nabla^4 \U$ is nowhere vanishing,  that is $(\bar \nabla^4 \U )^{-1}$ exists.
Specifically, the chiral action may be identified with the functional 
\bea
S_{\rm c} [\cL_{\rm c}]=  \int\text{d}^{4|8}z \, E \, \frac{\U}{\bar \nabla^4 \U}
\mathcal{L}_{\rm c} ~,
\label{5.6}
\eea
which 
possesses the two fundamental properties:
(i) it is locally superconformal under the conditions \eqref{ChiralLagrangian};
and (ii) it is independent of $\U$, 
\bea
\d_\U  \int\text{d}^{4|8}z \, E \, \frac{\U}{\bar \nabla^4 \U}
\mathcal{L}_{\rm c} =0~,
\eea
for an arbitrary variation $\d \U$.
Using the representation \eqref{5.6} for the chiral action \eqref{chiralAc}, it holds that
\bea
\int\text{d}^{4|8}z \, E \, \mathcal{L}  = \int\rd^4x\rd^4\q\, \cE \,\cL_{\rm c} ~, \qquad 
\cL_{\rm c} =  \bar \nabla^4 \cL~.
\eea

There is an alternative definition of the chiral action that follows from the superform approach to 
the construction of supersymmetric invariants \cite{Castellani,Ectoplasm,GGKS}.
It is based on the use of the  following  super 4-form \cite{GKT-M}:
\bea
\Xi_4&=&
-4E_\bd^{j}\wedge E^\bd_{j}\wedge E_\ad^{i}\wedge E^\ad_{i}\,
\cL_{\rm c}
-2E_\bd^{j}\wedge E^{\bd}_{ j}\wedge E_\ad^{i}\wedge E^a\, 
(\ts_a)^{\ad\a}\de_{\a i}\cL_{\rm c}
\non\\
&&
-\frac{\ri}{2}E_\bd^{j}\wedge E_\ad^{i}\wedge E^b\wedge  E^a\,(\ts_{ab})^{\ad\bd}\de^{ij}\cL_{\rm c}
\non\\
&&
-\frac{\ri}{4}E_\ad^{i}\wedge E^\ad_{i}\wedge E^b\wedge  E^a\Big(
(\s_{ab})_{\a\b}\de^{\a\b}
-8({\ts}_{ab})_{\ad\bd}\bar{W}^{\ad\bd}\Big)\cL_{\rm c}
\non\\
&&
-\frac{\ri}{36}\ve_{abcd}E_\ad^{i}\wedge E^c\wedge E^b\wedge  E^a\Big(
(\ts^d)^{\ad\a}\de_\a^{j}\de_{ij}
-6(\ts^d)^{\bd\a}\bar{W}_{\ad\bd}\de_{\a i}\Big){\cL}_{\rm c}
\non\\
&&
+\frac{1}{24}\ve_{abcd}E^d\wedge E^c\wedge E^b\wedge E^a
\Big(
\de^4
+\bar{W}^{\ad\bd}\bar{W}_{\ad\bd}\Big)\cL_{\rm c}
~.
\label{Sigma_4}
\eea
This superform is closed, 
 \bea
 \rd \, \X_4 =0~.
 \eea
 It proves to be primary\footnote{The superform may be degauged to $\sSU(2)$ superspace. Then the condition \eqref{337} is equivalent to the super-Weyl invariance of $\X_4$. } 
 \bea
 K^B \X_4 =0~.
 \label{337}
 \eea
 The chiral action \eqref{chiralAc} can be recast
  as an integral of $\Xi_4$ over a spacetime $\cM^4$,
\begin{subequations}\label{2.24}
 \bea
 S_{\rm c} [\cL_{\rm c}]&=& \int_{\cM^4} \Xi_4
 ~, \label{2.24.a}
\eea
where $\cM^4$ is the bosonic body of the curved superspace  $\cM^{4|4}$
obtained by switching off  the Grassmann variables. 
 It turns out that 
 \eqref{2.24.a} leads to the following representation 
 \cite{ButterN=2} (see also \cite{Butter:2012xg}):
 \bea
 S_{\rm c} 
&=&
\int {\rm d}^4x\, e \,\bigg(
~
 \nabla^4 
 + \bar{W}^{\ad\bd} \bar{W}_{\ad\bd} 
 -  \frac{\ri}{12} \bar{\psi}_d{}^l_\dd \Big((\tilde{\s}^d)^{\dd \a} \nabla_\a^q \nabla_{lq} -6(\s^d)_{\a \ad} \bar{W}^{\ad \dd} \nabla^\a_l \Big)
\non\\
&&
+\frac{1}{4} \bar{\psi}_c{}_\gd^k \bar{\psi}_d{}^l_\dd  \Big( 
(\tilde{\s}^{cd})^{\gd \dd} \nabla_{kl}
-\hf \ve^{\gd \dd} \ve_{kl} (\s^{cd})_{\b \g} \nabla^{\b \g} 
- 4\ve^{\gd\dd} \ve_{kl} (\tilde{\s}^{cd})_{\ad\bd} \bar{W}^{\ad\bd} \Big)
\non\\
&&
- \frac{1}{4} \ve^{abcd} (\tilde{\s}_a)^{\bd \a} \bar{\psi}_b{}_\bd^j \bar{\psi}_c{}_\gd^k \bar{\psi}_d{}^\gd_k \nabla_{\a j} 
- \frac{\ri}{4} \ve^{abcd} \bar{\psi}_a{}_{\ad}^i \bar{\psi}_b{}^{\ad}_i \bar{\psi}_c{}_{\bd}^j \bar{\psi}_d{}^{\bd}_j \bigg) {\cL}_{\rm c}\Big|_{\theta=0} ~,~~~~~~~~~~~
\label{2.24.b}
\eea
\end{subequations}
where $e:= \det (e_m{}^a)$.
This result agrees with the action of a chiral multiplet coupled to conformal supergravity \cite{deRoo:1980mm}.

%%%%%%%%%%%%%%%%%%%%%%%%%%%%

\subsection{Projective action}

Consider a Lagrangian $\cL^{(2)}$ that is a real weight-2 projective multiplet.
Associated with $\cL^{(2)}$ is the action
\bea
S[ \cL^{(2)} ]&=&
\frac{1}{2\pi} \oint_\g (v, \rd v)
\int \rd^{4|8}z\, E\, 
\frac{ \U^{(n)} }{ \nabla^{(4)} \U^{(n)} }
\cL^{(2)}~, \qquad  (v, \rd v) := v^i \rd v_i ~,~~~
\label{InvarAc}
\eea
where $\U^{(n)} (z,v)$ is a primary weight-$n$ isotwistor superfield and the operator 
$\nabla^{(4)}$ is defined in \eqref{3.42}.
This action proves to have the following fundamental properties:  (i) it is locally superconformal; and (ii) it is independent of $\U^{(n)}$, 
\bea
\d_{\U^{(n)}} \oint_\g (v, \rd v)
\int \rd^{4|8}z\, E\, 
\frac{ \U^{(n)} }{ \nabla^{(4)} \U^{(n)} }
\cL^{(2)} =0~.
\eea

In the $n=0$ case we can specialise $\U^{(0)}$ to be $W_0 \bar W_0$, where $W_0$ is the chiral field strength of a vector multiplet, see section \ref{section8.1}, such that 
the descendant 
\bea
{\S_0}^{ij} := \frac 14 \nabla^{ij} W_0 = \frac 14 \bar \nabla^{ij} \bar W_0
\eea
is nowhere vanishing, that is $(\S_0{}^{ij} \S_{0\,ij} )^{-1}$ exists.  Then \eqref{InvarAc} turns into \cite{KLRT-M2}
\bea
S[ \cL^{(2)} ]&=&
\frac{1}{2\pi} \oint_\g (v, \rd v)
\int \rd^{4|8}z\, E\, 
\frac{ W_0\bar W_0 }{(\S_0{}^{(2)})^2 }\cL^{(2)}~.
\label{InvarAc-vec}
\eea

An important remark is in order. In the case of Minkowski superspace, it may be seen that the definition of the projective action 
\eqref{InvarAc} is equivalent to \eqref{PAP}.

There is a remarkable relationship between the projective and the chiral actions
\cite{KT-M2009,K2008} derived originally in $\sSU(2)$ superspace. 
It makes use of the vector multiplet introduced above.
For every chiral Lagrangian $\cL_{\rm c}$ with the properties \eqref{ChiralLagrangian},
the chiral action 
\bea
S_{\rm chiral}= \int \rd^4 x \,{\rm d}^4\q  \, \cE \, \cL_{\rm c} &+& {\rm c.c.}
\eea
can be represented as a projective action
\bea
S_{\rm chiral}&=&\frac{1}{2\pi} \oint_\g (v, \rd v)
\int \rd^{4|8}z\, E\,
\frac{W_0{\bar W_0}   }{({\S_0}^{(2)})^2 }\cL^{(2)}_{\rm c}
~, 
\non \\
\quad 
\cL^{(2)}_{\rm c} &=&
 -\frac{1}{4} { V} \,\Big(
 \nabla^{(2)}
 \frac{\cL_{\rm c}}{ W}
+
\bar \nabla^{(2)}
\frac{{\bar \cL}_{\rm c} }{\bar { W}} \Big)
\equiv V {\mathfrak G}^{(2)} 
~.  \label{720}
\eea
Here $V(z,v)$  is a tropical prepotential for the vector multiplet 
with the chiral field strength $ W_0$, see the next section.

On the other hand, the projective action \eqref{InvarAc-vec} can be rewritten as a special chiral action \cite{KT-M2009} 
\bea
S[\cL^{(2)}]&=&\int {\rm d}^4x \,{\rm d}^4 \q \, \cE \, W_0 {\mathbb W}~,
\qquad
{\mathbb W}  = 
 \frac{1}{8\pi}  \oint (v, \rd v)
 \bar \nabla^{(-2)} 
\Big(  \frac{\cL^{(2)}}{\S_0{}^{(2)}} \Big)~,
\label{721}
\eea
with the operator $\nabla^{(-2)} $ being defined in \eqref{6.8}. 
The composite superfield  ${\mathbb V}$ 
can be interpreted as a  tropical prepotential for the vector multiplet 
described by the reduced chiral superfield $\mathbb W$.

An important example of a dynamical system described by the projective action is provided by the off-shell sigma model
\eqref{superconformal_sigma}, in which $\U^{(1)}$ and $\breve{\U}^{(1)}$ are now covariant arctic and antarctic multiplets, respectively. This most general locally superconformal sigma model was studied in detail in \cite{Butter:2014xua}, where its component reduction was worked out.

%%%%%%%%%%%%%%%%%%%%%%%%%%%%%%%%%%%%%%%%%%%%%%%%%%%%%%%

\section{Vector and tensor multiplets}

Of special importance in $\cN=2$ supersymmetry are vector and tensor multiplets. 
Here we review their fundamental properties in the framework of conformal superspace. 

%%%%%%%%%%%%%%%%%%%%%%%%%%%%%%%%%%%%%

\subsection{Vector multiplet} \label{section8.1}
 
In rigid supersymmetry, the off-shell $\cN=2$ vector multiplet 
was formulated by Grimm, Sohnius and Wess \cite{GSW}.
In conformal superspace, it can be described by a
field strength $W$,   
which has the superconformal properties 
\begin{subequations}\label{ReducedChiral}
\bea
K^A W=0~,  \quad {\mathbb D} W = W~,\quad
\bar \nabla ^\ad_i W= 0 \label{ReducedChiral.a}
\eea
and satisfies  the Bianchi identity
\bea
\S^{ij} := \frac{1}{ 4} \nabla^{ij}W&=&\frac{1}{ 4} 
\bar \nabla^{ij} \bar{W} ~. \label{ReducedChiral.b}
\eea
\end{subequations}
Covariantly chiral scalars satisfying the reality condition \eqref{ReducedChiral.b} are called reduced chiral. The constraint \eqref{ReducedChiral.b} uniquely determines the dimension of $W$.

There are several ways to realise $W$ as a gauge invariant field strength.
One possibility is to introduce a curved superspace extension 
of Mezincescu's prepotential \cite{Mezincescu} (see also \cite{HST}),  $V_{ij}=V_{ji}$,
which is a primary  unconstrained real $\sSU(2)$ triplet of dimension $-2$. The expression for $W$ in terms of $V_{ij}$ \cite{Butter:2010jm} is
\begin{align}
W = \frac{1}{4} \bar \nabla^4 \nabla^{ij}  V_{ij}~,
\end{align}
where the chiral operator $\bar \nabla^4$ is defined in \eqref{3.25}.
It may be shown that  that $V_{ij}$ is defined only up to gauge transformations of the form
\begin{align}
\delta V^{ij} &= \nabla^{\alpha}{}_k \Lambda_\alpha{}^{kij}
     + \bar\nabla_{\ad}{}_k \bar\Lambda^\ad{}^{kij}, \quad
     \Lambda_\alpha{}^{kij} = \Lambda_\alpha{}^{(kij)}~,
     \quad \bar\Lambda^\ad{}_{kij} := \overline{ \Lambda_\alpha{}^{kij} }~,
\label{pre-gauge1}
\end{align}
with the primary gauge parameter $ \Lambda_\alpha{}^{kij} $ being completely arbitrary modulo the algebraic condition given.
The superconformal properties of $ \Lambda_\alpha{}^{kij} $ are determined by those of 
$V^{ij}$.

 Let us show how Mezincescu's prepotential for the vector multiplet can be introduced 
within standard superspace.
For this a simple generalisation of  the rigid supersymmetric analysis in \cite{HST} can be used.
One begins with the first-order  action
\begin{align}
S &= \frac{1}{4} \int \rd^4 x \,{\rm d}^4\q \,\cE\, \cW \cW
+    {\rm c.c.}
     - \frac{\rm i}{8} \int \rd^{4|8} z \,
\,E\, \Big(\cW \nabla^{ij}  V_{ij} - \bar \cW \bar \nabla^{ij}  V_{ij}\Big)~,
\label{F1}
\end{align}
where $\cW$ is a covariantly chiral superfield, and $V^{ij}=V^{ji} $ is an {\it unconstrained} real
SU(2) triplet  acting as a Lagrange multiplier. 
Varying (\ref{F1}) with respect to $V_{ij}$ gives $\cW =W$, where $W$ obeys the Bianchi identity \eqref{ReducedChiral.b}.
As a result, the second term in (\ref{F1}) drops out and we end up with the 
$\cN=2$ super-Maxwell action
\bea
S = \frac{1}{4} \int \rd^4 x \,{\rm d}^4\q \,\cE\, W W +{\rm c.c.}
\eea
On the other hand,
because the action (\ref{F1}) is quadratic in $\cW$,
we may easily integrate $\cW$ out using its equation of motion
\begin{align}\label{eq_WMezincescu}
\cW =  {\rm i} W_{\rm D}~,  \qquad W_{\rm D}:= \frac{1}{4}\bar{\nabla}^4 \nabla^{ij}  V_{ij} ~.
\end{align}
This leads to the dual action
\begin{align}
S = \frac{1}{4} \int \rd^4 x \,{\rm d}^4\q \,\cE\, W_{\rm D} W_{\rm D} +{\rm c.c.}
\end{align}
The dual field strength  $W_{\rm D}$ must be both reduced chiral and given by \eqref{eq_WMezincescu}.

Within the curved projective-superspace approach of \cite{KLRT-M1,KLRT-M2,K2008}, the constraints on
$W$ can be solved  in terms of a covariant  real weight-0 tropical prepotential $V(v^i)$, 
$\breve{V} =V$. 
The solution  \cite{KT-M2009} is
\bea
&&W  = \frac{1}{8\pi}  \oint_\g (v, \rd v)
 \bar \nabla^{(-2)} V(v)~, 
\quad  \bar \nabla^{(-2)} := \frac{1}{(v,u)^2}u_{i}u_{j} \bar \nabla^{ij}~. 
\label{6.8}
\eea
where $\g$ is an appropriately chosen contour.
We recall that  $v^i \in {\mathbb C}^2 \setminus \{0\}$ denotes
the homogeneous coordinates for ${\mathbb C}P^1$.
The right-hand side of the expression for $W$  involves a constant isotwistor $u_i$, which is chosen to obey the constraint $(v,u):= v^i u_i \neq 0$, but otherwise is completely arbitrary. 
Using the analyticity constraints  \eqref{3.29} obeyed by $V$,
one can check that $W$ is invariant 
under arbitrary {\it projective transformations} \eqref{projectiveGaugeVar}.
The field strength \eqref{6.8} proves to be  invariant under gauge transformations 
\bea
V  \to V+ \l + \breve{\l}~,
\label{89}
\eea
where the gauge parameter $\l(v)$ is a covariant weight-0 arctic multiplet.

It is worth discussing how  the Mezincescu prepotential $V_{ij}$ emerges 
within projective superspace, see \cite{Butter:2010jm} for more details.  
One begins with the expression for
$W$ in terms of   $V(v)$, eq. \eqref{6.8}. 
In accordance with eq. \eqref{iso3},
the analyticity conditions on $V$ may be solved 
in terms of  an unconstrained isotwistor superfield $U^{(-4)}$,
which is real under smile-conjugation
\begin{align}
V(v) &= \frac{1}{16} \bar \nabla^{(2)} \nabla^{(2)}
U^{(-4)} (v)
 =
     \frac{1}{16} 
\nabla^{(2)}     \bar \nabla^{(2)} 
     U^{(-4)} (v)~.
\end{align}
Using this construction, one may rewrite $W$ as
\begin{align}
W &= \frac{1}{128\pi} \oint_\g (v, {\rm d}v)
\bar \nabla^{(-2)}
\bar \nabla^{(2)} \nabla^{(2)}
     U^{(-4)} 
     =
     \frac{1}{8\pi} \bar \nabla^4 \oint_\g (v, {\rm d}v) 
   \nabla^{(2)}
     U^{(-4)}
\end{align}
where 
the chiral operator $\bar \nabla^4$ is defined in \eqref{3.25}.
This may be rewritten as
\begin{align}\label{e.7}
W &= \frac{1}{8\pi} \, \bar \nabla^4
\nabla^{ij}
\,\oint_\g (v ,{\rm d}v) \, v_i v_j\, U^{(-4)}(v)
   =  \frac{1}{4} \bar \nabla^4 \nabla^{ij}
   V_{ij}~,
\end{align}
where we have defined the Mezincescu prepotential
\begin{align}
V_{ij} = \frac{1}{2\pi} \oint_\g (v, {\rm d}v) \, v_i v_j\, U^{(-4)} (v)~.
\end{align}

Given a system of $n$ Abelian vector multiplets with chiral field strengths $W_I$, 
let $\cF (W)$ be a holomorphic function of degree $+2$, 
\bea
W_I \frac{\pa}{\pa W_I} \cF(W) = 2 \cF(W)~.
\eea
Then the following action
\bea
S =  \int \rd^4 x \,{\rm d}^4\q \,\cE\, \cF (W) +{\rm c.c.}
\label{815}
\eea
is locally superconformal. The component reduction of this model was described by Butter and Novak \cite{Butter:2012xg}, and their results agree with \cite{deWit:1984rvr}.
The model has led to the notion of {\it special K\"ahler geometry} \cite{deWit:1984wbb}, see \cite{FVP} for a review. A rigid supersymmetric limit of \eqref{815} corresponds to {\it rigid special K\"ahler geometry} \cite{ST,Gates:1983py}.

%%%%%%%%%%%%%%%%%%%%%%%%%%%%%%%%%%%%%
%%%%%%%%%%%%%%%%%%%%%%%%%%%%%%%%%%%%%

\subsection{Tensor multiplet}

In rigid supersymmetry, the massless $\cN=2$ tensor multiplet was introduced by Wess
\cite{Wess}. It was rediscovered by de Wit and van Holten \cite{deWit:1979xpv}.
The tensor multiplet can be described in conformal  superspace by
its gauge invariant field strength $G^{ij}$,  which is 
a real $\cO(2)$ multiplet.  It obeys the constraints 
\bea
\nabla^{(i}_\a G^{jk)} =  {\bar \nabla}^{(i}_\ad G^{jk)} = 0~,
\eea
which generalise those given in \cite{BS1,BS2,SSW}.
These constraints are solved in terms of a chiral
prepotential $\Psi$ with the superconformal properties
\bea
K^A \J=0~,  \quad {\mathbb D} \J = \J~,\quad
\bar \nabla ^\ad_i \J= 0 ~.
\eea
The solution to the tensor multiplet constraints was given in 
\cite{HST,GS82,Siegel83,Muller86}. In conformal superspace the solution is
\begin{align}
\label{eq_Gprepotential}
G^{ij} = \frac{1}{4}\nabla^{ij} \Psi
+\frac{1}{4}\bar \nabla^{ij} {\bar \Psi}~.
\end{align}
The chiral prepotential is invariant under gauge  transformations
\bea
\Psi \rightarrow \Psi + \ri \Lambda~,
\label{819}
\eea
where the gauge parameter $\Lambda$ is a {\it reduced chiral} superfield with the properties \eqref{ReducedChiral}.

Consider a system of $(n+1)$ tensor multiplets, $n>0$, and let $G^{(2)}_I$
be  their gauge-invariant field strengths,  $I=0,1,\dots, n$. 
Its dynamics  can be described by  a Lagrangian of the form
\bea
\cL^{(2)} = \cL (G^{(2)}_I)~, \qquad G^{(2)}_I \frac{\pa }{\pa G^{(2)}_I} \cL =\cL~
\eea
where $\cL$ is a real homogeneous function of degree  $+1$.
Of special significance is the special choice of $\cL$ defined by the Lagrangian
\be
\cL^{(2)} = \frac{1}{{\rm i} \,G^{(2)}_0} \Big( \cF( G^{(2)}_{ I}) 
- \bar{\cF} ( G^{(2)}_{ I}) \Big)~.
\label{cmap1}
\ee
Here 
and $\cF(z^I)$  is a holomorphic homogeneous function of second degree, 
$\cF({\mathfrak c}\,z^I)= {\mathfrak c}^2\cF(z^I)$. 
This model provides a manifestly supersymmetric description of the c-map
 \cite{cmap1,cmap2}.
The rigid c-map is described by the model \eqref{328}.

For a single tensor multiplet there is only one superconformal model, which is described by the Lagrangian
\bea
\cL^{(2)}_{\rm IT}=  -{G}^{(2)}  \ln \frac{{ G}^{(2)}}{{\rm i} \U^{(1)} \breve{\U}{}^{(1)}}      
~,
\label{8222}
\eea
with  $\U^{(1)}$ a weight-one arctic multiplet
(both $\U^{(1)}$ and its smile-conjugate $\breve{\U}{}^{(1)}$ are pure gauge degrees of freedom). It describes an improved tensor multiplet.
Historically, the improved tensor multiplet was 
independently constructed in the following works (submitted to the journal Nuclear Physics within a one day time difference): (i) Ref.  \cite{KLR} provided its construction in terms of $\cN=1$ superfields in the rigid supersymmetric case; and (ii)  and Ref. \cite{deWPV} proposed this multiplet within the $\cN=2$ superconformal tensor calculus. 
The rigid supersymmetric version of \eqref{8222} was proposed in the first projective-superspace paper \cite{KLR}.

The improved tensor multiplet can be coupled to weight-0 polar hypermultiplets.
The corresponding locally superconformal $\s$-model \cite{K2008} is
\bea
\cL^{(2)} = - {G}^{(2)}   \ln \frac{{ G}^{(2)}}{{\rm i}  \breve{\U}{}^{(1)}  \U^{(1)}   }       
+ G^{(2)}  K (\U , \breve \U) ~, 
\label{8223}
\eea
where the K\"ahler potential is the same as in \eqref{nozeta}.
The rigid supersymmetric limit of this $\s$-model was studied in \cite{KLvU}. 
The above $\s$-model has a dual formulation in terms of polar hypermultiplets:
\bea
\cL^{(2)} =   {\rm i}  \breve{\U}{}^{(1)}  \U^{(1)}   \re^{ K (\U , \breve \U ) }  ~.
\label{8224}
\eea  
The locally $\cN=2$ superconformal $\s$-models \eqref{8223} and \eqref{8224} have a striking resemblance to their $\cN=1$ counterparts, see e.g. \cite{FGKV}.

%%%%%%%%%%%%%%%%%%%%%%%%%%%%%%%%%%%%%%
%%%%%%%%%%%%%%%%%%%%%%%%%%%%%%%%%%%%%%

\subsection{Linear multiplet action} 

The linear multiplet action is a BV-type superconformal invariant based on the Lagrangian
\bea
\cL^{(2)} = V G^{(2)}~.
\eea
There are three equivalent representations for the linear multiplet action: 
\bea
S[V G^{(2)} ]= \int \rd^4 x \,{\rm d}^4\q \, \cE \, W \J +{\rm c.c.} 
= \int \rd^{4|8} z \,     E \, V_{ij}G^{ij}
~.
\label{821}
\eea
The action is invariant under the  gauge transformations for the vector and tensor multiplets. The invariance under \eqref{89} follows from the identity 
\bea
S[ ( \l + \breve{\l}) G^{(2)} ] = 0~,
\eea
where $\l$ is an arctic multiplet. The invariance under \eqref{819} follows from the Bianchi identity  \eqref{ReducedChiral.b}.

We have seen that every chiral action can be represented as a projective action, eq. 
\eqref{720}. On the other hand every projective action can be recast as a chiral action, 
eq. \eqref{721}.  These results show that the linear multiplet action \eqref{821} is universal.

%%%%%%%%%%%%%%%%%%%%%%%%%%%%%%
%%%%%%%%%%%%%%%%%%%%%%%%%%%%%%

\subsection{Composite reduced chiral superfields}

The above discussion has an important implication. 
Given a composite real weight-0 tropical multiplet ${\mathbb V}$, the following descendant 
\bea
{\mathbb W}  &=& 
 \frac{1}{8\pi}  \oint_\g (v, \rd v)
 \bar \nabla^{(-2)} {\mathbb V}(v)
 \label{822}
 \eea
 is a primary reduced chiral superfield, with $\g$ being an appropriately chosen contour. This observation has been used in \cite{Butter:2010jm} to derive a number of composite reduced chiral superfields.
 
Our first example is 
\bea
 {\mathbb V}=  \ln \frac{{ G}^{(2)}}{{\rm i} \U^{(1)} \breve{\U}{}^{(1)}}      ~.
 \eea
It can be seen that  the arctic multiplet $\U^{(1)}$ and its conjugate $\breve{\U}{}^{(1)}$
do not contribute to the contour integral, and so they will be ignored below.
Evaluating the contour integral in \eqref{822} gives
\bea
\mathbb W := -\frac{G}{8} \bar \nabla_{ij} \left(\frac{G^{ij}}{G^2} \right) ~,
\qquad G:=\sqrt{\frac{1}{2} G^{ij}G_{ij}}~,
\label{824}
\eea
see  \cite{Butter:2010jm}  for the technical details. 
This composite multiplet was discovered originally (in a different but equivalent form) 
in \cite{deWPV} using the superconformal tensor calculus.
It was  later reconstructed  in curved superspace by M\"uller
 \cite{Muller86} with the aid of the results in  \cite{deWPV} and \cite{Siegel85}.
 Its contour origin was explored in the globally supersymmetric case  by Siegel \cite{Siegel85}. 

Our second example is 
\bea
{\mathbb V}= \frac{H^{(2n)}}{(G^{(2)})^{n}} ~,
\eea
where $H^{(2n)}$ is a real $\cO(2n)$ multiplet, see eqs. \eqref{54} and \eqref{55}. 
Evaluating the contour integral in \eqref{822} gives
\begin{align}\label{eq_In}
\mathbb W_n
     &= -\frac{(2n)!}{2^{2n+2}\, (n+1)! (n-1)!}\,
     {G} \, \bar \nabla_{ij} \mathcal R^{ij}_n~,
\end{align}
where
\begin{align}\label{eq_O2nR}
\mathcal R_n^{ij} =
     \left(\delta^{ij}_{kl} - \frac{1}{2 {G}^2} {G}^{ij} {G}_{kl} \right)
     H^{kl \,i_1 \cdots i_{2n-2}}
      {G}_{i_1 i_2} \cdots {G}_{i_{2n-3} i_{2n-2}}  {G}^{-2n} ~.
\end{align}
The expression for $\mathbb W_n$ has an overall structure similar to
 \eqref{824}, except the argument
$\mathcal R_n^{ij}$ of the derivative is much more complicated.

%%%%%%%%%%%%%%%%%%%%%%%%%%%%%%%%
%%%%%%%%%%%%%%%%%%%%%%%%%%%%%%%%%

\section{Off-shell formulations for supergravity} 

Within the conformal approach to locally supersymmetric theories \cite{KakuTownsend}, Poincar\'e and AdS supergravity  may be realised as conformal supergravity coupled to a compensating multiplet. Two compensating massless multiplets are typically required in the case of $\cN=2$ supergravity, see \cite{deWPV,FVP}
for comprehensive discussions. 
In this section we describe two off-shell formulations for $\cN=2$ supergravity. 

%%%%%%%%%%%%%%%%%%%%%%%%%%%%%%%%%%

\subsection{Supergravity with vector and tensor multiplet compensators}

The minimal formulation for $\cN=2$  supergravity with vector and tensor 
compensators \cite{deWPV} admits a simple superspace description. 
Using the techniques developed above, 
the gauge-invariant supergravity action can be written as 
\bea
S_{\rm SUGRA}  &=& \frac{1}{ \k^2} \int \rd^4 x \,{\rm d}^4\q \, \cE \, \Big\{
\J {\mathbb W} - \frac{1}{4} W^2 +\x \J W \Big\}          +{\rm c.c.}   \non \\
 &=& \frac{1}{ \k^2} \int \rd^4 x \,{\rm d}^4\q \, \cE \, \Big\{
\J {\mathbb W} - \frac{1}{4} W^2 \Big\} +{\rm c.c.}
     ~+~ \frac{\x}{ \k^2} \int \rd^{4|8} z
      \,
     E \, G^{ij}V_{ij}~,~~~~~~~
\eea
where $\k$ is the gravitational constant, $\x$ the cosmological constant,  
$\mathbb W$ is given by the expression \eqref{824},
 and
$V_{ij}$ is the Mezincescu prepotential.
Within the projective-superspace approach of \cite{KLRT-M1,K2008,KLRT-M2}, 
this action is equivalently given by (\ref{InvarAc})
with the following Lagrangian \cite{K2008}
\bea
\k^2 \,\cL^{(2)}_{\rm SUGRA} =       {G}^{(2)}  \ln \frac{{ G}^{(2)}}{{\rm i} \U^{(1)} \breve{\U}{}^{(1)}}      
-       \hf {V}{\S}^{(2)} +\x V G^{(2)}~,
\label{6.14}
\eea
with $V$ the tropical prepotential for the vector multiplet, and $\U^+$ a weight-one arctic multiplet
(both $\U^{(1)}$ and its smile-conjugate $\breve{\U}{}^{(1)}$ are pure gauge degrees of freedom).
The fact that the vector  and the tensor  multiplets are compensators 
means that their field strengths  $W$ and $G^{ij}$ should possess non-vanishing  expectation values, that is 
$W\neq 0$ and $ G\equiv\sqrt{\frac{1}{2} G^{ij}G_{ij}} \neq 0$.

The equation of motion for the gravitational superfield \cite{BK10,Butter:2010jm} is
\begin{subequations} 
\bea
G-W \bar W= 0~,
\label{6.10}
\eea
and it is consistent with the conditions $W\neq 0$ and $ G\neq 0$.
The equations of motion for the compensators are \cite{Butter:2010jm}
\bea
\S^{ij} -\x G^{ij} &=&0 ~, 
\label{6.14a}\\
\mathbb W + \x W &=& 0~.
\label{6.15b}
\eea
\end{subequations}

The equations \eqref{6.14a} and \eqref{6.15b} can be degauged to $\sU(2)$ superspace, which results in the following equations
\bsubeq\label{6.15-degauged}
\bea
\frac{1}{4} \Big(
 \mathfrak{D}^{ij}
+4 S^{ij}
\Big)W
=\frac{1}{4}
\Big(
 \bar{\mathfrak{D}}^{ij}
+4\bar{S}^{ij}
\Big)\bar{W}
&=&\x G^{ij}
~, \\
\frac{G}{8}\big(
\bar{\mathfrak{D}}^{ij}
+4\bar{S}^{ij}
\big)
\frac{G_{ij}}{G^2}
&=&
\x W  
~.
\eea
\esubeq
The super-Weyl and local $\sU(1)_R$ gauge freedom can be used to impose the gauge condition $W=\bar{W}=1$.  The integrability conditions of these constraints are $\mathfrak{D}_AW=\mathfrak{D}_A\bar{W}=0$ which imply
\bea
G_{\a\ad}{}^{ij}
=0
~,~~~~~~
\Phi_a=G_a
~.
\eea
After employing the redefinitions $\cD_\a^i = \mathfrak{D}_\a^i$ and $\cD_a=\mathfrak{D}_a-\ri \,G_a \,{\mathbb Y}$, the resulting geometry coincides with $\sSU(2)$ superspace for which the $\sU(1)_R$ connection is pure gauge and can be set to zero.
The equation of motion \eqref{6.10} implies $G=1$ and $\cD_AG=0$. The latter can be shown to imply the condition $\cD_AG^{ij}=0$, which breaks the local $\sSU(2)_R$ to a residual $\sU(1)$ subgroup. It  consists of those transformations that keep $G^{ij}$ invariant. 
Integrability of the constraint  $\cD_AG^{ij}=0$
implies
\bea
Y_{\a\b}=0~,~~~G_{\a\ad}=0~,~~~S_{(i}{}^kG_{j)k}=0~,~~~\bar{S}_{(i}{}^kG_{j)k}=0
~.
\eea
The remaining supergravity equations \eqref{6.15-degauged} turn into 
\bea
S^{ij}=\bar{S}^{ij}=\x G^{ij}
~,~~~
S^2:=\frac{1}{2}S^{ij}S_{ij}=\x^2
~,~~~
\cD_A S^{ij}=0
~.
\label{6.15-degauged-2}
\eea
All the remaining information about the dynamics of supergravity is encoded in the super-Weyl tensor $W_{\a\b}$. 

A maximally supersymmetric solution of \eqref{6.15-degauged-2} 
 is characterised by the condition $W_{\a\b} =0$ and the resulting superspace geometry is uniquely determined to be 
\begin{subequations}
\bea
\{\cD_\a^i,\cD_\b^j\}&=&
4S^{ij}M_{\a\b}
+2 \ve_{\a\b}\ve^{ij}S^{kl}J_{kl}~,
\quad
\{\cD_\a^i,\cDB^\bd_j\}=
-2\ri\d^i_j(\s^c)_\a{}^\bd\cD_c
~,~~~~
\label{AdS-N2-1}
\\
{[}\cD_a,\cD_\b^j{]}&=&
{\ri\over 2} ({\s}_a)_{\b\gd}S^{jk}\cDB^\gd_k~,
\qquad \qquad \quad ~~~
[\cD_a,\cD_b]= - S^2
M_{ab}~.~
\label{AdS-N2-2}
\eea
\end{subequations} 
This geometry corresponds to the four-dimensional  $\cN=2$ AdS superspace 
\bea
{\rm AdS}^{4|8} = \frac{{\rm OSp}(2|4)}{{\rm SO}(3,1) \times {\rm SO} (2)}~.
\eea
The most general $\cN=2$ supersymmetric nonlinear $\s$-models in ${\rm AdS}_4$ were studied 
in \cite{KT-M-4D-conf-flat,BKsigma1,BKsigma2,BKLT-M}. They have important distinct features as compared with the $\cN=2$ supersymmetric nonlinear $\s$-models in 
Minkowski space. Specifically, the target space must be a non-compact hyperk\"ahler manifold endowed with a Killing vector field which generates an
$\sSO(2)$ group of rotations on the two-sphere of complex structures.

As a generalisation of \eqref{6.14}, we consider the model for matter-coupled supergravity \cite{K2008} 
\bea
\k^2 \cL^{(2)} =       -   \hf {V}{\S}^{(2)} 
+ {G}^{(2)} \Big(  \ln \frac{{ G}^{(2)}}{{\rm i}  \breve{\U}{}^{(1)}  \re^{-\x V} \U^{(1)}   }       
+ \k^2 K (\U^I , \breve \U^{\bar J} ) \Big)~,
\label{matter-coupled-SUGRA}
\eea
where $K(\F, \bar \F)$ is the K\"ahler potential of a K\"ahler manifold, and 
$\U^I$ are covariant weight-0 arctic multiplets.

%%%%%%%%%%%%%%%%%%%%%%%%%%%%%%%%%%%%

\subsection{Supergravity with vector and hyper multiplet compensators}

The compensators for this supergravity formulation are a vector multiplet and a polar hypermultiplet. The Lagrangian has the form 
\bea
\k^2 \,\cL^{(2)}_{\rm SUGRA} =          
-       \hf {V}\,{\S}^{(2)} - \ri \,\breve{\U}{}^{(1)} \re^{-\x V} \U^{(1)}~,
\eea 
with $\x$ a cosmological constant. The action is invariant under the gauge transformations
\bea
\d_\l V=  \l + \breve{\l}~, \qquad \d_\l \U^{(1)}=  {\x}\l  \U^{(1)}~.
\eea
The equation of motion for $\breve{\U}{}^{(1)} $ implies that 
\bea
 \re^{- \x V_+(\z) } \U^{(1)} (v) = \U^i  v_i ~,\qquad V_+ (\z) = \hf V_0 +\sum_{k=1}^\infty V_k\z^k~,
 \eea
 where $\U^i$ obeys the equations 
 \bea
 \underline{\nabla}_\a^{(i} \U^{j)} = 0 ~,\qquad   \underline{\bar \nabla}_\ad^{(i} \U^{j)} = 0 ~,
\eea
which defines the on-shell Fayet-Sohnius hypermultiplet. Here $\underline{\nabla}_A$ denotes the gauge and conformal covariant derivative, which is obtained form $\nabla_A$ by adding a $\sU(1)$ connection.
We point out that $V (\z) = V_+ (\z) + V_-(\z)$, where $V_-(\z)$ is the smile-conjugate of $V_+(\z)$.
The equation of motion for $V$ is 
\bea
{\S}^{(2)} - \x \ri \,\breve{\U}{}^{(1)} \re^{-\x V} \U^{(1)} =0 \quad \Longleftrightarrow \quad
\S^{ij} +\x \ri  \bar \U^{(i} \U^{j)}=0~.
\eea
Finally, the equation of motion for the gravitational superfield $H$ is (see \cite{KT,BK11}
for the derivation) 
\bea
\bar W W - \hf {\bar \U}_i \U^i =0~.
\eea

The supergravity-matter system \eqref{matter-coupled-SUGRA}
has a dual formulation \cite{K07} described by the Lagrangian
\bea
\k^2 \cL^{(2)} =       -   \hf {V}{\S}^{(2)} 
- {{\rm i}  \breve{\U}{}^{(1)}  \re^{-\x V -\k^2 K (\U , \breve \U ) }  \U^{(1)}   }       ~.
\eea
If the cosmological constant vanishes, $\x=0$, this supergravity-matter system turns into the one introduced in \cite{KLRT-M1}.

%%%%%%%%%%%%%%%%%%%%%%%%%%%%%%%%%%
%%%%%%%%%%%%%%%%%%%%%%%%%%%%%%%%%%

\section{Conformal supergravity, topological invariants and super-Weyl anomalies}

In this section we describe a powerful formalism to generate locally superconformal higher-derivative invariants developed in \cite{BdeWKL}. Its applications include the superfield construction of the $\cN=2$ Gauss-Bonnet  term and the general structure of super-Weyl anomalies in $\cN=2$ superconformal field theories. To start with, we review the $\cN=2$ conformal supergravity theory. 

%%%%%%%%%%%%%%%%%%%%%%%%%%%%%%

\subsection{Conformal supergravity} 

The action for $\cN=2$ conformal supergravity \cite{BdeRdeW} is 
\bea
S_{\rm CSG}  = \frac 14 \int \rd^4x 
\rd^4 \q \, {\mathcal{E}}	 \, W^{\a\b} W_{\a\b} + \text{c.c.}
\label{10.1}
\eea
The corresponding equation of motion is 
\bea
\nabla_{\a \b} W^{\a\b}
	= \bar \nabla^{\ad \bd} \bar W_{\ad \bd} =0
\label{10.2}
\eea
and states that the super-Bach multiplet \eqref{super-Bach} vanishes. 
This equation is obtained by varying $S_{\rm CSG}  $ with respect to a gravitational superfield $H = \bar H$ which is the only unconstrained prepotential of $\cN=2$ conformal supergravity modulo purely gauge degrees of freedom, see the discussions in \cite{KT,BdeWKL} and references therein. Performing this variation, we find 
\bea
\d_H \int \rd^4x 
\rd^4 \q \, {\mathcal E} \, W^{\a\b} W_{\a\b}  = 2 \int \rd^{4|8}z\, E\, \d H \nabla_{\a \b} W^{\a\b}~,
\label{10.3}
\eea
where the variation $\d H$ is a real primary superfield of dimension $-2$.
Since the Bach multiplet $\mathfrak B$, eq. \eqref{super-Bach},  and the variation $\d H$  are real, the functional 
\bea
{\mathfrak P} =-\frac{\ri}{2}  \int \rd^4x 
 \rd^4 \q \, {\mathcal{E}} \, W^{\a\b} W_{\a\b} + \text{c.c.}
\eea
is a topological invariant being proportional to the Pontryagin term. 
As a consequence of \eqref{ConsEq}, the right-hand side of \eqref{10.3} is invariant under gauge transformations of the form \cite{KT,BK10,BdeWKL}
\bea 
\d_\O H = \frac{1}{12} \nabla^{ij} \O_{ij} + \frac{1}{12} \bar \nabla_{ij} \bar \O^{ij} ~,
\label{10.5}
\eea
where the complex gauge parameter $\O_{ij}= \O_{ji}$ is unconstrained and 
has the superconformal properties
\bea
K^A  \O_{ij}= 0~, \qquad {\mathbb D}\O_{ij}= - 3\O_{ij}~, \qquad 
{\mathbb Y} \O_{ij} =-2 \O_{ij}~.
\eea
This gauge invariance expresses the fact that the action \eqref{10.1} is locally superconformal.

Any conformally flat superspace, 
$W_{\a \b} = 0$, is a solution of the equation \eqref{10.2}.
It is instructive to linearise the conformal supergravity action \eqref{10.1} 
about such a background.
From the linearised prepotential $H$, we construct the linearised super-Weyl tensor
\begin{align}
\mathfrak{W}_{\a \b} = \bar{\nabla}^{4} \nabla_{\a \b} H ~,
\label{10.6}
\end{align}
which is primary, $ K^{C} \mathfrak{W}_{\a \b} = 0$, and covariantly chiral, $\bar{\nabla}^{\ad}_i \mathfrak{W}_{\a \b} = 0$. It proves to be  invariant under the gauge transformations \eqref{10.5}, $\d_{\O} \mathfrak{W}_{\a \b} = 0$, and obeys the Bianchi identity
\begin{align}\label{linBI}
\nabla^{\a \b} \mathfrak{W}_{\a \b} = \bar{\nabla}^{\ad \bd} \bar{\mathfrak{W}}_{\ad \bd}~.
\end{align}
Thus, the action for linearised conformal supergravity is simply
\begin{align}\label{10.8}
S_{\rm LCSG}  = \frac 14 \int \rd^4x\rd^4 \q \, {\mathcal{E}}\, \mathfrak{W}^{\a\b} \mathfrak{W}_{\a\b} + \text{c.c.}
\end{align}
If the background superspace is flat, the field strength \eqref{10.6} 
reduces to that described in \cite{HST}, and 
the action \eqref{10.8} turns into  the one given in \cite{BdeRdeW,HST}.

The model \eqref{10.8} is known to possess $\sU(1)$ duality invariance \cite{KR21-2}. 
The formalism of $\sU(1)$ duality rotations has been used \cite{KR21-2} 
to construct nonlinear extensions of \eqref{10.8}.

%%%%%%%%%%%%%%%%%%%%%%%%%%%%%%%%%%%%%

\subsection{Logarithm construction and  the Gauss-Bonnet invariant}

We now turn to describing 
the logarithm construction of \cite{BdeWKL} and its use in defining the $\cN=2$  
supersymmetric extension of the Gauss-Bonnet term. 

Let $\bar{\Phi}$ be a primary antichiral scalar  with the superconformal properties:
\begin{align}
	K^{A} \bar{\Phi} = 0 ~, \quad \nabla_\a^i \bar{\Phi} = 0~, \quad \mathbb{D} \bar{\Phi} = w \bar \F \quad \implies \quad \mathbb{Y} \bar{\Phi} = 2w\bar \F ~,
\end{align}
where $w \neq 0$, but it is otherwise arbitrary. We assume $\bar \F $ to be nowhere vanishing such that ${\bar \F}^{-1} $ exists. 
Then, 
it may be shown that 
$\bar{\nabla}^4 \, \text{ln} \, \bar{\Phi}$ is a primary chiral superfield of dimension 2,
\begin{align}
	K^A \bar{\nabla}^4 \, \text{ln} \, \bar{\Phi} = 0 ~, \quad \bar{\nabla}^{\ad}_i \bar{\nabla}^4 \, \text{ln} \, \bar{\Phi} = 0~, \quad \mathbb{D} \bar{\nabla}^4 \, \text{ln} \, \bar{\Phi} = 2 \bar{\nabla}^4 \, \text{ln} \, \bar{\Phi} ~.
\end{align}

By following the degauging procedure to $\sU(2)$ superspace, which was detailed in section \ref{DegaugingSection}, it may be shown that
\begin{align}
	\label{9.7}
	\bar{\nabla}^{4} \, \text{ln} \, \bar{\Phi} = \bar{\D} \, \text{ln} \, \bar{\Phi} + \frac{w}{2} \Big (  \bar{Y}^{\ad \bd} \bar{Y}_{\ad \bd} + \bar{S}_{ij} \bar{S}^{ij} + \frac 1 6 \bar{\mathfrak{D}}_{ij} \bar{S}^{ij} \Big ) \equiv  \bar{\D} \, \text{ln} \, \bar{\Phi} + \frac{w}{2} \X
	 ~,
\end{align}
where $\bar{\D}$ denotes the chiral projecting operator \eqref{6.17}.
It is important to note that since
$\bar{\nabla}^{4} \, \text{ln} \, \bar{\Phi} $ and $\bar{\D} \, \text{ln} \, \bar{\Phi}$
 are both manifestly chiral, $\X$ shares this property
\begin{align}
\bar{\mathfrak{D}}^{\ad}_i \X = 0 ~.
\end{align}
At the same time, we emphasise that while the right hand side of \eqref{9.7} is primary, each individual term possesses an inhomogeneous contribution under the super-Weyl transformations of $\sU(2)$ superspace
\begin{align}
	\d_{\S} \bar{\D} \, \text{ln} \, \bar{\Phi} = 2 \S \bar{\D} \, \text{ln} \, \bar{\Phi} + w \bar{\D} \S ~, \qquad \d_{\S} \X = 2 \S \X
	 - 2 \bar{\D} \S ~.
\end{align}
In the case of $\sSU(2)$ superspace, these transformation laws turn into 
\bea
\d_{\s} \bar{\D} \, \text{ln} \, \bar{\Phi} = 2 \s \bar{\D} \, \text{ln} \, \bar{\Phi} + w \bar{\D} \s ~, \qquad \d_{\s} \X = 2 \s \X
	 - 2 \bar{\D} \s ~.
\eea
Our analysis leads to an important conclusion. Specifically, for every primary dimensionless chiral scalar $\J$, the following functional 
\bea
 \int \rd^4 x \,{\rm d}^4\q \, \cE \, \J \bar{\nabla}^{4} \, \text{ln} \, \bar{\Phi} 
= \int \rd^{4|8} z\, E \, \J \ln \bar \F 
 + \frac{w}{2}  \int \rd^4 x \,{\rm d}^4\q \, \cE \, \J \X
 \label{11.16}
 \eea
 is locally superconformal. Here the expression on the right is given in $\sU(2) $ superspace (its form is preserved upon degauging to $\sSU(2)$ superspace).

In Ref.  \cite{BdeWKL} 
the superconformal chiral action
\begin{align}
	\label{chiralAction}
	S_\c^- = -  \int \rd^4 x \,{\rm d}^4\q \, \cE \, \big(W^{\a \b} W_{\a \b} - 2 w^{-1} \bar{\nabla}^4 \, \text{ln} \, \bar{\Phi} \big) 
\end{align}
was identified with the $\cN=2$ Gauss-Bonnet topological invariant. More precisely, 
it may be shown that, at the component level, $S_\c^-$  is a combination 
of the Gauss-Bonnet and Pontryagin invariants. Under suitable boundary conditions on $\bar \F$, the functional \eqref{chiralAction} proves to be independent of $\bar \F$. 
This follows from \eqref{11.16} in conjunction with the identity in $\sU(2)$ superspace
\bea
{\bar {\mathfrak D}}^{\ad}_i  \s=0
\quad \Longrightarrow \quad 
\int {\rm d}^{4|8}z \,E\, \s=0~,
\eea
for any covariantly chiral scalar $\s$. 
Therefore, we obtain
\begin{align}
	\label{GB}
	S_\c^- 
	&= - \int \rd^4 x \,{\rm d}^4\q \, \cE \, \big(W^{\a \b} W_{\a \b} - 
	\X
	 \big) ~.
\end{align}
The topological nature of \eqref{chiralAction} was established 
in  \cite{BdeWKL} at the component level. A solid superspace proof is still absent. 

%%%%%%%%%%%%%%%%%%%%%%%%%%%%%%%

\subsection{Super-Weyl anomalies}

Consider a superconformal field theory coupled to 
 supergravity. The classical action 
of such a theory is invariant under the super-Weyl transformations, and it 
is independent of the supergravity compensators. 
In other words, the superconformal field theory couples to the Weyl multiplet.

In the quantum theory, integrating out the matter fields leads to an effective action 
that is no longer a functional of the Weyl multiplet only. 
There are two different contributions to the $\cN=2$ super-Weyl anomaly. 
One of them is given in terms of the supergravity multiplet. 
In the framework of $\sSU(2)$ superspace, the super-Weyl variation of the effective action $\G$ has the form
\cite{K2013}
\bea
\d_\s \G = (c-a) \int \rd^4x\, \rd^4\q\, \cE\, \s W^{\a\b}W_{\a\b} 
+ a \int \rd^4x\, \rd^4\q\, \cE\, \s \X ~+~{\rm c.c.} ~,
\label{11.20}
\eea
for some  anomaly coefficients $a$ and $c$. 
One can check that the super-Weyl variation \eqref{11.20}
obeys the Wess-Zumino consistency condition
\bea
(\d_{\s_1} \d_{\s_2} -\d_{\s_2} \d_{\s_1} ) \G=0~.
\eea
This property guarantees the existence of $\G$. 
The other sector of the $\cN=2$ super-Weyl anomaly
is determined by local couplings in a superconformal field theory.  
According to \cite{GHKSST,SchT2}, it is given by 
\bea
\d_\s \G = \int {\rm d}^{4|8}z  \,E\, \big(\s +\bar \s \big)
K(X, \bar X) ~,
\label{1022}
\eea
where the K\"ahler potential $K(X, \bar X) $ is the same as in \eqref{73}.
Since the chiral scalars $X^I$ are inert under the super-Weyl transformations, the anomaly clearly satisfies the Wess-Zumino consistency condition. 
The right-hand side of \eqref{1022} is not invariant under K\"ahler transformations 
However, the $\cN=2$ super-Weyl anomaly is invariant under a joint K\"ahler-Weyl transformation. A detailed analysis 
of the anomaly \eqref{1022} is given in the original publications
 \cite{GHKSST,SchT2}.

The super-Weyl anomaly \eqref{11.20} is generated by the $\cN=2$ {\it dilaton action} \cite{K2013}
\bea
S_{\rm D}  &=& \frac{1}{4} f^2\int \rd^4x\, \rd^4\q\, \cE\, \cZ^2 
+
\int \rd^4x\, \rd^4\q\, \cE\,  \Big\{  (c-a)\, W^{\a\b}W_{\a\b} 
+ a\,  \X \Big\}\ln \cZ   ~+~{\rm c.c.} \non \\
&&  +2 a \int {\rm d}^{4|8}z \,E\, \ln \cZ \ln \bar \cZ~.
\eea
where $f$ is a constant parameter, and $\cZ$ is the chiral field strength of a vector multiplet such that $\cZ^{-1}$ exists.  One may check that the super-Weyl variation $\d_\s S_{\rm D} $
coincides with the right-hand side of \eqref{11.20}.

%%%%%%%%%%%%%%%%%%%%%%%%%%%%%%%%%%%%%%%%%%%%%%%%%%%%%%%

\noindent
{\bf Acknowledgements:}\\ 
We thank S. James Gates Jr. and Konstantinos Koutrolikos for their kind invitation to contribute this chapter to the {\it Handbook of Quantum Gravity}.
The work of SK  is supported in part by the Australian 
Research Council, project No. DP200101944.
The work of ER is supported by the Hackett Postgraduate Scholarship UWA,
under the Australian Government Research Training Program. 
The work of GT-M is supported by the Australian Research Council (ARC)
Future Fellowship FT180100353, and by the Capacity Building Package of the University of Queensland.

%%%%%%%%%%%%%%%%%%%%%%%%%%%%%%%%%%%%%%%%%%%%%%%%%%%%%%%

\end{document}